%% file: root.tex
\documentclass{lmcs}
\pdfoutput=1

\usepackage[utf8]{inputenc}

\usepackage{lastpage}
\lmcsdoi{18}{1}{36}
\lmcsheading{}{\pageref{LastPage}}{}{}%
{Apr.~29,~2021}{Mar.~01,~2022}{}

\usepackage{isabelle,isabellesym}
\usepackage{latexsym}
\usepackage{amssymb}
\usepackage{framed}
\usepackage{mdframed}
\usepackage{mathtools}
\usepackage{pdfsetup}

\newenvironment{framedni}{\begin{framed}\parindent0pt}{\end{framed}}

\newenvironment{quoteni}%
{\list{}{\leftmargin=0mm\rightmargin=0mm}\item[]}%
{\endlist}

\isabellestyle{it}

\renewcommand{\isacharunderscore}{\_}
\renewcommand{\isacharunderscorekeyword}{\_}
\renewcommand{\isadigit}[1]{{\rm #1}}

\renewcommand{\isastyle}{\isastyleminor}

\allowdisplaybreaks

\begin{document}

\title{Verified Approximation Algorithms}

\titlecomment{This is a revised version of the conference publication
  \cite{ijcar/EssmannNR20}; the sections on Center Selection and Set
  Cover have been added.}

\author{Robin Eßmann\rsuper{a}}
\address{Technische Universit\"at M\"unchen, Germany}
\author{Tobias Nipkow\rsuper{a}}
\author{Simon Robillard\rsuper{b}}
\address{LIRMM, Université de Montpellier \& CNRS, Montpellier, France}
\author{Ujkan Sulejmani\rsuper{a}}

\begin{abstract}
We present the first formal verification of approximation algorithms
for NP-complete optimization problems:
vertex cover, independent set, set cover, center selection, load balancing, and bin packing.
We uncover incompletenesses in existing proofs and improve the approximation ratio in one case.
All proofs are uniformly invariant based.
\end{abstract}

\maketitle

\section{Introduction}

Approximation algorithms for NP-complete problems \cite{Vazirani} are a rich area of research
untouched by automated verification. We present the first formal verifications of
five classical and one lesser known approximation algorithm. Three of these algorithms
had been verified on paper by program verification experts \cite{BerghammerM03,BerghammerR03}.
We found that their claimed invariants need additional conjuncts before they are strong enough
to be real invariants. That is, their proofs are incomplete. The other
three algorithms only come with informal textbook proofs \cite{KleinbergT06}.

To put an end to this situation we formalized the correctness proofs of six
approximation algorithms for fundamental NP-complete problems
in the theorem prover Isabelle/HOL \cite{Concrete,LNCS2283}. We verified
(all proofs are online \cite{Approximation_Algorithms-AFP}) the following algorithms
and approximation ratios:
\begin{itemize}
\item the classic $k$-approxi\-mation algorithm for minimal vertex covers of rank $k$ hypergraphs;
\item Wei's \cite{Wei} $\Delta$-approximation algorithm for maximal independent sets of
  graphs with maximum degree $\Delta$;
\item the greedy $\frac32$ (resp.\ 2) approximation algorithm for the load balancing problem 
   of sorted (resp.\ unsorted) job loads;
\item the greedy 2-approximation algorithm for the center selection problem;
\item the greedy $H(d^*)$-approximation algorithm for set covers
  where $d^*$ is the size of the largest set and $H(n)$ is the $n$th harmonic number;
\item the $\frac32$-approxima\-tion algorithm for bin packing by Berghammer and Reuter \cite{BerghammerR03}.
\end{itemize}
All algorithms are expressed in a simple imperative \textit{WHILE}-language.
In each case we show that the approximation algorithm computes a valid solution
that is at most a certain factor worse than an optimum solution. The polynomial
running time of the approximation algorithm is easy to see in each case.

Isabelle not only helped finding mistakes in
pen-and-paper proofs but also encouraged proof refactoring
that led to simpler proofs, and in one case, to a stronger result:
The invariant given by Berghammer and Müller for Wei's
algorithm~\cite{BerghammerM03} is sufficient to show that the algorithm
has an approximation ratio of $\Delta + 1$. We managed to
simplify their argument significantly which led to an improved approximation ratio of $\Delta$.

Last but not least all our proofs are uniformly invariant-based.
This is in contrast to the three textbook proofs from \cite{KleinbergT06}
that rely on special \emph{ad hoc} arguments about the programs at hand.
We show that this is completely unnecessary and that invariant based proofs can even be simpler.

\input{Paper_VC.tex}

\input{Paper_MIS.tex}

\input{Paper_LB.tex}

\input{Paper_CS.tex}

\input{Paper_SC.tex}

\input{Paper_BP.tex}

\section{Conclusion and Future Work}

In this first application of theorem proving to approximation algorithms
we have verified a number of classical approximation algorithms for fundamental NP-complete problems,
have corrected purported invariants from the literature and could even strengthen the approximation ratio
in one case. Moreover we have given simple invariant-based proofs for algorithms where
the proofs in the literature (or their direct formalization) are more complicated.

Although we have demonstrated the benefits of formal verification of approximation algorithms, we have only scratched the surface of this rich theory. The next step is to explore the subject more systematically. As a large fraction of the theory of approximation algorithms is based on linear programming, this is a promising and challenging direction to explore. Some linear programming theory has been formalized in Isabelle already \cite{DBLP:conf/frocos/BotteschHT19,Linear_Programming-AFP}.
Approximation algorithms can also be formulated as relational programs, and verified accordingly. This approach was explored in~\cite{berghammer2016relational}, with some support from theorem provers, but has yet to be fully formalized.

\paragraph{Acknowledgement}
Tobias Nipkow is supported by DFG grant NI 491/16-1.

\bibliographystyle{alphaurl}
\bibliography{root}

\end{document}

%% file: Paper_VC.tex
\begin{isabellebody}%
\setisabellecontext{Paper{\isacharunderscore}{\kern0pt}VC}%
\isadelimtheory
\endisadelimtheory
\isatagtheory
\endisatagtheory
{\isafoldtheory}%
\isadelimtheory
\endisadelimtheory
\isadelimproof
\endisadelimproof
\isatagproof
\endisatagproof
{\isafoldproof}%
\isadelimproof
\endisadelimproof
\begin{isamarkuptext}%
\section{Isabelle/HOL and Imperative Programs}

Isabelle/HOL is largely based on standard mathematical notation but with some differences
and extensions.

Type variables are denoted by \isa{{\isacharprime}{\kern0pt}a}, \isa{{\isacharprime}{\kern0pt}b}, etc. The notation $t$~\isa{{\isacharcolon}{\kern0pt}{\isacharcolon}{\kern0pt}}~$\tau$
means that term $t$ has type $\tau$. Except for function types \isa{{\isacharprime}{\kern0pt}a\ {\isasymRightarrow}\ {\isacharprime}{\kern0pt}b},
type constructors follow postfix syntax, e.g.,
\isa{{\isacharprime}{\kern0pt}a\ set} is the type of sets of elements of type \isa{{\isacharprime}{\kern0pt}a}.
The image of a function \isa{f} over a set \isa{S} is denoted by \isa{f\ {\isacharbackquote}{\kern0pt}\ S}.
Function \mbox{\isa{some\ {\isacharcolon}{\kern0pt}{\isacharcolon}{\kern0pt}\ {\isacharprime}{\kern0pt}a\ set\ {\isasymRightarrow}\ {\isacharprime}{\kern0pt}a}} picks an arbitrary element from a set;
the result is unspecified if the set is empty.

The types \isa{nat} and \isa{real} represent the sets $\mathbb{N}$ and $\mathbb{R}$.
In this paper we drop the coercion function \isa{real\ {\isacharcolon}{\kern0pt}{\isacharcolon}{\kern0pt}\ nat\ {\isasymRightarrow}\ real}.
The set \isa{{\isacharbraceleft}{\kern0pt}m{\isachardot}{\kern0pt}{\isachardot}{\kern0pt}n{\isacharbraceright}{\kern0pt}} is the closed interval $[m,n]$.

The Isabelle/HOL distribution comes with a simple implementation of Hoare logic
where programs are annotated with pre- and post-conditions and invariants
as in this example, where all variables are of type \isa{nat}:
\begin{framedni}
\isa{{\isacharbraceleft}{\kern0pt}m\ {\isacharequal}{\kern0pt}\ {\isadigit{0}}\ {\isasymand}\ p\ {\isacharequal}{\kern0pt}\ {\isadigit{0}}{\isacharbraceright}{\kern0pt}}\\
\isa{WHILE\ m\ {\isasymnoteq}\ a\ \ \ INV\ {\isacharbraceleft}{\kern0pt}p\ {\isacharequal}{\kern0pt}\ m{\isacharasterisk}{\kern0pt}b{\isacharbraceright}{\kern0pt}\ \ \ DO\ p\ {\isacharcolon}{\kern0pt}{\isacharequal}{\kern0pt}\ p{\isacharplus}{\kern0pt}b{\isacharsemicolon}{\kern0pt}\ m\ {\isacharcolon}{\kern0pt}{\isacharequal}{\kern0pt}\ m{\isacharplus}{\kern0pt}{\isadigit{1}}\ OD}\\
\isa{{\isacharbraceleft}{\kern0pt}p\ {\isacharequal}{\kern0pt}\ a{\isacharasterisk}{\kern0pt}b{\isacharbraceright}{\kern0pt}}
\end{framedni}

\noindent
The box around the program means that it has been verified. All our proofs employ a verification condition generator
and essentially reduce to showing the preservation of the invariants.

Behind the scenes, a Hoare triple is just a HOL formula based on a formalized program semantics.
\isa{{\isacharbraceleft}{\kern0pt}P{\isacharbraceright}{\kern0pt}\ c\ {\isacharbraceleft}{\kern0pt}Q{\isacharbraceright}{\kern0pt}} is syntactic sugar for the following partial correctness formula: if execution of \isa{c}
starting in a state that satisfies \isa{P} terminates, then the final state satisfies \isa{Q}.
All expressions in a Hoare triple (pre- and post-conditions, invariants, tests in loops
and conditionals, and right-hand sides of assignments) are arbitrary HOL expressions
and can talk about both program variables and elements from the context.

\section{Vertex Cover}

We verify the proof in \cite{BerghammerM03} that the classic greedy algorithm
for vertex cover is a 2-approximation algorithm. In fact, we generalize the setup
from graphs to hypergraphs. A hypergraph is simply a set of edges \isa{E},
where an edge is a set of vertices of type \isa{{\isacharprime}{\kern0pt}a}.
A vertex cover for \isa{E} is a set of vertices \isa{C} that intersects with every edge of \isa{E}:
\begin{quoteni}
\isa{vc\ {\isacharcolon}{\kern0pt}{\isacharcolon}{\kern0pt}\ {\isacharprime}{\kern0pt}a\ set\ set\ {\isasymRightarrow}\ {\isacharprime}{\kern0pt}a\ set\ {\isasymRightarrow}\ bool}\\
\isa{vc\ E\ C\ {\isacharequal}{\kern0pt}\ {\isacharparenleft}{\kern0pt}{\isasymforall}e{\isasymin}E{\isachardot}{\kern0pt}\ e\ {\isasyminter}\ C\ {\isasymnoteq}\ {\isasymemptyset}{\isacharparenright}{\kern0pt}}
\end{quoteni}

A matching (\isa{matching\ {\isacharcolon}{\kern0pt}{\isacharcolon}{\kern0pt}\ {\isacharprime}{\kern0pt}a\ set\ set\ {\isasymRightarrow}\ bool}) is a set of pairwise disjoint sets. The following is a key
property that relates \isa{vc} and \isa{matching}:
\begin{quoteni}
\isa{finite\ C\ {\isasymand}\ matching\ M\ {\isasymand}\ M\ {\isasymsubseteq}\ E\ {\isasymand}\ vc\ E\ C\ {\isasymlongrightarrow}\ {\isacharbar}{\kern0pt}M{\isacharbar}{\kern0pt}\ {\isasymle}\ {\isacharbar}{\kern0pt}C{\isacharbar}{\kern0pt}}
\end{quoteni}\vspace{-2ex}%
\end{isamarkuptext}\isamarkuptrue%
\begin{isamarkuptext}%
We fix a rank-\isa{k} hypergraph \isa{E\ {\isacharcolon}{\kern0pt}{\isacharcolon}{\kern0pt}\ {\isacharprime}{\kern0pt}a\ set\ set} with the assumptions \isa{{\isasymemptyset}\ {\isasymnotin}\ E}, \isa{finite\ E} and \mbox{\isa{e\ {\isasymin}\ E\ {\isasymlongrightarrow}\ finite\ e\ {\isasymand}\ {\isacharbar}{\kern0pt}e{\isacharbar}{\kern0pt}\ {\isasymle}\ k}}.
(Via an Isabelle ``locale''. We use the same mechanism in all of our proofs.)

We have verified the well known greedy algorithm that computes a vertex cover \isa{C} for \isa{E}.
It keeps picking an arbitrary edge that is not covered by \isa{C} yet until all vertices are covered.
The final \isa{C} has at most \isa{k} times as many vertices as any vertex cover of \isa{E}
(which is essentially optimal \cite{BansalK10}).
\begin{framedni}
\isa{{\isacharbraceleft}{\kern0pt}True{\isacharbraceright}{\kern0pt}}\\
\isa{C\ {\isacharcolon}{\kern0pt}{\isacharequal}{\kern0pt}}\isa{{\isasymemptyset}}\isa{{\isacharsemicolon}{\kern0pt}\ F\ {\isacharcolon}{\kern0pt}{\isacharequal}{\kern0pt}\ E{\isacharsemicolon}{\kern0pt}}\\
\isa{WHILE} \isa{F\ {\isasymnoteq}\ {\isasymemptyset}} \ \ \isa{INV\ {\isacharbraceleft}{\kern0pt}invar\ C\ F{\isacharbraceright}{\kern0pt}}\\
\isa{DO\ C\ {\isacharcolon}{\kern0pt}{\isacharequal}{\kern0pt}\ C\ {\isasymunion}\ some\ F{\isacharsemicolon}{\kern0pt}\ F\ {\isacharcolon}{\kern0pt}{\isacharequal}{\kern0pt}\ F\ {\isacharminus}{\kern0pt}\ {\isacharbraceleft}{\kern0pt}e{\isacharprime}{\kern0pt}\ {\isasymin}\ F\ {\isacharbar}{\kern0pt}\ some\ F\ {\isasyminter}\ e{\isacharprime}{\kern0pt}\ {\isasymnoteq}\ {\isasymemptyset}{\isacharbraceright}{\kern0pt}\ OD}\\
\isa{{\isacharbraceleft}{\kern0pt}}\isa{vc\ E\ C\ {\isasymand}\ {\isacharparenleft}{\kern0pt}{\isasymforall}C{\isacharprime}{\kern0pt}{\isachardot}{\kern0pt}\ finite\ C{\isacharprime}{\kern0pt}\ {\isasymand}\ vc\ E\ C{\isacharprime}{\kern0pt}\ {\isasymlongrightarrow}\ {\isacharbar}{\kern0pt}C{\isacharbar}{\kern0pt}\ {\isasymle}\ k\ {\isacharasterisk}{\kern0pt}\ {\isacharbar}{\kern0pt}C{\isacharprime}{\kern0pt}{\isacharbar}{\kern0pt}{\isacharparenright}{\kern0pt}}\isa{{\isacharbraceright}{\kern0pt}}
\end{framedni}

\noindent
where \isa{invar} is the following invariant:
\begin{quoteni}
\isa{invar\ {\isacharcolon}{\kern0pt}{\isacharcolon}{\kern0pt}\ {\isacharprime}{\kern0pt}a\ set\ {\isasymRightarrow}\ {\isacharprime}{\kern0pt}a\ set\ set\ {\isasymRightarrow}\ bool}\\
\isa{invar\ C\ F\ {\isacharequal}{\kern0pt}\isanewline
{\isacharparenleft}{\kern0pt}F\ {\isasymsubseteq}\ E\ {\isasymand}\ vc\ {\isacharparenleft}{\kern0pt}E\ {\isacharminus}{\kern0pt}\ F{\isacharparenright}{\kern0pt}\ C\ {\isasymand}\ finite\ C\ {\isasymand}\ {\isacharparenleft}{\kern0pt}{\isasymexists}M{\isachardot}{\kern0pt}\ inv{\isacharunderscore}{\kern0pt}matching\ C\ F\ M{\isacharparenright}{\kern0pt}{\isacharparenright}{\kern0pt}}
\medskip

\noindent
\isa{inv{\isacharunderscore}{\kern0pt}matching\ C\ F\ M\ {\isacharequal}{\kern0pt}\isanewline
{\isacharparenleft}{\kern0pt}matching\ M\ {\isasymand}\ M\ {\isasymsubseteq}\ E\ {\isasymand}\ {\isacharbar}{\kern0pt}C{\isacharbar}{\kern0pt}\ {\isasymle}\ k\ {\isacharasterisk}{\kern0pt}\ {\isacharbar}{\kern0pt}M{\isacharbar}{\kern0pt}\ {\isasymand}\ {\isacharparenleft}{\kern0pt}{\isasymforall}e{\isasymin}M{\isachardot}{\kern0pt}\ {\isasymforall}f{\isasymin}F{\isachardot}{\kern0pt}\ e\ {\isasyminter}\ f\ {\isacharequal}{\kern0pt}\ {\isasymemptyset}{\isacharparenright}{\kern0pt}{\isacharparenright}{\kern0pt}}
\end{quoteni}
The key step in the program proof is that the invariant is invariant:
\begin{lem}
\isa{F\ {\isasymnoteq}\ {\isasymemptyset}\ {\isasymand}\ invar\ C\ F\ {\isasymlongrightarrow}\isanewline
invar\ {\isacharparenleft}{\kern0pt}C\ {\isasymunion}\ some\ F{\isacharparenright}{\kern0pt}\ {\isacharparenleft}{\kern0pt}F\ {\isacharminus}{\kern0pt}\ {\isacharbraceleft}{\kern0pt}e{\isacharprime}{\kern0pt}\ {\isasymin}\ F\ {\isacharbar}{\kern0pt}\ some\ F\ {\isasyminter}\ e{\isacharprime}{\kern0pt}\ {\isasymnoteq}\ {\isasymemptyset}{\isacharbraceright}{\kern0pt}{\isacharparenright}{\kern0pt}}
\end{lem}

Our invariant is stronger than the one in \cite{BerghammerM03} which lacks \isa{F\ {\isasymsubseteq}\ E}.
Indeed, without this property, the claimed invariant is not invariant (as acknowledged by Müller-Olm).%
\end{isamarkuptext}\isamarkuptrue%
\isadelimtheory
\endisadelimtheory
\isatagtheory
\endisatagtheory
{\isafoldtheory}%
\isadelimtheory
\endisadelimtheory
\end{isabellebody}%

%% file: Paper_MIS.tex
\begin{isabellebody}%
\setisabellecontext{Paper{\isacharunderscore}{\kern0pt}MIS}%
\isadelimtheory
\endisadelimtheory
\isatagtheory
\endisatagtheory
{\isafoldtheory}%
\isadelimtheory
\endisadelimtheory
\begin{isamarkuptext}%
\section{Independent Set}

As in the previous section, a graph is a set of edges. An independent set
of a graph \isa{E} is a subset of its vertices such that no two vertices are adjacent.

\begin{quoteni}
\isa{iv\ {\isacharcolon}{\kern0pt}{\isacharcolon}{\kern0pt}\ {\isacharprime}{\kern0pt}a\ set\ set\ {\isasymRightarrow}\ {\isacharprime}{\kern0pt}a\ set\ {\isasymRightarrow}\ bool}\\
\isa{iv\ E\ S\ {\isacharequal}{\kern0pt}\ {\isacharparenleft}{\kern0pt}S\ {\isasymsubseteq}\ {\isasymUnion}\ E\ {\isasymand}\ {\isacharparenleft}{\kern0pt}{\isasymforall}v\isactrlsub {\isadigit{1}}\ v\isactrlsub {\isadigit{2}}{\isachardot}{\kern0pt}\ v\isactrlsub {\isadigit{1}}\ {\isasymin}\ S\ {\isasymand}\ v\isactrlsub {\isadigit{2}}\ {\isasymin}\ S\ {\isasymlongrightarrow}\ {\isacharbraceleft}{\kern0pt}v\isactrlsub {\isadigit{1}}{\isacharcomma}{\kern0pt}\ v\isactrlsub {\isadigit{2}}{\isacharbraceright}{\kern0pt}\ {\isasymnotin}\ E{\isacharparenright}{\kern0pt}{\isacharparenright}{\kern0pt}}
\end{quoteni}

We fix a finite graph \isa{E\ {\isacharcolon}{\kern0pt}{\isacharcolon}{\kern0pt}\ {\isacharprime}{\kern0pt}a\ set\ set} such that all edges of \isa{E} are sets of cardinality 2.
The set of vertices \isa{{\isasymUnion}\ E} is denoted \isa{V}, and the maximum degree (number of neighbors) of any
vertex in \isa{V} is denoted \isa{{\isasymDelta}}.
We show that the greedy algorithm proposed by Wei is a \isa{{\isasymDelta}}-approximation algorithm. The proof is inspired by
one given in~\cite{BerghammerM03}. In particular, the proof relies on an auxiliary
variable \isa{P}, which is not needed for the execution of the algorithm, but is used for
bookkeeping in the proof. In~\cite{BerghammerM03}, \isa{P} is initially a program variable and
is later removed from the program and turned into an existentially quantified variable in the invariant.
We directly use the latter representation.

\begin{framedni}
\isa{{\isacharbraceleft}{\kern0pt}\ True\ {\isacharbraceright}{\kern0pt}}\\
\isa{S\ {\isacharcolon}{\kern0pt}{\isacharequal}{\kern0pt}} \isa{{\isasymemptyset}}\isa{{\isacharsemicolon}{\kern0pt}\ X\ {\isacharcolon}{\kern0pt}{\isacharequal}{\kern0pt}}\isa{{\isasymemptyset}}\isa{{\isacharsemicolon}{\kern0pt}}\\
\isa{WHILE} \isa{X\ {\isasymnoteq}\ V} \ \isa{INV\ {\isacharbraceleft}{\kern0pt}} \isa{{\isasymexists}P{\isachardot}{\kern0pt}\ inv{\isacharunderscore}{\kern0pt}partition\ S\ X\ P} \isa{{\isacharbraceright}{\kern0pt}}\\
\isa{DO\ x\ {\isacharcolon}{\kern0pt}{\isacharequal}{\kern0pt}\ some\ {\isacharparenleft}{\kern0pt}V\ {\isacharminus}{\kern0pt}\ X{\isacharparenright}{\kern0pt}{\isacharsemicolon}{\kern0pt}\ S\ {\isacharcolon}{\kern0pt}{\isacharequal}{\kern0pt}\ S\ {\isasymunion}\ {\isacharbraceleft}{\kern0pt}x{\isacharbraceright}{\kern0pt}{\isacharsemicolon}{\kern0pt}\ X\ {\isacharcolon}{\kern0pt}{\isacharequal}{\kern0pt}\ X\ {\isasymunion}\ neighbors\ x\ {\isasymunion}\ {\isacharbraceleft}{\kern0pt}x{\isacharbraceright}{\kern0pt}\ OD}\\
\isa{{\isacharbraceleft}{\kern0pt}} \isa{iv\ E\ S\ {\isasymand}\ {\isacharparenleft}{\kern0pt}{\isasymforall}S{\isacharprime}{\kern0pt}{\isachardot}{\kern0pt}\ iv\ E\ S{\isacharprime}{\kern0pt}\ {\isasymlongrightarrow}\ {\isacharbar}{\kern0pt}S{\isacharprime}{\kern0pt}{\isacharbar}{\kern0pt}\ {\isasymle}\ {\isacharbar}{\kern0pt}S{\isacharbar}{\kern0pt}\ {\isacharasterisk}{\kern0pt}\ {\isasymDelta}{\isacharparenright}{\kern0pt}} \isa{{\isacharbraceright}{\kern0pt}}
\end{framedni}

To keep the size of definitions manageable, we split the invariant in two. The
first part is not concerned with \isa{P}, but suffices to prove the functional
correctness of the algorithm, i.e., that it outputs an independent set of the graph:

\begin{quoteni}
\isa{inv{\isacharunderscore}{\kern0pt}iv\ {\isacharcolon}{\kern0pt}{\isacharcolon}{\kern0pt}\ {\isacharprime}{\kern0pt}a\ set\ {\isasymRightarrow}\ {\isacharprime}{\kern0pt}a\ set\ {\isasymRightarrow}\ bool}\\
\isa{inv{\isacharunderscore}{\kern0pt}iv\ S\ X\ {\isacharequal}{\kern0pt}\isanewline
{\isacharparenleft}{\kern0pt}iv\ E\ S\ {\isasymand}\ X\ {\isasymsubseteq}\ V\ {\isasymand}\ {\isacharparenleft}{\kern0pt}{\isasymforall}v\isactrlsub {\isadigit{1}}{\isasymin}V\ {\isacharminus}{\kern0pt}\ X{\isachardot}{\kern0pt}\ {\isasymforall}v\isactrlsub {\isadigit{2}}{\isasymin}S{\isachardot}{\kern0pt}\ {\isacharbraceleft}{\kern0pt}v\isactrlsub {\isadigit{1}}{\isacharcomma}{\kern0pt}\ v\isactrlsub {\isadigit{2}}{\isacharbraceright}{\kern0pt}\ {\isasymnotin}\ E{\isacharparenright}{\kern0pt}\ {\isasymand}\ S\ {\isasymsubseteq}\ X{\isacharparenright}{\kern0pt}}
\end{quoteni}
This invariant is taken almost verbatim from~\cite{BerghammerM03}, except that in~\cite{BerghammerM03}
it says that \isa{S} is an independent set of the subgraph generated by \isa{X}. This is later
used to show that the \isa{x} picked at each iteration from \isa{V\ {\isacharminus}{\kern0pt}\ X} is not already in \isa{S}.
Defining subgraphs adds unnecessary complexity to the invariant. We simply
state \isa{S\ {\isasymsubseteq}\ X}, together with the fact that \isa{S} is an independent set of the whole graph.

We now extend the invariant with properties of the auxiliary variable \isa{P}.

\begin{quoteni}
\isa{inv{\isacharunderscore}{\kern0pt}partition\ {\isacharcolon}{\kern0pt}{\isacharcolon}{\kern0pt}\ {\isacharprime}{\kern0pt}a\ set\ {\isasymRightarrow}\ {\isacharprime}{\kern0pt}a\ set\ {\isasymRightarrow}\ {\isacharprime}{\kern0pt}a\ set\ set\ {\isasymRightarrow}\ bool}\\
\isa{inv{\isacharunderscore}{\kern0pt}partition\ S\ X\ P\ {\isacharequal}{\kern0pt}\isanewline
{\isacharparenleft}{\kern0pt}inv{\isacharunderscore}{\kern0pt}iv\ S\ X\ {\isasymand}\isanewline
\isaindent{{\isacharparenleft}{\kern0pt}}{\isasymUnion}\ P\ {\isacharequal}{\kern0pt}\ X\ {\isasymand}\ {\isacharparenleft}{\kern0pt}{\isasymforall}p{\isasymin}P{\isachardot}{\kern0pt}\ {\isasymexists}s{\isasymin}V{\isachardot}{\kern0pt}\ p\ {\isacharequal}{\kern0pt}\ {\isacharbraceleft}{\kern0pt}s{\isacharbraceright}{\kern0pt}\ {\isasymunion}\ neighbors\ s{\isacharparenright}{\kern0pt}\ {\isasymand}\ {\isacharbar}{\kern0pt}P{\isacharbar}{\kern0pt}\ {\isacharequal}{\kern0pt}\ {\isacharbar}{\kern0pt}S{\isacharbar}{\kern0pt}\ {\isasymand}\ finite\ P{\isacharparenright}{\kern0pt}}
\end{quoteni}
We can view the set \isa{P} as an auxiliary program variable. In order to satisfy the invariant, \isa{P}
would be initially empty and the loop body would include the assignment
\isa{P\ {\isacharcolon}{\kern0pt}{\isacharequal}{\kern0pt}\ P\ {\isasymunion}\ {\isacharbraceleft}{\kern0pt}neighbors\ x\ {\isasymunion}\ {\isacharbraceleft}{\kern0pt}x{\isacharbraceright}{\kern0pt}{\isacharbraceright}{\kern0pt}}. Intuitively, \isa{P} contains the sets of vertices that are added
to \isa{X} at each iteration (or more precisely, an over-approximation, since some vertices in
\isa{neighbors\ x} may have been added to \isa{X} in a previous iteration).
Instead of adding an unnecessary variable to the program, we only use the existentially quantified
invariant. The assignments described above correspond directly to instantiations of the quantifier
that are needed to solve proof obligations. This is illustrated with the following lemma, which
corresponds to the preservation of the invariant:
\begin{lem}
\isa{{\isacharparenleft}{\kern0pt}{\isasymexists}P{\isachardot}{\kern0pt}\ inv{\isacharunderscore}{\kern0pt}partition\ S\ X\ P{\isacharparenright}{\kern0pt}\ {\isasymand}\ x\ {\isasymin}\ V\ {\isacharminus}{\kern0pt}\ X\ {\isasymlongrightarrow}\isanewline
{\isacharparenleft}{\kern0pt}{\isasymexists}P{\isacharprime}{\kern0pt}{\isachardot}{\kern0pt}\ inv{\isacharunderscore}{\kern0pt}partition\ {\isacharparenleft}{\kern0pt}S\ {\isasymunion}\ {\isacharbraceleft}{\kern0pt}x{\isacharbraceright}{\kern0pt}{\isacharparenright}{\kern0pt}\ {\isacharparenleft}{\kern0pt}X\ {\isasymunion}\ neighbors\ x\ {\isasymunion}\ {\isacharbraceleft}{\kern0pt}x{\isacharbraceright}{\kern0pt}{\isacharparenright}{\kern0pt}\ P{\isacharprime}{\kern0pt}{\isacharparenright}{\kern0pt}}
\end{lem}
\noindent
The existential quantifier in the antecedent yields a witness \isa{P}. After instantiating the
quantifier in the succedent with \isa{P\ {\isasymunion}\ {\isacharbraceleft}{\kern0pt}neighbors\ x\ {\isasymunion}\ {\isacharbraceleft}{\kern0pt}x{\isacharbraceright}{\kern0pt}{\isacharbraceright}{\kern0pt}}, the goal can be solved
straightforwardly.
Finally, the following lemma combines the invariant and the negated post-condition to prove the
approximation ratio:
\begin{lem}
\isa{inv{\isacharunderscore}{\kern0pt}partition\ S\ V\ P\ {\isasymlongrightarrow}\ {\isacharparenleft}{\kern0pt}{\isasymforall}S{\isacharprime}{\kern0pt}{\isachardot}{\kern0pt}\ iv\ E\ S{\isacharprime}{\kern0pt}\ {\isasymlongrightarrow}\ {\isacharbar}{\kern0pt}S{\isacharprime}{\kern0pt}{\isacharbar}{\kern0pt}\ {\isasymle}\ {\isacharbar}{\kern0pt}S{\isacharbar}{\kern0pt}\ {\isacharasterisk}{\kern0pt}\ {\isasymDelta}{\isacharparenright}{\kern0pt}}
\end{lem}
\noindent
To prove it, we observe that any set \isa{p\ {\isasymin}\ P} consists of a vertex \isa{x} and its neighbors, therefore
an independent set \isa{S{\isacharprime}{\kern0pt}} can contain at most \isa{{\isasymDelta}} of the vertices in \isa{p}, thus
\isa{{\isacharbar}{\kern0pt}S{\isacharprime}{\kern0pt}{\isacharbar}{\kern0pt}\ {\isasymle}\ {\isacharbar}{\kern0pt}P{\isacharbar}{\kern0pt}\ {\isacharasterisk}{\kern0pt}\ {\isasymDelta}}. Furthermore, as indicated by the invariant,
\isa{{\isacharbar}{\kern0pt}P{\isacharbar}{\kern0pt}\ {\isacharequal}{\kern0pt}\ {\isacharbar}{\kern0pt}S{\isacharbar}{\kern0pt}}.

Compared to the proof in~\cite{BerghammerM03}, our invariant describes the contents of the set \isa{P}
more precisely, and thus yields a better approximation ratio.
In~\cite{BerghammerM03}, the invariant merely indicates that \isa{X\ {\isacharequal}{\kern0pt}\ {\isasymUnion}\ P}, together with two
cardinality properties: \isa{{\isasymforall}p{\isasymin}P{\isachardot}{\kern0pt}\ {\isacharbar}{\kern0pt}p{\isacharbar}{\kern0pt}\ {\isasymle}\ {\isasymDelta}\ {\isacharplus}{\kern0pt}\ {\isadigit{1}}} and \isa{{\isacharbar}{\kern0pt}P{\isacharbar}{\kern0pt}\ {\isasymle}\ {\isacharbar}{\kern0pt}S{\isacharbar}{\kern0pt}}.
Taken with the negated post-condition, this invariant can be used to show that for any independent
set \isa{S{\isacharprime}{\kern0pt}}, we have \isa{{\isacharbar}{\kern0pt}S{\isacharprime}{\kern0pt}{\isacharbar}{\kern0pt}\ {\isasymle}\ {\isacharbar}{\kern0pt}S{\isacharbar}{\kern0pt}\ {\isacharasterisk}{\kern0pt}\ {\isacharparenleft}{\kern0pt}{\isasymDelta}\ {\isacharplus}{\kern0pt}\ {\isadigit{1}}{\isacharparenright}{\kern0pt}}. The proof of this lemma makes use of the
following (in)equalities: \isa{{\isacharbar}{\kern0pt}S{\isacharprime}{\kern0pt}{\isacharbar}{\kern0pt}\ {\isasymle}\ {\isacharbar}{\kern0pt}V{\isacharbar}{\kern0pt}}, \isa{{\isacharbar}{\kern0pt}V{\isacharbar}{\kern0pt}\ {\isacharequal}{\kern0pt}\ {\isacharbar}{\kern0pt}{\isasymUnion}\ P{\isacharbar}{\kern0pt}},
\isa{{\isacharbar}{\kern0pt}{\isasymUnion}\ P{\isacharbar}{\kern0pt}\ {\isasymle}\ {\isacharbar}{\kern0pt}P{\isacharbar}{\kern0pt}\ {\isacharasterisk}{\kern0pt}\ {\isacharparenleft}{\kern0pt}{\isasymDelta}\ {\isacharplus}{\kern0pt}\ {\isadigit{1}}{\isacharparenright}{\kern0pt}} and finally \isa{{\isacharbar}{\kern0pt}P{\isacharbar}{\kern0pt}\ {\isacharasterisk}{\kern0pt}\ {\isacharparenleft}{\kern0pt}{\isasymDelta}\ {\isacharplus}{\kern0pt}\ {\isadigit{1}}{\isacharparenright}{\kern0pt}\ {\isasymle}\ {\isacharbar}{\kern0pt}S{\isacharbar}{\kern0pt}\ {\isacharasterisk}{\kern0pt}\ {\isacharparenleft}{\kern0pt}{\isasymDelta}\ {\isacharplus}{\kern0pt}\ {\isadigit{1}}{\isacharparenright}{\kern0pt}}.
Note that this only relies on the trivial fact that an independent set cannot contain more vertices
than the graph. By contrast, our own argument takes into account information regarding the edges of
the graph.

Although this proof results in a weaker approximation ratio than our own, it yields a useful insight:
an approximation ratio is given by the cardinality of the largest set \isa{p\ {\isasymin}\ P} (i.e., the largest
number of vertices added to \isa{X} during any given iteration). In the worst case, this is equal to
\isa{{\isasymDelta}\ {\isacharplus}{\kern0pt}\ {\isadigit{1}}}, but in practice the number may be smaller. This suggests a
variant of the algorithm that stores that value in a variable \isa{r}, as described in~\cite{BerghammerM03}. At every iteration, the variable
\isa{r} is assigned the value \isa{max\ r\ {\isacharbar}{\kern0pt}{\isacharbraceleft}{\kern0pt}x{\isacharbraceright}{\kern0pt}\ {\isasymunion}\ neighbors\ x\ {\isacharminus}{\kern0pt}\ X{\isacharbar}{\kern0pt}}. Ultimately, the
algorithm returns both the independent set \isa{S} and the value \isa{r}, with the guarantee that
\isa{{\isacharbar}{\kern0pt}S{\isacharprime}{\kern0pt}{\isacharbar}{\kern0pt}\ {\isasymle}\ {\isacharbar}{\kern0pt}S{\isacharbar}{\kern0pt}\ {\isacharasterisk}{\kern0pt}\ r} for any independent set \isa{S{\isacharprime}{\kern0pt}}.

We also formalized this variant and proved the aforementioned property. The proof follows the idea outlined above,
but does away with the variable \isa{P} entirely: instead, the invariant simply maintains that
\isa{inv{\isacharunderscore}{\kern0pt}iv\ S\ X\ {\isasymand}\ {\isacharbar}{\kern0pt}X{\isacharbar}{\kern0pt}\ {\isasymle}\ {\isacharbar}{\kern0pt}S{\isacharbar}{\kern0pt}\ {\isacharasterisk}{\kern0pt}\ r}, and the proof of preservation is adapted accordingly. Indeed, this
demonstrates that the argument used in~\cite{BerghammerM03} does not require an auxiliary variable nor an
existentially quantified invariant. For the proof of the approximation ratio \isa{{\isasymDelta}}, a similar
simplification is not as easy to obtain, because the argument relies on a global property of the
graph (a constraint that edges impose on independent sets) that is not easy to summarize in an inductive invariant.

So far, we have only considered an algorithm where the vertex \isa{x} is picked non-deterministically.
An obvious heuristic is to pick, at every iteration, the vertex with the smallest number of
neighbors among \isa{V\ {\isacharminus}{\kern0pt}\ X}. Halldórsson and Radhakrishnan~\cite{halldorsson1997greed} prove
that this heuristic achieves an approximation ratio of \isa{{\isacharparenleft}{\kern0pt}{\isasymDelta}\ {\isacharplus}{\kern0pt}\ {\isadigit{2}}{\isacharparenright}{\kern0pt}\ {\isacharslash}{\kern0pt}\ {\isadigit{3}}}. They proceed by considering the
sequence of graph reductions corresponding to the execution of the algorithm. Formalizing this argument
would require an inductive invariant to keep track of the sequences of values of \isa{X} and \isa{x} at each
iteration and the relations between those values, providing a much more precise record than the auxiliary
variable \isa{P}. In addition, the proof itself is far more complex than the arguments
presented here, relying on case analysis for different types of graphs. This is beyond the scope of our paper.%
\end{isamarkuptext}\isamarkuptrue%
\isadelimtheory
\endisadelimtheory
\isatagtheory
\endisatagtheory
{\isafoldtheory}%
\isadelimtheory
\endisadelimtheory
\end{isabellebody}%

%% file: Paper_LB.tex
\begin{isabellebody}%
\setisabellecontext{Paper{\isacharunderscore}{\kern0pt}LB}%
\isadelimtheory
\endisadelimtheory
\isatagtheory
\endisatagtheory
{\isafoldtheory}%
\isadelimtheory
\endisadelimtheory
\isadelimproof
\endisadelimproof
\isatagproof
\endisatagproof
{\isafoldproof}%
\isadelimproof
\endisadelimproof
\isadelimproof
\endisadelimproof
\isatagproof
\endisatagproof
{\isafoldproof}%
\isadelimproof
\endisadelimproof
\isadelimproof
\endisadelimproof
\isatagproof
\endisatagproof
{\isafoldproof}%
\isadelimproof
\endisadelimproof
\isadelimproof
\endisadelimproof
\isatagproof
\endisatagproof
{\isafoldproof}%
\isadelimproof
\endisadelimproof
\isadelimproof
\endisadelimproof
\isatagproof
\endisatagproof
{\isafoldproof}%
\isadelimproof
\endisadelimproof
\begin{isamarkuptext}%
\section{Load Balancing}

Our starting point for the load balancing problem is \cite[Chapter 11.1]{KleinbergT06}. We need to distribute \isa{n\ {\isacharcolon}{\kern0pt}{\isacharcolon}{\kern0pt}\ nat} jobs on \isa{m\ {\isacharcolon}{\kern0pt}{\isacharcolon}{\kern0pt}\ nat} machines with \isa{{\isadigit{0}}\ {\isacharless}{\kern0pt}\ m}. A job \isa{j\ {\isasymin}\ {\isacharbraceleft}{\kern0pt}{\isadigit{1}}{\isachardot}{\kern0pt}{\isachardot}{\kern0pt}n{\isacharbraceright}{\kern0pt}} has a load \isa{t{\isacharparenleft}{\kern0pt}j{\isacharparenright}{\kern0pt}\ {\isacharcolon}{\kern0pt}{\isacharcolon}{\kern0pt}\ nat}.
Variables \isa{m}, \isa{n}, and \isa{t} are fixed throughout this section.
A solution is described by a function \isa{A} that maps machines to sets of jobs:
$\isa{k\ {\isasymin}\ {\isacharbraceleft}{\kern0pt}{\isadigit{1}}{\isachardot}{\kern0pt}{\isachardot}{\kern0pt}m{\isacharbraceright}{\kern0pt}}$ has job \isa{j} assigned to it iff $\isa{j\ {\isasymin}\ A{\isacharparenleft}{\kern0pt}k{\isacharparenright}{\kern0pt}}$.
The sum of job loads on a machine is given by a function $\isa{T}$ that is derived from
\isa{t} and \isa{A}: \isa{T\ k\ {\isacharequal}{\kern0pt}\ {\isacharparenleft}{\kern0pt}{\isasymSum}j{\isasymin}A\ k{\isachardot}{\kern0pt}\ t\ j{\isacharparenright}{\kern0pt}}. Predicate \isa{lb} defines when \isa{T} and \isa{A} are a partial solution for \isa{j\ {\isasymle}\ n} jobs:
\begin{quoteni}
\isa{lb\ {\isacharcolon}{\kern0pt}{\isacharcolon}{\kern0pt}\ {\isacharparenleft}{\kern0pt}nat\ {\isasymRightarrow}\ nat{\isacharparenright}{\kern0pt}\ {\isasymRightarrow}\ {\isacharparenleft}{\kern0pt}nat\ {\isasymRightarrow}\ nat\ set{\isacharparenright}{\kern0pt}\ {\isasymRightarrow}\ nat\ {\isasymRightarrow}\ bool}\\
\isa{lb\ T\ A\ j\ {\isacharequal}{\kern0pt}\isanewline
{\isacharparenleft}{\kern0pt}{\isacharparenleft}{\kern0pt}{\isasymforall}x{\isasymin}{\isacharbraceleft}{\kern0pt}{\isadigit{1}}{\isachardot}{\kern0pt}{\isachardot}{\kern0pt}m{\isacharbraceright}{\kern0pt}{\isachardot}{\kern0pt}\ {\isasymforall}y{\isasymin}{\isacharbraceleft}{\kern0pt}{\isadigit{1}}{\isachardot}{\kern0pt}{\isachardot}{\kern0pt}m{\isacharbraceright}{\kern0pt}{\isachardot}{\kern0pt}\ x\ {\isasymnoteq}\ y\ {\isasymlongrightarrow}\ A\ x\ {\isasyminter}\ A\ y\ {\isacharequal}{\kern0pt}\ {\isasymemptyset}{\isacharparenright}{\kern0pt}\ {\isasymand}\isanewline
\isaindent{{\isacharparenleft}{\kern0pt}}{\isacharparenleft}{\kern0pt}{\isasymUnion}\isactrlbsub x{\isasymin}{\isacharbraceleft}{\kern0pt}{\isadigit{1}}{\isachardot}{\kern0pt}{\isachardot}{\kern0pt}m{\isacharbraceright}{\kern0pt}\isactrlesub \ A\ x{\isacharparenright}{\kern0pt}\ {\isacharequal}{\kern0pt}\ {\isacharbraceleft}{\kern0pt}{\isadigit{1}}{\isachardot}{\kern0pt}{\isachardot}{\kern0pt}j{\isacharbraceright}{\kern0pt}\ {\isasymand}\ {\isacharparenleft}{\kern0pt}{\isasymforall}x{\isasymin}{\isacharbraceleft}{\kern0pt}{\isadigit{1}}{\isachardot}{\kern0pt}{\isachardot}{\kern0pt}m{\isacharbraceright}{\kern0pt}{\isachardot}{\kern0pt}\ {\isacharparenleft}{\kern0pt}{\isasymSum}y{\isasymin}A\ x{\isachardot}{\kern0pt}\ t\ y{\isacharparenright}{\kern0pt}\ {\isacharequal}{\kern0pt}\ T\ x{\isacharparenright}{\kern0pt}{\isacharparenright}{\kern0pt}}
\end{quoteni}
It consists of three conjuncts. The first ensures that the sets returned by \isa{A} are pairwise disjoint, thus, no job appears in more than one machine. The second conjunct ensures that every job \isa{x\ {\isasymin}\ {\isacharbraceleft}{\kern0pt}{\isadigit{1}}{\isachardot}{\kern0pt}{\isachardot}{\kern0pt}j{\isacharbraceright}{\kern0pt}} is contained in at least one machine. It also ensures that only jobs \isa{{\isacharbraceleft}{\kern0pt}{\isadigit{1}}{\isachardot}{\kern0pt}{\isachardot}{\kern0pt}j{\isacharbraceright}{\kern0pt}} have been added. The final conjunct ensures that \isa{T} is correctly defined to be the total load on a machine. To ensure that jobs are distributed evenly, we need to consider the machine with maximum load. This load is referred to as the \isa{makespan} of a solution:
\begin{quoteni}
\isa{makespan\ {\isacharcolon}{\kern0pt}{\isacharcolon}{\kern0pt}\ {\isacharparenleft}{\kern0pt}nat\ {\isasymRightarrow}\ nat{\isacharparenright}{\kern0pt}\ {\isasymRightarrow}\ nat}\\
\isa{makespan\ T\ {\isacharequal}{\kern0pt}\ Max\ {\isacharparenleft}{\kern0pt}T\ {\isacharbackquote}{\kern0pt}\ {\isacharbraceleft}{\kern0pt}{\isadigit{1}}{\isachardot}{\kern0pt}{\isachardot}{\kern0pt}m{\isacharbraceright}{\kern0pt}{\isacharparenright}{\kern0pt}}
\end{quoteni}

The greedy approximation algorithm outlined in \cite{KleinbergT06} relies on the ability to determine the machine \isa{k\ {\isasymin}\ {\isacharbraceleft}{\kern0pt}{\isadigit{1}}{\isachardot}{\kern0pt}{\isachardot}{\kern0pt}m{\isacharbraceright}{\kern0pt}} that has a minimum combined load. As the goal is to approximate the optimum in polynomial time, a linear scan through \isa{T} suffices to find the machine with minimum load. However, other methods may be considered to further improve time complexity. To determine the machine with minimum load, we will use the following function:
\begin{quoteni}
\isa{min{\isacharunderscore}{\kern0pt}arg\ {\isacharcolon}{\kern0pt}{\isacharcolon}{\kern0pt}\ {\isacharparenleft}{\kern0pt}nat\ {\isasymRightarrow}\ nat{\isacharparenright}{\kern0pt}\ {\isasymRightarrow}\ nat\ {\isasymRightarrow}\ nat}\\
\isa{min{\isacharunderscore}{\kern0pt}arg\ T\ {\isadigit{0}}\ {\isacharequal}{\kern0pt}\ {\isadigit{1}}}\\
\isa{min{\isacharunderscore}{\kern0pt}arg\ T\ {\isacharparenleft}{\kern0pt}x\ {\isacharplus}{\kern0pt}\ {\isadigit{1}}{\isacharparenright}{\kern0pt}\ {\isacharequal}{\kern0pt}\isanewline
{\isacharparenleft}{\kern0pt}\textsf{let}\ k\ {\isacharequal}{\kern0pt}\ min{\isacharunderscore}{\kern0pt}arg\ T\ x\ \textsf{in}\ \textsf{if}\ T\ {\isacharparenleft}{\kern0pt}x\ {\isacharplus}{\kern0pt}\ {\isadigit{1}}{\isacharparenright}{\kern0pt}\ {\isacharless}{\kern0pt}\ T\ k\ \textsf{then}\ x\ {\isacharplus}{\kern0pt}\ {\isadigit{1}}\ \textsf{else}\ k{\isacharparenright}{\kern0pt}}
\end{quoteni}

We will focus on the approximation factor of $\frac32$, which can be proved if the job loads are assumed to be sorted in descending order. The proof for the approximation factor of \isa{{\isadigit{2}}} if jobs are unsorted is very similar and we describe the differences at the end.
We say that \isa{j} jobs are sorted in descending order if \isa{sorted} holds:
\begin{quoteni}
\isa{sorted\ {\isacharcolon}{\kern0pt}{\isacharcolon}{\kern0pt}\ nat\ {\isasymRightarrow}\ bool}\\
\isa{sorted\ j\ {\isacharequal}{\kern0pt}\ {\isacharparenleft}{\kern0pt}{\isasymforall}x{\isasymin}{\isacharbraceleft}{\kern0pt}{\isadigit{1}}{\isachardot}{\kern0pt}{\isachardot}{\kern0pt}j{\isacharbraceright}{\kern0pt}{\isachardot}{\kern0pt}\ {\isasymforall}y{\isasymin}{\isacharbraceleft}{\kern0pt}{\isadigit{1}}{\isachardot}{\kern0pt}{\isachardot}{\kern0pt}x{\isacharbraceright}{\kern0pt}{\isachardot}{\kern0pt}\ t\ x\ {\isasymle}\ t\ y{\isacharparenright}{\kern0pt}}
\end{quoteni}

Below we prove the following conditional Hoare triple that expresses the approximation factor and functional correctness of the algorithm given in \cite{KleinbergT06}:
\begin{framedni}
\isa{sorted\ n} \isa{{\isasymlongrightarrow}}\\
\isa{{\isacharbraceleft}{\kern0pt}True{\isacharbraceright}{\kern0pt}}\\
\isa{T\ {\isacharcolon}{\kern0pt}{\isacharequal}{\kern0pt}\ {\isacharparenleft}{\kern0pt}{\isasymlambda}{\isacharunderscore}{\kern0pt}{\isachardot}{\kern0pt}\ {\isadigit{0}}{\isacharparenright}{\kern0pt}{\isacharsemicolon}{\kern0pt}\ A\ {\isacharcolon}{\kern0pt}{\isacharequal}{\kern0pt}\ {\isacharparenleft}{\kern0pt}{\isasymlambda}{\isacharunderscore}{\kern0pt}{\isachardot}{\kern0pt}\ {\isasymemptyset}{\isacharparenright}{\kern0pt}{\isacharsemicolon}{\kern0pt}\ j\ {\isacharcolon}{\kern0pt}{\isacharequal}{\kern0pt}\ {\isadigit{0}}{\isacharsemicolon}{\kern0pt}}\\
\isa{WHILE\ j\ {\isacharless}{\kern0pt}\ n\ INV\ {\isacharbraceleft}{\kern0pt}inv\isactrlsub {\isadigit{2}}\ T\ A\ j{\isacharbraceright}{\kern0pt}}\\
\isa{DO\ i\ {\isacharcolon}{\kern0pt}{\isacharequal}{\kern0pt}\ min{\isacharunderscore}{\kern0pt}arg\ T\ m{\isacharsemicolon}{\kern0pt}\ j\ {\isacharcolon}{\kern0pt}{\isacharequal}{\kern0pt}\ j\ {\isacharplus}{\kern0pt}\ {\isadigit{1}}{\isacharsemicolon}{\kern0pt}}\\
\hphantom{\isa{DO}}\isa{A\ {\isacharcolon}{\kern0pt}{\isacharequal}{\kern0pt}\ A{\isacharparenleft}{\kern0pt}i\ {\isacharcolon}{\kern0pt}{\isacharequal}{\kern0pt}\ A{\isacharparenleft}{\kern0pt}i{\isacharparenright}{\kern0pt}\ {\isasymunion}\ {\isacharbraceleft}{\kern0pt}j{\isacharbraceright}{\kern0pt}{\isacharparenright}{\kern0pt}{\isacharsemicolon}{\kern0pt}\ T\ {\isacharcolon}{\kern0pt}{\isacharequal}{\kern0pt}\ T{\isacharparenleft}{\kern0pt}i\ {\isacharcolon}{\kern0pt}{\isacharequal}{\kern0pt}\ T{\isacharparenleft}{\kern0pt}i{\isacharparenright}{\kern0pt}\ {\isacharplus}{\kern0pt}\ t{\isacharparenleft}{\kern0pt}j{\isacharparenright}{\kern0pt}{\isacharparenright}{\kern0pt}}\\
\isa{OD}\\
\isa{{\isacharbraceleft}{\kern0pt}lb\ T\ A\ n\ {\isasymand}}\\
\hphantom{\isa{{\isacharbraceleft}{\kern0pt}}}\isa{{\isacharparenleft}{\kern0pt}{\isasymforall}T{\isacharprime}{\kern0pt}\ A{\isacharprime}{\kern0pt}{\isachardot}{\kern0pt}\ lb\ T{\isacharprime}{\kern0pt}\ A{\isacharprime}{\kern0pt}\ n\ {\isasymlongrightarrow}\ makespan\ T\ {\isasymle}\ {\isadigit{3}}\ {\isacharslash}{\kern0pt}\ {\isadigit{2}}\ {\isacharasterisk}{\kern0pt}\ makespan\ T{\isacharprime}{\kern0pt}{\isacharparenright}{\kern0pt}{\isacharbraceright}{\kern0pt}}
\end{framedni}
\noindent
Property \isa{sorted\ n} does not need to be part of the precondition because it does not mention
any program variable.
Therefore we can make \isa{sorted\ n} an assumption of the whole Hoare triple,
which simplifies the proof.
The notation \isa{f{\isacharparenleft}{\kern0pt}a\ {\isacharcolon}{\kern0pt}{\isacharequal}{\kern0pt}\ b{\isacharparenright}{\kern0pt}} denotes an updated version of function \isa{f}
that maps \isa{a} to \isa{b} and behaves like \isa{f} otherwise.
Thus an assignment \isa{f\ {\isacharcolon}{\kern0pt}{\isacharequal}{\kern0pt}\ f{\isacharparenleft}{\kern0pt}i\ {\isacharcolon}{\kern0pt}{\isacharequal}{\kern0pt}\ b{\isacharparenright}{\kern0pt}} is nothing but the conventional imperative array update
notation \isa{f{\isacharbrackleft}{\kern0pt}i{\isacharbrackright}{\kern0pt}\ {\isacharcolon}{\kern0pt}{\isacharequal}{\kern0pt}\ b}.

Functional correctness follows because each iteration extends a partial solution for \isa{j} jobs
to one for \isa{j\ {\isacharplus}{\kern0pt}\ {\isadigit{1}}} jobs:
\begin{lem}\label{lem:add_job}
\isa{lb\ T\ A\ j\ {\isasymand}\ x\ {\isasymin}\ {\isacharbraceleft}{\kern0pt}{\isadigit{1}}{\isachardot}{\kern0pt}{\isachardot}{\kern0pt}m{\isacharbraceright}{\kern0pt}\ {\isasymlongrightarrow}\isanewline
lb\ {\isacharparenleft}{\kern0pt}T{\isacharparenleft}{\kern0pt}x\ {\isacharcolon}{\kern0pt}{\isacharequal}{\kern0pt}\ T\ x\ {\isacharplus}{\kern0pt}\ t\ {\isacharparenleft}{\kern0pt}j\ {\isacharplus}{\kern0pt}\ {\isadigit{1}}{\isacharparenright}{\kern0pt}{\isacharparenright}{\kern0pt}{\isacharparenright}{\kern0pt}\ {\isacharparenleft}{\kern0pt}A{\isacharparenleft}{\kern0pt}x\ {\isacharcolon}{\kern0pt}{\isacharequal}{\kern0pt}\ A\ x\ {\isasymunion}\ {\isacharbraceleft}{\kern0pt}j\ {\isacharplus}{\kern0pt}\ {\isadigit{1}}{\isacharbraceright}{\kern0pt}{\isacharparenright}{\kern0pt}{\isacharparenright}{\kern0pt}\ {\isacharparenleft}{\kern0pt}j\ {\isacharplus}{\kern0pt}\ {\isadigit{1}}{\isacharparenright}{\kern0pt}}
\end{lem}
\noindent
Moreover, it is easy to see that the initialization establishes \isa{lb\ T\ A\ j}.

To prove the approximation factor in both the sorted and unsorted case, the following lower bound is important:
\begin{lem}\label{lem:job_dist_lower_bound_makespan}
\isa{lb\ T\ A\ j\ {\isasymlongrightarrow}\ {\isacharparenleft}{\kern0pt}$\sum_{x\ {\isacharequal}{\kern0pt}\ {\isadigit{1}}}^{j}$\ t\ x{\isacharparenright}{\kern0pt}\ {\isacharslash}{\kern0pt}\ m\ {\isasymle}\ makespan\ T}
\end{lem}
\noindent
This is a result of $\sum_{x = 1}^{m} T(x) = \sum_{x = 1}^{j} t(x)$ together with
this general property of sums: \isa{finite\ A\ {\isasymand}\ A\ {\isasymnoteq}\ {\isasymemptyset}\ {\isasymlongrightarrow}\ {\isacharparenleft}{\kern0pt}{\isasymSum}a{\isasymin}A{\isachardot}{\kern0pt}\ f\ a{\isacharparenright}{\kern0pt}\ {\isasymle}\ {\isacharbar}{\kern0pt}A{\isacharbar}{\kern0pt}\ {\isacharasterisk}{\kern0pt}\ Max\ {\isacharparenleft}{\kern0pt}f\ {\isacharbackquote}{\kern0pt}\ A{\isacharparenright}{\kern0pt}}.

A similar observation applies to individual jobs. Any job must be a lower bound on some machine, as it is assigned to one and, by extension, it must also be a lower bound of the makespan:
\begin{lem}\label{lem:max_job_lower_bound_makespan}
\isa{lb\ T\ A\ j\ {\isasymlongrightarrow}\ Max\isactrlsub {\isadigit{0}}\ {\isacharparenleft}{\kern0pt}t\ {\isacharbackquote}{\kern0pt}\ {\isacharbraceleft}{\kern0pt}{\isadigit{1}}{\isachardot}{\kern0pt}{\isachardot}{\kern0pt}j{\isacharbraceright}{\kern0pt}{\isacharparenright}{\kern0pt}\ {\isasymle}\ makespan\ T}
\end{lem}
\noindent
As any job load is a lower bound on the makespan over the machines, the job with maximum load must also be a lower bound. Note that \isa{Max\isactrlsub {\isadigit{0}}} returns 0 for the empty set.

When jobs are sorted in descending order, a stricter lower bound for an individual job can be established. We observe that an added job is at most as large as the jobs preceding it. Therefore, if a machine contains at least two jobs, this added job is only at most \emph{half} as large as the makespan. We can use this observation by assuming the machines to be filled with more than \isa{m} jobs, as this will ensure that some machine must contain at least two jobs.
\begin{lem}\label{lem:sorted_job_lower_bound_makespan}
\isa{lb\ T\ A\ j\ {\isasymand}\ m\ {\isacharless}{\kern0pt}\ j\ {\isasymand}\ sorted\ j\ {\isasymlongrightarrow}\ {\isadigit{2}}\ {\isacharasterisk}{\kern0pt}\ t\ j\ {\isasymle}\ makespan\ T}
\end{lem}
\noindent
Note that this lower bound only holds if there are strictly more jobs than machines. One must, however, also consider how the algorithm behaves in the other case. One may intuitively see that the algorithm will be able to distribute the jobs such that every machine will only have at most one job assigned to it, making the algorithm trivially optimal. To prove this, we need to show the following behavior of \isa{min{\isacharunderscore}{\kern0pt}arg}:
\begin{lem}\label{lem:min_zero}
\leavevmode
\makeatletter
\@nobreaktrue
\makeatother
\begin{enumerate}
\item \isa{x\ {\isasymin}\ {\isacharbraceleft}{\kern0pt}{\isadigit{1}}{\isachardot}{\kern0pt}{\isachardot}{\kern0pt}m{\isacharbraceright}{\kern0pt}\ {\isasymand}\ T\ x\ {\isacharequal}{\kern0pt}\ {\isadigit{0}}\ {\isasymlongrightarrow}\ T\ {\isacharparenleft}{\kern0pt}min{\isacharunderscore}{\kern0pt}arg\ T\ m{\isacharparenright}{\kern0pt}\ {\isacharequal}{\kern0pt}\ {\isadigit{0}}}
\item \isa{x\ {\isasymin}\ {\isacharbraceleft}{\kern0pt}{\isadigit{1}}{\isachardot}{\kern0pt}{\isachardot}{\kern0pt}m{\isacharbraceright}{\kern0pt}\ {\isasymand}\ T\ x\ {\isacharequal}{\kern0pt}\ {\isadigit{0}}\ {\isasymlongrightarrow}\ min{\isacharunderscore}{\kern0pt}arg\ T\ m\ {\isasymle}\ x}
\end{enumerate}
\end{lem}
\noindent
Both properties can be shown by induction on the number of machines \isa{m}.

As the proof in \cite{KleinbergT06} is only informal, Kleinberg and Tardos do not provide any loop invariant. We propose the following invariant for sorted jobs:
\begin{quoteni}
\isa{inv\isactrlsub {\isadigit{2}}\ {\isacharcolon}{\kern0pt}{\isacharcolon}{\kern0pt}\ {\isacharparenleft}{\kern0pt}nat\ {\isasymRightarrow}\ nat{\isacharparenright}{\kern0pt}\ {\isasymRightarrow}\ {\isacharparenleft}{\kern0pt}nat\ {\isasymRightarrow}\ nat\ set{\isacharparenright}{\kern0pt}\ {\isasymRightarrow}\ nat\ {\isasymRightarrow}\ bool}\\
\isa{inv\isactrlsub {\isadigit{2}}\ T\ A\ j\ {\isacharequal}{\kern0pt}}\\
\isa{{\isacharparenleft}{\kern0pt}lb\ T\ A\ j\ {\isasymand}\ j\ {\isasymle}\ n\ {\isasymand}}\\
\hphantom{(}\isa{{\isacharparenleft}{\kern0pt}{\isasymforall}T{\isacharprime}{\kern0pt}\ A{\isacharprime}{\kern0pt}{\isachardot}{\kern0pt}\ lb\ T{\isacharprime}{\kern0pt}\ A{\isacharprime}{\kern0pt}\ j\ {\isasymlongrightarrow}\ makespan\ T\ {\isasymle}\ {\isadigit{3}}\ {\isacharslash}{\kern0pt}\ {\isadigit{2}}\ {\isacharasterisk}{\kern0pt}\ makespan\ T{\isacharprime}{\kern0pt}{\isacharparenright}{\kern0pt}\ {\isasymand}}\\
\hphantom{(}\isa{{\isacharparenleft}{\kern0pt}{\isasymforall}x\ {\isachargreater}{\kern0pt}\ j{\isachardot}{\kern0pt}\ T\ x\ {\isacharequal}{\kern0pt}\ {\isadigit{0}}{\isacharparenright}{\kern0pt}\ {\isasymand}\ {\isacharparenleft}{\kern0pt}j\ {\isasymle}\ m\ {\isasymlongrightarrow}\ makespan\ T\ {\isacharequal}{\kern0pt}\ Max\isactrlsub {\isadigit{0}}\ {\isacharparenleft}{\kern0pt}t\ {\isacharbackquote}{\kern0pt}\ {\isacharbraceleft}{\kern0pt}{\isadigit{1}}{\isachardot}{\kern0pt}{\isachardot}{\kern0pt}j{\isacharbraceright}{\kern0pt}{\isacharparenright}{\kern0pt}{\isacharparenright}{\kern0pt}{\isacharparenright}{\kern0pt}}
\end{quoteni}
The final two conjuncts relate to the trivially optimal behavior of the algorithm if \isa{j\ {\isasymle}\ m}. The penultimate conjunct shows that only as many machines can be occupied as there are available jobs. 
The final conjunct ensures that every job is distributed on its own machine, making the makespan equivalent to the job with maximum load.

It should be noted that if the makespan is sufficiently large, an added job may not increase the makespan at all, as the machine with minimum load combined with the job may not exceed the previous makespan. As such, we will also consider the possibility that an added job can simply be ignored without affecting the overall makespan.
\begin{lem}\label{lem:remove_small_job}
\begin{isabelle}%
makespan\ {\isacharparenleft}{\kern0pt}T{\isacharparenleft}{\kern0pt}x\ {\isacharcolon}{\kern0pt}{\isacharequal}{\kern0pt}\ T\ x\ {\isacharplus}{\kern0pt}\ y{\isacharparenright}{\kern0pt}{\isacharparenright}{\kern0pt}\ {\isasymnoteq}\ T\ x\ {\isacharplus}{\kern0pt}\ y\ {\isasymlongrightarrow}\isanewline
makespan\ {\isacharparenleft}{\kern0pt}T{\isacharparenleft}{\kern0pt}x\ {\isacharcolon}{\kern0pt}{\isacharequal}{\kern0pt}\ T\ x\ {\isacharplus}{\kern0pt}\ y{\isacharparenright}{\kern0pt}{\isacharparenright}{\kern0pt}\ {\isacharequal}{\kern0pt}\ makespan\ T%
\end{isabelle}
\end{lem}
To make use of this observation, we need to be able to relate the makespan of a solution with the added job to the makespan of a solution without it. One can easily show the following by removing \isa{j\ {\isacharplus}{\kern0pt}\ {\isadigit{1}}} from the solution:
\begin{lem}\label{lem:smaller_optimum}
\isa{lb\ T\ A\ {\isacharparenleft}{\kern0pt}j\ {\isacharplus}{\kern0pt}\ {\isadigit{1}}{\isacharparenright}{\kern0pt}\ {\isasymlongrightarrow}\isanewline
{\isacharparenleft}{\kern0pt}{\isasymexists}T{\isacharprime}{\kern0pt}\ A{\isacharprime}{\kern0pt}{\isachardot}{\kern0pt}\ lb\ T{\isacharprime}{\kern0pt}\ A{\isacharprime}{\kern0pt}\ j\ {\isasymand}\ makespan\ T{\isacharprime}{\kern0pt}\ {\isasymle}\ makespan\ T{\isacharparenright}{\kern0pt}}
\end{lem}

We can now prove the preservation of \isa{inv\isactrlsub {\isadigit{2}}}. Let \isa{i\ {\isacharequal}{\kern0pt}\ min{\isacharunderscore}{\kern0pt}arg\ T\ m} be the machine with minimum load. We define:
\begin{quote}
\isa{T\isactrlsub g\ {\isacharcolon}{\kern0pt}{\isacharequal}{\kern0pt}\ T\ {\isacharparenleft}{\kern0pt}i\ {\isacharcolon}{\kern0pt}{\isacharequal}{\kern0pt}\ T{\isacharparenleft}{\kern0pt}i{\isacharparenright}{\kern0pt}\ {\isacharplus}{\kern0pt}\ t{\isacharparenleft}{\kern0pt}j\ {\isacharplus}{\kern0pt}\ {\isadigit{1}}{\isacharparenright}{\kern0pt}{\isacharparenright}{\kern0pt}} \qquad
\isa{A\isactrlsub g\ {\isacharcolon}{\kern0pt}{\isacharequal}{\kern0pt}\ A\ {\isacharparenleft}{\kern0pt}i\ {\isacharcolon}{\kern0pt}{\isacharequal}{\kern0pt}\ A{\isacharparenleft}{\kern0pt}i{\isacharparenright}{\kern0pt}\ {\isasymunion}\ {\isacharbraceleft}{\kern0pt}j\ {\isacharplus}{\kern0pt}\ {\isadigit{1}}{\isacharbraceright}{\kern0pt}{\isacharparenright}{\kern0pt}}
\end{quote}
We begin with a case distinction. If \isa{j\ {\isacharplus}{\kern0pt}\ {\isadigit{1}}\ {\isasymle}\ m}, we can make use of the additional conjuncts to prove the trivially optimal behavior. We first note \textit{in-range}: $\isa{j\ {\isacharplus}{\kern0pt}\ {\isadigit{1}}\ {\isasymin}\ {\isacharbraceleft}{\kern0pt}{\isadigit{1}}{\isachardot}{\kern0pt}{\isachardot}{\kern0pt}m{\isacharbraceright}{\kern0pt}}$. Moreover, from the penultimate conjunct, $\isa{T{\isacharparenleft}{\kern0pt}j\ {\isacharplus}{\kern0pt}\ {\isadigit{1}}{\isacharparenright}{\kern0pt}\ {\isacharequal}{\kern0pt}\ {\isadigit{0}}}$. Combining this with Lemma~\ref{lem:min_zero}.1, we can see that $\isa{T{\isacharparenleft}{\kern0pt}i{\isacharparenright}{\kern0pt}\ {\isacharequal}{\kern0pt}\ {\isadigit{0}}}$. Therefore \isa{T\isactrlsub g{\isacharparenleft}{\kern0pt}i{\isacharparenright}{\kern0pt}\ {\isacharequal}{\kern0pt}\ t{\isacharparenleft}{\kern0pt}j\ {\isacharplus}{\kern0pt}\ {\isadigit{1}}{\isacharparenright}{\kern0pt}} and with the final conjunct of the assumed invariant, the makespan of \isa{T\isactrlsub g} remains equivalent to the job with maximum load. To prove that the penultimate conjunct is preserved, we again use \textit{in-range}, $\isa{T{\isacharparenleft}{\kern0pt}j\ {\isacharplus}{\kern0pt}\ {\isadigit{1}}{\isacharparenright}{\kern0pt}\ {\isacharequal}{\kern0pt}\ {\isadigit{0}}}$, and Lemma~\ref{lem:min_zero}.2 to prove that $\isa{i\ {\isasymle}\ j\ {\isacharplus}{\kern0pt}\ {\isadigit{1}}}$. Moreover, \isa{T\isactrlsub g} only differs from \isa{T} by the modification of machine \isa{i}. Thus, the penultimate conjunct for \isa{j\ {\isacharplus}{\kern0pt}\ {\isadigit{1}}} jobs is preserved as well. From Lemma~\ref{lem:max_job_lower_bound_makespan} we can then see that, as the makespan of \isa{T\isactrlsub g} is equivalent to the job with maximum load, it must be trivially optimal. Functional correctness can be shown using Lemma~\ref{lem:add_job}, and proving the preservation of the remaining conjunct is trivial. We now come to the case $\isa{j\ {\isacharplus}{\kern0pt}\ {\isadigit{1}}\ {\isachargreater}{\kern0pt}\ m}$. We first show that the penultimate conjunct is preserved (the final conjunct can be ignored, as $\isa{{\isasymnot}\ j\ {\isacharplus}{\kern0pt}\ {\isadigit{1}}\ {\isasymle}\ m}$). This follows from the correctness of \isa{min{\isacharunderscore}{\kern0pt}arg}, as the index returned by it has to be in $\isa{{\isacharbraceleft}{\kern0pt}{\isadigit{1}}{\isachardot}{\kern0pt}{\isachardot}{\kern0pt}m{\isacharbraceright}{\kern0pt}}$ as long as $\isa{m\ {\isachargreater}{\kern0pt}\ {\isadigit{0}}}$. Therefore, we can simply show this from the penultimate conjunct of the assumed invariant. We now come to the proof of the approximation factor:
\begin{quoteni}
\isa{{\isasymforall}T{\isacharprime}{\kern0pt}\ A{\isacharprime}{\kern0pt}{\isachardot}{\kern0pt}\ lb\ T{\isacharprime}{\kern0pt}\ A{\isacharprime}{\kern0pt}\ {\isacharparenleft}{\kern0pt}j\ {\isacharplus}{\kern0pt}\ {\isadigit{1}}{\isacharparenright}{\kern0pt}\ {\isasymlongrightarrow}\ makespan\ T\isactrlsub g\ {\isasymle}\ {\isadigit{3}}\ {\isacharslash}{\kern0pt}\ {\isadigit{2}}\ {\isacharasterisk}{\kern0pt}\ makespan\ T{\isacharprime}{\kern0pt}}
\end{quoteni}
To prove it, we fix \isa{T\isactrlsub {\isadigit{1}}} and \isa{A\isactrlsub {\isadigit{1}}} such that \isa{lb\ T\isactrlsub {\isadigit{1}}\ A\isactrlsub {\isadigit{1}}\ {\isacharparenleft}{\kern0pt}j\ {\isacharplus}{\kern0pt}\ {\isadigit{1}}{\isacharparenright}{\kern0pt}}. Using Lemma~\ref{lem:smaller_optimum}, one can now obtain \isa{T\isactrlsub {\isadigit{0}}} and \isa{A\isactrlsub {\isadigit{0}}} such that \isa{lb\ T\isactrlsub {\isadigit{0}}\ A\isactrlsub {\isadigit{0}}\ j} and \textit{MK}: \isa{makespan\ T\isactrlsub {\isadigit{0}}\ {\isasymle}\ makespan\ T\isactrlsub {\isadigit{1}}}. From the assumed loop invariant, we can now show:
\begin{align*}
\isa{makespan\ T} &\le \frac32 \isa{makespan\ T\isactrlsub {\isadigit{0}}}\tag*{by \isa{inv\isactrlsub {\isadigit{2}}}-def}\\
             &\le \frac32 \isa{makespan\ T\isactrlsub {\isadigit{1}}}\tag*{by \textit{MK}}
\end{align*}
To prove the makespan for \isa{j\ {\isacharplus}{\kern0pt}\ {\isadigit{1}}} jobs, there are now two cases to consider: The added job \isa{j\ {\isacharplus}{\kern0pt}\ {\isadigit{1}}} contributes to the makespan or it does not. The case in which it does not can be shown by combining the previous calculation with Lemma~\ref{lem:remove_small_job}. For the first case, we may then assume that \isa{makespan\ T\isactrlsub g\ {\isacharequal}{\kern0pt}\ T{\isacharparenleft}{\kern0pt}i{\isacharparenright}{\kern0pt}\ {\isacharplus}{\kern0pt}\ t{\isacharparenleft}{\kern0pt}j\ {\isacharplus}{\kern0pt}\ {\isadigit{1}}{\isacharparenright}{\kern0pt}}. Like in Lemma~\ref{lem:job_dist_lower_bound_makespan}, we note that \textit{sum-eq}: \isa{{\isacharparenleft}{\kern0pt}$\sum_{x\ {\isacharequal}{\kern0pt}\ {\isadigit{1}}}^{m}$\ T\ x{\isacharparenright}{\kern0pt}\ {\isacharequal}{\kern0pt}\ {\isacharparenleft}{\kern0pt}$\sum_{x\ {\isacharequal}{\kern0pt}\ {\isadigit{1}}}^{j}$\ t\ x{\isacharparenright}{\kern0pt}}. Moreover, \textit{min-avg}: \isa{m\ {\isacharasterisk}{\kern0pt}\ T\ {\isacharparenleft}{\kern0pt}min{\isacharunderscore}{\kern0pt}arg\ T\ m{\isacharparenright}{\kern0pt}\ {\isasymle}\ {\isacharparenleft}{\kern0pt}$\sum_{i\ {\isacharequal}{\kern0pt}\ {\isadigit{1}}}^{m}$\ T\ i{\isacharparenright}{\kern0pt}}. This allows us to calculate the following lower bound for \isa{T{\isacharparenleft}{\kern0pt}i{\isacharparenright}{\kern0pt}}:
\begin{align*}
\isa{m\ {\isacharasterisk}{\kern0pt}\ T{\isacharparenleft}{\kern0pt}i{\isacharparenright}{\kern0pt}} &\le \sum_{i = 1}^{m} T(i) = \sum_{i = 1}^{j} t(i)\tag*{by \textit{min-avg} and \textit{sum-eq}}\\
\iff  T(i) &\le \frac{\sum_{i = 1}^{j} t(i)}{m}\tag*{because \isa{m\ {\isachargreater}{\kern0pt}\ {\isadigit{0}}}}\\
           &\le \isa{makespan\ T\isactrlsub {\isadigit{0}}} \le \isa{makespan\ T\isactrlsub {\isadigit{1}}}\tag*{by Lemma~\ref{lem:job_dist_lower_bound_makespan} and \textit{MK}}
\end{align*}
From Lemma~\ref{lem:sorted_job_lower_bound_makespan} we can also show that \isa{t{\isacharparenleft}{\kern0pt}j\ {\isacharplus}{\kern0pt}\ {\isadigit{1}}{\isacharparenright}{\kern0pt}} is a lower bound for $\frac12$ of the makespan of \isa{T\isactrlsub {\isadigit{1}}}. Therefore:
\begin{align*}
\isa{makespan\ T\isactrlsub g\ {\isacharequal}{\kern0pt}\ T{\isacharparenleft}{\kern0pt}i{\isacharparenright}{\kern0pt}\ {\isacharplus}{\kern0pt}\ t{\isacharparenleft}{\kern0pt}j\ {\isacharplus}{\kern0pt}\ {\isadigit{1}}{\isacharparenright}{\kern0pt}} &\le \isa{makespan\ T\isactrlsub {\isadigit{1}}} + \frac{\isa{makespan\ T\isactrlsub {\isadigit{1}}}}{2}\\
                                &= \frac32 \isa{makespan\ T\isactrlsub {\isadigit{1}}}
\end{align*}
The proof of functional correctness and remaining conjuncts is again trivial.

Let us now consider the unsorted case where one can still show an approximation factor of 2. The algorithm is identical but the invariant is simpler:
\begin{quoteni}
\isa{inv\isactrlsub {\isadigit{1}}\ T\ A\ j\ {\isacharequal}{\kern0pt}\isanewline
{\isacharparenleft}{\kern0pt}lb\ T\ A\ j\ {\isasymand}\ j\ {\isasymle}\ n\ {\isasymand}\ {\isacharparenleft}{\kern0pt}{\isasymforall}T{\isacharprime}{\kern0pt}\ A{\isacharprime}{\kern0pt}{\isachardot}{\kern0pt}\ lb\ T{\isacharprime}{\kern0pt}\ A{\isacharprime}{\kern0pt}\ j\ {\isasymlongrightarrow}\ makespan\ T\ {\isasymle}\ {\isadigit{2}}\ {\isacharasterisk}{\kern0pt}\ makespan\ T{\isacharprime}{\kern0pt}{\isacharparenright}{\kern0pt}{\isacharparenright}{\kern0pt}}
\end{quoteni}
The proof for this invariant is a simpler version of the proof above: We do not need the
initial case distinction (case \isa{j\ {\isacharplus}{\kern0pt}\ {\isadigit{1}}\ {\isasymle}\ m} need not be considered separately),
and we use Lemma~\ref{lem:max_job_lower_bound_makespan} instead of Lemma~\ref{lem:sorted_job_lower_bound_makespan} to obtain a bound for \isa{t{\isacharparenleft}{\kern0pt}j\ {\isacharplus}{\kern0pt}\ {\isadigit{1}}{\isacharparenright}{\kern0pt}}.%
\end{isamarkuptext}\isamarkuptrue%
\isadelimtheory
\endisadelimtheory
\isatagtheory
\endisatagtheory
{\isafoldtheory}%
\isadelimtheory
\endisadelimtheory
\end{isabellebody}%

%% file: Paper_CS.tex
\begin{isabellebody}%
\setisabellecontext{Paper{\isacharunderscore}{\kern0pt}CS}%
\isadelimtheory
\endisadelimtheory
\isatagtheory
\endisatagtheory
{\isafoldtheory}%
\isadelimtheory
\endisadelimtheory
\isadelimproof
\endisadelimproof
\isatagproof
\endisatagproof
{\isafoldproof}%
\isadelimproof
\endisadelimproof
\begin{isamarkuptext}%
\section{Center Selection}

This section is based on \cite[Chapter 11.2]{KleinbergT06}.
Given a finite, non-empty set of sites $\isa{S\ {\isacharcolon}{\kern0pt}{\isacharcolon}{\kern0pt}\ {\isacharprime}{\kern0pt}a\ set}$ (i.e., points) in a metric space, our objective is to select an optimal set of centers \isa{C\ {\isasymsubseteq}\ S} of size \isa{k\ {\isachargreater}{\kern0pt}\ {\isadigit{0}}}, such that
\begin{quoteni}
\isa{radius\ C\ {\isacharequal}{\kern0pt}\ Max\ {\isacharparenleft}{\kern0pt}distance\ C\ {\isacharbackquote}{\kern0pt}\ S{\isacharparenright}{\kern0pt}}
\end{quoteni}
is minimized, where \isa{distance\ C\ s\ {\isacharequal}{\kern0pt}\ Min\ {\isacharparenleft}{\kern0pt}dist\ s\ {\isacharbackquote}{\kern0pt}\ C{\isacharparenright}{\kern0pt}}.

We call a site \isa{s} a $\textit{candidate}$ (with respect to some \isa{r}) if \isa{distance\ C\ s\ {\isachargreater}{\kern0pt}\ {\isadigit{2}}r} and call it $\textit{included}$ if \isa{distance\ C\ s\ {\isasymle}\ {\isadigit{2}}r} (i.e., if it already lies within \isa{{\isadigit{2}}r} of some center). Now consider the following two observations for the optimal radius $r^*$:
\begin{enumerate}
 \item Selecting as a center a candidate site \isa{s} w.r.t.\ $r^*$ guarantees that at least one more site will now lie within $2r^*$ of a center---namely \isa{s} itself---and rightfully discards sites which are already included.
  \item Should there exist a candidate site w.r.t.\ $r^*$, then a furthest site will also be a candidate (w.r.t.\ $r^*$).
\end{enumerate}

The strategy is then to construct \isa{C} by repeatedly selecting a furthest site. Of course, it's not immediatly clear why this approach should work for the $2r^*$ boundary but not, say, for $1.5r^*$. We resolve this matter in the proof's details.

Before we proceed, we note that our formal proof diverges from that of Kleinberg and Tardos. While they begin by constructing an algorithm around the first observation and proceed to establish, rather informally, a semantic correspondence between it and the algorithm presented in this paper, we follow a more direct approach---that is, we only reason about the presented algorithm. This reduces the first algorithm to a pedagogical tool.

The algorithm repeatedly selects a site that lies furthest from the set of all hitherto selected centers.
The following Hoare triple expresses functional correctness and the approximation factor of 2.

\begin{framedni}
\isa{{\isacharbraceleft}{\kern0pt}k\ {\isasymle}\ {\isacharbar}{\kern0pt}S{\isacharbar}{\kern0pt}{\isacharbraceright}{\kern0pt}}\\
\isa{C\ {\isacharcolon}{\kern0pt}{\isacharequal}{\kern0pt}\ {\isacharparenleft}{\kern0pt}SOME\ s{\isachardot}{\kern0pt}\ s\ {\isasymin}\ S{\isacharparenright}{\kern0pt}{\isacharsemicolon}{\kern0pt}}\\
\isa{WHILE\ {\isacharbar}{\kern0pt}C{\isacharbar}{\kern0pt}\ {\isacharless}{\kern0pt}\ k\ INV\ {\isacharbraceleft}{\kern0pt}invar\ C{\isacharbraceright}{\kern0pt}\ DO}\\
\hphantom{\isa{DO}}\isa{C\ {\isacharcolon}{\kern0pt}{\isacharequal}{\kern0pt}\ C\ {\isasymunion}\ {\isacharbraceleft}{\kern0pt}furthest{\isacharunderscore}{\kern0pt}from\ C{\isacharbraceright}{\kern0pt}{\isacharsemicolon}{\kern0pt}}\\
\isa{OD}\\
\isa{{\isacharbraceleft}{\kern0pt}{\isacharbar}{\kern0pt}C{\isacharbar}{\kern0pt}\ {\isacharequal}{\kern0pt}\ k\ {\isasymand}\ {\isacharparenleft}{\kern0pt}{\isasymforall}C{\isacharprime}{\kern0pt}{\isachardot}{\kern0pt}\ {\isadigit{0}}\ {\isacharless}{\kern0pt}\ {\isacharbar}{\kern0pt}C{\isacharprime}{\kern0pt}{\isacharbar}{\kern0pt}\ {\isasymand}\ {\isacharbar}{\kern0pt}C{\isacharprime}{\kern0pt}{\isacharbar}{\kern0pt}\ {\isasymle}\ k\ {\isasymlongrightarrow}\ radius\ C\ {\isasymle}\ {\isadigit{2}}\ {\isacharasterisk}{\kern0pt}\ radius\ C{\isacharprime}{\kern0pt}{\isacharparenright}{\kern0pt}{\isacharbraceright}{\kern0pt}}
\end{framedni}
\noindent 
where 
\isa{furthest{\isacharunderscore}{\kern0pt}from\ C\ {\isacharequal}{\kern0pt}\ {\isacharparenleft}{\kern0pt}SOME\ s{\isachardot}{\kern0pt}\ s\ {\isasymin}\ S\ {\isasymand}\ distance\ C\ s\ {\isacharequal}{\kern0pt}\ Max\ {\isacharparenleft}{\kern0pt}distance\ C\ {\isacharbackquote}{\kern0pt}\ S{\isacharparenright}{\kern0pt}{\isacharparenright}{\kern0pt}}.
\medskip

We now present the proof, beginning with the invariant. To motivate the choice of invariant, consider the special case when there exists a candidate site (w.r.t.\ some fixed \isa{r}) in each loop iteration. Let \isa{r} be the radius of an arbitrary set, \isa{C} be the set of previously selected centers, \isa{s\ {\isasymin}\ S} the site to be added to \isa{C}, and \isa{s{\isacharprime}{\kern0pt}\ {\isasymin}\ S} a candidate site w.r.t.\ \isa{r}. (Recall: candidate means \isa{distance\ C\ s{\isacharprime}{\kern0pt}\ {\isachargreater}{\kern0pt}\ {\isadigit{2}}\ {\isacharasterisk}{\kern0pt}\ r}). By choice of \isa{s}, we have \isa{distance\ C\ s\ {\isasymge}\ distance\ C\ s{\isacharprime}{\kern0pt}} and therefore \isa{distance\ C\ s\ {\isachargreater}{\kern0pt}\ {\isadigit{2}}\ {\isacharasterisk}{\kern0pt}\ r}. Since \isa{distance\ C\ s\ {\isacharequal}{\kern0pt}\ Min\ {\isacharparenleft}{\kern0pt}dist\ C\ {\isacharbackquote}{\kern0pt}\ S{\isacharparenright}{\kern0pt}}, we have \isa{{\isasymforall}c\ {\isasymin}\ C{\isachardot}{\kern0pt}\ dist\ c\ s\ {\isasymge}\ distance\ C\ s\ {\isachargreater}{\kern0pt}\ {\isadigit{2}}\ {\isacharasterisk}{\kern0pt}\ r}. Inductively, it follows (which we prove later) that
\begin{quoteni}
\isa{{\isasymforall}c\isactrlsub {\isadigit{1}}\ {\isasymin}\ C{\isachardot}{\kern0pt}\ {\isasymforall}c\isactrlsub {\isadigit{2}}\ {\isasymin}\ C{\isachardot}{\kern0pt}\ c\isactrlsub {\isadigit{1}}\ {\isasymnoteq}\ c\isactrlsub {\isadigit{2}}\ {\isasymlongrightarrow}\ dist\ c\isactrlsub {\isadigit{1}}\ c\isactrlsub {\isadigit{2}}\ {\isachargreater}{\kern0pt}\ {\isadigit{2}}\ {\isacharasterisk}{\kern0pt}\ r}.
\end{quoteni}

Now consider the case when in some loop iteration no more candidate sites exist, i.e., all sites are included. As before, let \isa{C} be the set of previously selected centers. By assumption, we have \isa{{\isasymforall}s\ {\isasymin}\ S{\isachardot}{\kern0pt}\ distance\ C\ s\ {\isasymle}\ {\isadigit{2}}\ {\isacharasterisk}{\kern0pt}\ r}.

This case distinction forms the invariant, which is defined as:
\begin{quoteni}
\begin{isabelle}%
invar\ C\ {\isacharequal}{\kern0pt}\isanewline
{\isacharparenleft}{\kern0pt}C\ {\isasymnoteq}\ {\isasymemptyset}\ {\isasymand}\ {\isacharbar}{\kern0pt}C{\isacharbar}{\kern0pt}\ {\isasymle}\ k\ {\isasymand}\ C\ {\isasymsubseteq}\ S\ {\isasymand}\isanewline
\isaindent{{\isacharparenleft}{\kern0pt}}{\isacharparenleft}{\kern0pt}{\isasymforall}C{\isacharprime}{\kern0pt}{\isachardot}{\kern0pt}\ {\isacharparenleft}{\kern0pt}{\isasymforall}c\isactrlsub {\isadigit{1}}{\isasymin}C{\isachardot}{\kern0pt}\ {\isasymforall}c\isactrlsub {\isadigit{2}}{\isasymin}C{\isachardot}{\kern0pt}\ c\isactrlsub {\isadigit{1}}\ {\isasymnoteq}\ c\isactrlsub {\isadigit{2}}\ {\isasymlongrightarrow}\ {\isadigit{2}}\ {\isacharasterisk}{\kern0pt}\ radius\ C{\isacharprime}{\kern0pt}\ {\isacharless}{\kern0pt}\ dist\ c\isactrlsub {\isadigit{1}}\ c\isactrlsub {\isadigit{2}}{\isacharparenright}{\kern0pt}\ {\isasymor}\isanewline
\isaindent{{\isacharparenleft}{\kern0pt}{\isacharparenleft}{\kern0pt}{\isasymforall}C{\isacharprime}{\kern0pt}{\isachardot}{\kern0pt}\ }{\isacharparenleft}{\kern0pt}{\isasymforall}s{\isasymin}S{\isachardot}{\kern0pt}\ distance\ C\ s\ {\isasymle}\ {\isadigit{2}}\ {\isacharasterisk}{\kern0pt}\ radius\ C{\isacharprime}{\kern0pt}{\isacharparenright}{\kern0pt}{\isacharparenright}{\kern0pt}{\isacharparenright}{\kern0pt}%
\end{isabelle}
\end{quoteni}
It is obvious that the invariant holds initially for \isa{C\ {\isacharequal}{\kern0pt}\ {\isacharbraceleft}{\kern0pt}SOME\ s{\isachardot}{\kern0pt}\ s\ {\isasymin}\ S{\isacharbraceright}{\kern0pt}}.
Before we prove that the invariant holds in each iteration, we first prove a useful lemma.

\begin{lem}\label{lem:dist_ins}
\begin{isabelle}%
{\isacharparenleft}{\kern0pt}{\isasymforall}c\isactrlsub {\isadigit{1}}{\isasymin}C{\isachardot}{\kern0pt}\ {\isasymforall}c\isactrlsub {\isadigit{2}}{\isasymin}C{\isachardot}{\kern0pt}\ c\isactrlsub {\isadigit{1}}\ {\isasymnoteq}\ c\isactrlsub {\isadigit{2}}\ {\isasymlongrightarrow}\ x\ {\isacharless}{\kern0pt}\ dist\ c\isactrlsub {\isadigit{1}}\ c\isactrlsub {\isadigit{2}}{\isacharparenright}{\kern0pt}\ {\isasymand}\isanewline
x\ {\isacharless}{\kern0pt}\ distance\ C\ s\ {\isasymand}\ finite\ C\ {\isasymand}\ C\ {\isasymnoteq}\ {\isasymemptyset}\ {\isasymlongrightarrow}\isanewline
{\isacharparenleft}{\kern0pt}{\isasymforall}c\isactrlsub {\isadigit{1}}{\isasymin}C\ {\isasymunion}\ {\isacharbraceleft}{\kern0pt}s{\isacharbraceright}{\kern0pt}{\isachardot}{\kern0pt}\ {\isasymforall}c\isactrlsub {\isadigit{2}}{\isasymin}C\ {\isasymunion}\ {\isacharbraceleft}{\kern0pt}s{\isacharbraceright}{\kern0pt}{\isachardot}{\kern0pt}\ c\isactrlsub {\isadigit{1}}\ {\isasymnoteq}\ c\isactrlsub {\isadigit{2}}\ {\isasymlongrightarrow}\ x\ {\isacharless}{\kern0pt}\ dist\ c\isactrlsub {\isadigit{1}}\ c\isactrlsub {\isadigit{2}}{\isacharparenright}{\kern0pt}%
\end{isabelle}
\end{lem}
\proof
The case \isa{c\isactrlsub {\isadigit{1}}\ {\isasymin}\ C\ {\isasymand}\ c\isactrlsub {\isadigit{2}}\ {\isasymin}\ C} is true by assumption and the case \isa{c\isactrlsub {\isadigit{1}}\ {\isacharequal}{\kern0pt}\ s\ {\isasymand}\ c\isactrlsub {\isadigit{2}}\ {\isacharequal}{\kern0pt}\ s} is vacuously true. Assume \isa{c\isactrlsub {\isadigit{1}}\ {\isasymin}\ C\ {\isasymand}\ c\isactrlsub {\isadigit{2}}\ {\isacharequal}{\kern0pt}\ s}. Then
\begin{align*}
\isa{x} &< \isa{distance\ C\ c\isactrlsub {\isadigit{2}}} \tag*{by \isa{distance\ C\ s\ {\isachargreater}{\kern0pt}\ x}} \\
  &\leq \isa{dist\ c\isactrlsub {\isadigit{2}}\ c\isactrlsub {\isadigit{1}}} \tag*{by definition and \isa{C\ {\isasymnoteq}\ {\isacharbraceleft}{\kern0pt}{\isacharbraceright}{\kern0pt}} and \isa{finite\ C}} \\
  &= \isa{dist\ c\isactrlsub {\isadigit{1}}\ c\isactrlsub {\isadigit{2}}} \tag*{by \isa{dist{\isacharunderscore}{\kern0pt}commute}}
\end{align*}
By symmetry, the case \isa{c\isactrlsub {\isadigit{1}}\ {\isacharequal}{\kern0pt}\ s\ {\isasymand}\ c\isactrlsub {\isadigit{2}}\ {\isasymin}\ C} is also true.
\endproof

Now we prove the preservation of the invariant. Assuming \isa{invar\ C}, we show \isa{invar\ {\isacharparenleft}{\kern0pt}C\ {\isasymunion}\ {\isacharbraceleft}{\kern0pt}s{\isacharbraceright}{\kern0pt}{\isacharparenright}{\kern0pt}} where \isa{s\ {\isacharequal}{\kern0pt}\ furthest{\isacharunderscore}{\kern0pt}from\ C}. The first three conjuncts hold trivially. 

\proof
Let \isa{C{\isacharprime}{\kern0pt}} be an arbitrary set and let \isa{r\ {\isacharequal}{\kern0pt}\ radius\ C{\isacharprime}{\kern0pt}}. To prove the fourth conjunct, we distinguish two cases:

\textbf{Case 1} \isa{s} is a candidate site w.r.t.\ \isa{r}. This negates \isa{{\isasymforall}s{\isacharprime}{\kern0pt}\ {\isasymin}\ S{\isachardot}{\kern0pt}\ distance\ C\ s{\isacharprime}{\kern0pt}\ {\isasymle}\ {\isadigit{2}}\ {\isacharasterisk}{\kern0pt}\ r}. Consequently, \isa{{\isasymforall}c\isactrlsub {\isadigit{1}}\ {\isasymin}\ C{\isachardot}{\kern0pt}\ {\isasymforall}c\isactrlsub {\isadigit{2}}\ {\isasymin}\ C{\isachardot}{\kern0pt}\ c\isactrlsub {\isadigit{1}}\ {\isasymnoteq}\ c\isactrlsub {\isadigit{2}}\ {\isasymlongrightarrow}\ {\isadigit{2}}\ {\isacharasterisk}{\kern0pt}\ r\ {\isacharless}{\kern0pt}\ dist\ c\isactrlsub {\isadigit{1}}\ c\isactrlsub {\isadigit{2}}} must hold, since one of the two must be true by assumption \isa{invar\ C}. By substituting \isa{{\isadigit{2}}r} for \isa{x} in Lemma $\ref{lem:dist_ins}$, we conclude \isa{{\isasymforall}c\isactrlsub {\isadigit{1}}\ {\isasymin}\ {\isacharparenleft}{\kern0pt}C\ {\isasymunion}\ {\isacharbraceleft}{\kern0pt}s{\isacharbraceright}{\kern0pt}{\isacharparenright}{\kern0pt}{\isachardot}{\kern0pt}\ {\isasymforall}c\isactrlsub {\isadigit{2}}\ {\isasymin}\ {\isacharparenleft}{\kern0pt}C\ {\isasymunion}\ {\isacharbraceleft}{\kern0pt}s{\isacharbraceright}{\kern0pt}{\isacharparenright}{\kern0pt}{\isachardot}{\kern0pt}\ c\isactrlsub {\isadigit{1}}\ {\isasymnoteq}\ c\isactrlsub {\isadigit{2}}\ {\isasymlongrightarrow}\ {\isadigit{2}}\ {\isacharasterisk}{\kern0pt}\ r\ {\isacharless}{\kern0pt}\ dist\ c\isactrlsub {\isadigit{1}}\ c\isactrlsub {\isadigit{2}}}, hence \isa{invar\ {\isacharparenleft}{\kern0pt}C\ {\isasymunion}\ {\isacharbraceleft}{\kern0pt}s{\isacharbraceright}{\kern0pt}{\isacharparenright}{\kern0pt}}.

\textbf{Case 2} \isa{s} is not a candidate site w.r.t.\ \isa{r}, i.e., \isa{distance\ C\ s\ {\isasymle}\ {\isadigit{2}}\ {\isacharasterisk}{\kern0pt}\ r}. Then, for an arbitrary \isa{s{\isacharprime}{\kern0pt}\ {\isasymin}\ S}
\begin{align*}
  \isa{distance\ {\isacharparenleft}{\kern0pt}C\ {\isasymunion}\ {\isacharbraceleft}{\kern0pt}s{\isacharbraceright}{\kern0pt}{\isacharparenright}{\kern0pt}\ s{\isacharprime}{\kern0pt}} &\leq \isa{distance\ C\ s{\isacharprime}{\kern0pt}} \tag*{by monotonocity of \isa{Min}}\\
                           &\leq \isa{distance\ C\ s} \tag*{by choice of \isa{s}} \\
                           &\leq \isa{{\isadigit{2}}\ {\isacharasterisk}{\kern0pt}\ radius\ C{\isacharprime}{\kern0pt}} \tag*{by assumption}
\end{align*}
Hence \isa{invar\ {\isacharparenleft}{\kern0pt}C\ {\isasymunion}\ {\isacharbraceleft}{\kern0pt}s{\isacharbraceright}{\kern0pt}{\isacharparenright}{\kern0pt}}, i.e., the invariant is preserved.
\endproof

We now prove a lemma concerning the cardinality of the selected centers \isa{C} that will be useful in showing the postcondition. It is furthermore in this lemma that the question raised above about the factor \isa{{\isadigit{2}}} is answered. Concretely, we will show that the \isa{dist\ c\isactrlsub {\isadigit{1}}\ c\isactrlsub {\isadigit{2}}\ {\isachargreater}{\kern0pt}\ {\isadigit{2}}\ {\isacharasterisk}{\kern0pt}\ r} for all \isa{c\isactrlsub {\isadigit{1}}\ {\isasymnoteq}\ c\isactrlsub {\isadigit{2}}} in \isa{C} condition implies that any smaller set of centers must have radius larger than \isa{r}.

\begin{lem}\label{lem:inv_last_2}~\\
\isa{finite\ C\ {\isasymand}\ {\isacharbar}{\kern0pt}C{\isacharbar}{\kern0pt}\ {\isachargreater}{\kern0pt}\ n\ {\isasymand}\ C\ {\isasymsubseteq}\ S\ {\isasymand}}
\\
\isa{{\isacharparenleft}{\kern0pt}{\isasymforall}c\isactrlsub {\isadigit{1}}\ {\isasymin}\ C{\isachardot}{\kern0pt}\ {\isasymforall}c\isactrlsub {\isadigit{2}}\ {\isasymin}\ C{\isachardot}{\kern0pt}\ c\isactrlsub {\isadigit{1}}\ {\isasymnoteq}\ c\isactrlsub {\isadigit{2}}\ {\isasymlongrightarrow}\ dist\ c\isactrlsub {\isadigit{1}}\ c\isactrlsub {\isadigit{2}}\ {\isachargreater}{\kern0pt}\ {\isadigit{2}}\ {\isacharasterisk}{\kern0pt}\ r{\isacharparenright}{\kern0pt}\ {\isasymlongrightarrow}}
\\
\isa{{\isacharparenleft}{\kern0pt}{\isasymforall}C{\isacharprime}{\kern0pt}{\isachardot}{\kern0pt}\ {\isadigit{0}}\ {\isacharless}{\kern0pt}\ {\isacharbar}{\kern0pt}C{\isacharprime}{\kern0pt}{\isacharbar}{\kern0pt}\ {\isasymand}\ {\isacharbar}{\kern0pt}C{\isacharprime}{\kern0pt}{\isacharbar}{\kern0pt}\ {\isasymle}\ n\ {\isasymlongrightarrow}\ radius\ C{\isacharprime}{\kern0pt}\ {\isachargreater}{\kern0pt}\ r{\isacharparenright}{\kern0pt}}
\end{lem}
\proof
Assume, to the contrary, that there exists a set of centers \isa{C{\isacharprime}{\kern0pt}} with \isa{{\isadigit{0}}\ {\isacharless}{\kern0pt}\ {\isacharbar}{\kern0pt}C{\isacharprime}{\kern0pt}{\isacharbar}{\kern0pt}\ {\isasymle}\ n} and \isa{radius\ C{\isacharprime}{\kern0pt}\ {\isasymle}\ r}.
We show that \isa{{\isachardoublequote}{\kern0pt}{\isasymbar}C{\isacharprime}{\kern0pt}{\isasymbar}\ {\isachargreater}{\kern0pt}\ n{\isachardoublequote}{\kern0pt}}, a contradiction.
Since the selected centers in \isa{C} are themselves sites in \isa{S}, each must be by definition within \isa{radius\ C{\isacharprime}{\kern0pt}\ {\isasymle}\ r} of a center in \isa{C{\isacharprime}{\kern0pt}},
but the centers in \isa{C} are all more than \isa{{\isadigit{2}}\ {\isacharasterisk}{\kern0pt}\ r} apart from each other. There cannot be two distinct \isa{c\isactrlsub {\isadigit{1}}{\isacharcomma}{\kern0pt}\ c\isactrlsub {\isadigit{2}}\ {\isasymin}\ C}
with \isa{dist\ c\isactrlsub i\ c{\isacharprime}{\kern0pt}\ {\isasymle}\ r}, otherwise we would have:
\begin{align*}
\isa{{\isadigit{2}}\ {\isacharasterisk}{\kern0pt}\ r}\, &\isa{{\isacharless}{\kern0pt}}\, \isa{dist\ c\isactrlsub {\isadigit{1}}\ c\isactrlsub {\isadigit{2}}} \tag*{by assumption}\\
   &\isa{{\isasymle}}\, \isa{dist\ c\isactrlsub {\isadigit{1}}\ c{\isacharprime}{\kern0pt}\ {\isacharplus}{\kern0pt}\ dist\ c{\isacharprime}{\kern0pt}\ c\isactrlsub {\isadigit{2}}} \tag*{by the triangle inequality} \\
   &\isa{{\isasymle}}\, \isa{{\isadigit{2}}\ {\isacharasterisk}{\kern0pt}\ r} \tag*{by \isa{dist\ c\isactrlsub i\ c{\isacharprime}{\kern0pt}\ {\isasymle}\ r}}
\end{align*}
(Note that this argument would not work for any \isa{{\isasymalpha}\ {\isacharasterisk}{\kern0pt}\ r} with \isa{{\isasymalpha}\ {\isacharless}{\kern0pt}\ {\isadigit{2}}}.)
Therefore there is a unique center in \isa{C{\isacharprime}{\kern0pt}} for each center in \isa{C}, and hence \isa{n\ {\isacharequal}{\kern0pt}\ {\isacharbar}{\kern0pt}C{\isacharprime}{\kern0pt}{\isacharbar}{\kern0pt}\ {\isasymge}\ {\isacharbar}{\kern0pt}C{\isacharbar}{\kern0pt}}.
But we assume \isa{{\isacharbar}{\kern0pt}C{\isacharbar}{\kern0pt}\ {\isachargreater}{\kern0pt}\ n}, a contradiction.
\endproof

Finally we prove that the invariant and \isa{k\ {\isasymle}\ {\isacharbar}{\kern0pt}C{\isacharbar}{\kern0pt}} imply the postcondition:
\begin{quoteni}
\isa{{\isacharbar}{\kern0pt}C{\isacharbar}{\kern0pt}\ {\isacharequal}{\kern0pt}\ k\ {\isasymand}\ {\isacharparenleft}{\kern0pt}{\isasymforall}C{\isacharprime}{\kern0pt}{\isachardot}{\kern0pt}\ {\isacharbar}{\kern0pt}C{\isacharprime}{\kern0pt}{\isacharbar}{\kern0pt}\ {\isachargreater}{\kern0pt}\ {\isadigit{0}}\ {\isasymand}\ {\isacharbar}{\kern0pt}C{\isacharprime}{\kern0pt}{\isacharbar}{\kern0pt}\ {\isasymle}\ k\ {\isasymlongrightarrow}\ radius\ C\ {\isasymle}\ {\isadigit{2}}\ {\isacharasterisk}{\kern0pt}\ radius\ C{\isacharprime}{\kern0pt}{\isacharparenright}{\kern0pt}}
\end{quoteni}

\proof
Here \isa{{\isacharbar}{\kern0pt}C{\isacharbar}{\kern0pt}\ {\isacharequal}{\kern0pt}\ k} follows immediately. Let \isa{C{\isacharprime}{\kern0pt}} be arbitrary and assume \isa{{\isadigit{0}}\ {\isacharless}{\kern0pt}\ {\isacharbar}{\kern0pt}C{\isacharprime}{\kern0pt}{\isacharbar}{\kern0pt}\ {\isasymle}\ k}. We define \isa{r\ {\isacharequal}{\kern0pt}\ radius\ C{\isacharprime}{\kern0pt}}. By the invariant, \isa{{\isacharparenleft}{\kern0pt}{\isasymforall}s\ {\isasymin}\ S{\isachardot}{\kern0pt}\ distance\ C\ s\ {\isasymle}\ {\isadigit{2}}\ {\isacharasterisk}{\kern0pt}\ r{\isacharparenright}{\kern0pt}\ {\isasymor}\ {\isacharparenleft}{\kern0pt}{\isasymforall}c\isactrlsub {\isadigit{1}}\ {\isasymin}\ C{\isachardot}{\kern0pt}\ {\isasymforall}c\isactrlsub {\isadigit{2}}\ {\isasymin}\ C{\isachardot}{\kern0pt}\ c\isactrlsub {\isadigit{1}}\ {\isasymnoteq}\ c\isactrlsub {\isadigit{2}}\ {\isasymlongrightarrow}\ dist\ c\isactrlsub {\isadigit{1}}\ c\isactrlsub {\isadigit{2}}\ {\isachargreater}{\kern0pt}\ {\isadigit{2}}\ {\isacharasterisk}{\kern0pt}\ r{\isacharparenright}{\kern0pt}} holds for \isa{C{\isacharprime}{\kern0pt}}. We prove \isa{radius\ C\ {\isasymle}\ {\isadigit{2}}\ {\isacharasterisk}{\kern0pt}\ r} by case distinction on this disjunction.

\textbf{Case 1} Assumption: \isa{{\isasymforall}s\ {\isasymin}\ S{\isachardot}{\kern0pt}\ distance\ C\ s\ {\isasymle}\ {\isadigit{2}}\ {\isacharasterisk}{\kern0pt}\ r}. Then it trivially follows that \isa{Max\ {\isacharparenleft}{\kern0pt}distance\ C\ {\isacharbackquote}{\kern0pt}\ S{\isacharparenright}{\kern0pt}\ {\isacharequal}{\kern0pt}\ radius\ C\ {\isasymle}\ {\isadigit{2}}\ {\isacharasterisk}{\kern0pt}\ r}.

\textbf{Case 2} Assumption: \isa{{\isasymforall}c\isactrlsub {\isadigit{1}}\ {\isasymin}\ C{\isachardot}{\kern0pt}\ {\isasymforall}c\isactrlsub {\isadigit{2}}\ {\isasymin}\ C{\isachardot}{\kern0pt}\ c\isactrlsub {\isadigit{1}}\ {\isasymnoteq}\ c\isactrlsub {\isadigit{2}}\ {\isasymlongrightarrow}\ dist\ c\isactrlsub {\isadigit{1}}\ c\isactrlsub {\isadigit{2}}\ {\isachargreater}{\kern0pt}\ {\isadigit{2}}\ {\isacharasterisk}{\kern0pt}\ r}. We prove the conclusion by contradiction. Assume, to the contrary, that \isa{radius\ C\ {\isachargreater}{\kern0pt}\ {\isadigit{2}}\ {\isacharasterisk}{\kern0pt}\ r}. This implies that the site \isa{s\ {\isacharequal}{\kern0pt}\ furthest{\isacharunderscore}{\kern0pt}from\ C} is a candidate site w.r.t.\ \isa{r} since \isa{distance\ C\ s\ {\isacharequal}{\kern0pt}\ radius\ C\ {\isachargreater}{\kern0pt}\ {\isadigit{2}}\ {\isacharasterisk}{\kern0pt}\ r}. Furthermore, \isa{s} cannot be in \isa{C}, for if it were, then \isa{{\isadigit{0}}\ {\isacharequal}{\kern0pt}\ distance\ C\ s\ {\isacharequal}{\kern0pt}\ radius\ C\ {\isachargreater}{\kern0pt}\ {\isadigit{2}}\ {\isacharasterisk}{\kern0pt}\ r}, which is impossible. We can therefore insert it into \isa{C} to obtain a set satisfying \isa{{\isasymforall}c\isactrlsub {\isadigit{1}}\ {\isasymin}\ C\ {\isasymunion}\ {\isacharbraceleft}{\kern0pt}s{\isacharbraceright}{\kern0pt}{\isachardot}{\kern0pt}\ c\isactrlsub {\isadigit{2}}\ {\isasymin}\ C\ {\isasymunion}\ {\isacharbraceleft}{\kern0pt}s{\isacharbraceright}{\kern0pt}{\isachardot}{\kern0pt}\ c\isactrlsub {\isadigit{1}}\ {\isasymnoteq}\ c\isactrlsub {\isadigit{2}}\ {\isasymlongrightarrow}\ dist\ c\isactrlsub {\isadigit{1}}\ c\isactrlsub {\isadigit{2}}\ {\isachargreater}{\kern0pt}\ {\isadigit{2}}\ {\isacharasterisk}{\kern0pt}\ r} by Lemma $\ref{lem:dist_ins}$. We thus have, by Lemma $\ref{lem:inv_last_2}$, that \isa{{\isasymforall}C{\isachardot}{\kern0pt}\ {\isacharbar}{\kern0pt}C{\isacharbar}{\kern0pt}\ {\isasymle}\ k\ {\isasymand}\ {\isacharbar}{\kern0pt}C{\isacharbar}{\kern0pt}\ {\isachargreater}{\kern0pt}\ {\isadigit{0}}\ {\isasymlongrightarrow}\ radius\ C\ {\isachargreater}{\kern0pt}\ r}. In particular, \isa{radius\ C{\isacharprime}{\kern0pt}\ {\isachargreater}{\kern0pt}\ r\ {\isacharequal}{\kern0pt}\ radius\ C{\isacharprime}{\kern0pt}}, a contradiction.
\endproof%
\end{isamarkuptext}\isamarkuptrue%
\isadelimtheory
\endisadelimtheory
\isatagtheory
\endisatagtheory
{\isafoldtheory}%
\isadelimtheory
\endisadelimtheory
\end{isabellebody}%

%% file: Paper_SC.tex
\begin{isabellebody}%
\setisabellecontext{Paper{\isacharunderscore}{\kern0pt}SC}%
\isadelimtheory
\endisadelimtheory
\isatagtheory
\endisatagtheory
{\isafoldtheory}%
\isadelimtheory
\endisadelimtheory
\isadelimproof
\endisadelimproof
\isatagproof
\endisatagproof
{\isafoldproof}%
\isadelimproof
\endisadelimproof
\isadelimproof
\endisadelimproof
\isatagproof
\endisatagproof
{\isafoldproof}%
\isadelimproof
\endisadelimproof
\begin{isamarkuptext}%
\section{Set Cover}
As in the two previous sections, we base our formalization on \cite{KleinbergT06}, but this time Chapter 11.3.
We are given a finite set of elements $\isa{U\ {\isacharcolon}{\kern0pt}{\isacharcolon}{\kern0pt}\ {\isacharprime}{\kern0pt}a\ set}$ and an indexed collection $\isa{S\ {\isacharcolon}{\kern0pt}{\isacharcolon}{\kern0pt}\ nat\ {\isasymRightarrow}\ {\isacharprime}{\kern0pt}a\ set}$ of subsets of \isa{U}, defined over a segment $\isa{{\isacharbraceleft}{\kern0pt}{\isadigit{1}}{\isachardot}{\kern0pt}{\isachardot}{\kern0pt}m{\isacharbraceright}{\kern0pt}}$ of the natural numbers, such that $\isa{U\ {\isacharequal}{\kern0pt}\ {\isasymUnion}\ {\isacharparenleft}{\kern0pt}S\ {\isacharbackquote}{\kern0pt}\ {\isacharbraceleft}{\kern0pt}{\isadigit{1}}{\isachardot}{\kern0pt}{\isachardot}{\kern0pt}m{\isacharbraceright}{\kern0pt}{\isacharparenright}{\kern0pt}}$. A function \isa{w\ {\isacharcolon}{\kern0pt}{\isacharcolon}{\kern0pt}\ nat\ {\isasymRightarrow}\ real} associates a non-negative weight to each (index of a) subset. Our goal is to find a set cover \isa{C} of $\isa{U}$
\begin{quoteni}
\isa{sc\ {\isacharcolon}{\kern0pt}{\isacharcolon}{\kern0pt}\ nat\ set\ {\isasymRightarrow}\ {\isacharprime}{\kern0pt}a\ set\ {\isasymRightarrow}\ bool}\\
\isa{sc\ C\ U\ {\isacharequal}{\kern0pt}\ {\isacharparenleft}{\kern0pt}C\ {\isasymsubseteq}\ {\isacharbraceleft}{\kern0pt}{\isadigit{1}}{\isachardot}{\kern0pt}{\isachardot}{\kern0pt}m{\isacharbraceright}{\kern0pt}\ {\isasymand}\ {\isasymUnion}\ {\isacharparenleft}{\kern0pt}S\ {\isacharbackquote}{\kern0pt}\ C{\isacharparenright}{\kern0pt}\ {\isacharequal}{\kern0pt}\ U{\isacharparenright}{\kern0pt}}
\end{quoteni}
that minimizes the total weight \isa{W\ C\ {\isacharequal}{\kern0pt}\ {\isacharparenleft}{\kern0pt}{\isasymSum}i{\isasymin}C{\isachardot}{\kern0pt}\ w\ i{\isacharparenright}{\kern0pt}}.
The greedy approach takes the subset that covers the most elements relative to its weight,
i.e., \ the subset \isa{S\ i} that minimizes
\begin{quoteni}
\isa{cost\ R\ i} = $\displaystyle\frac{\isa{w\ i}}{\isa{{\isacharbar}{\kern0pt}S\ i\ {\isasyminter}\ R{\isacharbar}{\kern0pt}}}$
\end{quoteni}
where \isa{R} is the set of elements yet to be covered.
We will prove that this greedy algorithm has an approximation factor of $\isa{H\ d\isactrlsup {\isacharasterisk}{\kern0pt}}$ where $\isa{H\ n}$ is the \isa{n}-th harmonic number and \isa{d\isactrlsup {\isacharasterisk}{\kern0pt}} the cardinality of the largest subset.
The algorithm is described by the following Hoare triple:
\begin{framedni}
\isa{{\isacharbraceleft}{\kern0pt}True{\isacharbraceright}{\kern0pt}}\\
\isa{R\ {\isacharcolon}{\kern0pt}{\isacharequal}{\kern0pt}\ U{\isacharsemicolon}{\kern0pt}\ C\ {\isacharcolon}{\kern0pt}{\isacharequal}{\kern0pt}\ {\isasymemptyset}{\isacharsemicolon}{\kern0pt}}\\
\isa{WHILE\ R\ {\isasymnoteq}\ {\isasymemptyset}\ INV\ {\isacharbraceleft}{\kern0pt}inv\ C\ R{\isacharbraceright}{\kern0pt}}\\
\isa{DO\ i\ {\isacharcolon}{\kern0pt}{\isacharequal}{\kern0pt}\ min{\isacharunderscore}{\kern0pt}arg\ R\ m{\isacharsemicolon}{\kern0pt}}\\
\hphantom{\isa{DO}}\isa{R\ {\isacharcolon}{\kern0pt}{\isacharequal}{\kern0pt}\ R\ {\isacharminus}{\kern0pt}\ S\ i{\isacharsemicolon}{\kern0pt}}\\
\hphantom{\isa{DO}}\isa{C\ {\isacharcolon}{\kern0pt}{\isacharequal}{\kern0pt}\ C\ {\isasymunion}\ {\isacharbraceleft}{\kern0pt}i{\isacharbraceright}{\kern0pt}{\isacharsemicolon}{\kern0pt}}\\
\isa{OD}\\
\isa{{\isacharbraceleft}{\kern0pt}sc\ C\ U\ {\isasymand}\ {\isacharparenleft}{\kern0pt}{\isasymforall}C{\isacharprime}{\kern0pt}{\isachardot}{\kern0pt}\ sc\ C{\isacharprime}{\kern0pt}\ U\ {\isasymlongrightarrow}\ W\ C\ {\isasymle}\ H\ d\isactrlsup {\isacharasterisk}{\kern0pt}\ {\isacharasterisk}{\kern0pt}\ W\ C{\isacharprime}{\kern0pt}{\isacharparenright}{\kern0pt}{\isacharbraceright}{\kern0pt}}
\end{framedni}
\noindent
The invariant \isa{inv} is explained later.
The function \isa{min{\isacharunderscore}{\kern0pt}arg} (a variant of the load balancing \isa{min{\isacharunderscore}{\kern0pt}arg}) selects a subset with minimum cost while taking care that the overlap with \isa{R} is non-empty (unless there is no alternative):
\begin{quoteni}
\isa{min{\isacharunderscore}{\kern0pt}arg\ {\isacharcolon}{\kern0pt}{\isacharcolon}{\kern0pt}\ {\isacharprime}{\kern0pt}a\ set\ {\isasymRightarrow}\ nat\ {\isasymRightarrow}\ nat}\\
\isa{min{\isacharunderscore}{\kern0pt}arg\ R\ {\isadigit{0}}\ {\isacharequal}{\kern0pt}\ {\isadigit{1}}\isasep\isanewline%
min{\isacharunderscore}{\kern0pt}arg\ R\ {\isacharparenleft}{\kern0pt}x\ {\isacharplus}{\kern0pt}\ {\isadigit{1}}{\isacharparenright}{\kern0pt}\ {\isacharequal}{\kern0pt}\isanewline
{\isacharparenleft}{\kern0pt}\textsf{let}\ j\ {\isacharequal}{\kern0pt}\ min{\isacharunderscore}{\kern0pt}arg\ R\ x\isanewline
\isaindent{{\isacharparenleft}{\kern0pt}}\textsf{in}\ \textsf{if}\ S\ j\ {\isasyminter}\ R\ {\isacharequal}{\kern0pt}\ {\isasymemptyset}\ {\isasymor}\ S\ {\isacharparenleft}{\kern0pt}x\ {\isacharplus}{\kern0pt}\ {\isadigit{1}}{\isacharparenright}{\kern0pt}\ {\isasyminter}\ R\ {\isasymnoteq}\ {\isasymemptyset}\ {\isasymand}\ cost\ R\ {\isacharparenleft}{\kern0pt}x\ {\isacharplus}{\kern0pt}\ {\isadigit{1}}{\isacharparenright}{\kern0pt}\ {\isacharless}{\kern0pt}\ cost\ R\ j\isanewline
\isaindent{{\isacharparenleft}{\kern0pt}\textsf{in}\ }\textsf{then}\ x\ {\isacharplus}{\kern0pt}\ {\isadigit{1}}\ \textsf{else}\ j{\isacharparenright}{\kern0pt}}
\end{quoteni}
Like with the load balancing problem, a linear scan suffices here.
The following properties of \isa{min{\isacharunderscore}{\kern0pt}arg} can be shown by induction on $\isa{m}$:
\begin{lem}\label{lem:min_arg}
\leavevmode
\makeatletter
\@nobreaktrue
\makeatother
\begin{enumerate}
\item \isa{{\isadigit{0}}\ {\isacharless}{\kern0pt}\ m\ {\isasymlongrightarrow}\ min{\isacharunderscore}{\kern0pt}arg\ R\ m\ {\isasymin}\ {\isacharbraceleft}{\kern0pt}{\isadigit{1}}{\isachardot}{\kern0pt}{\isachardot}{\kern0pt}m{\isacharbraceright}{\kern0pt}}
\item \isa{S\ {\isacharparenleft}{\kern0pt}min{\isacharunderscore}{\kern0pt}arg\ R\ m{\isacharparenright}{\kern0pt}\ {\isasyminter}\ R\ {\isacharequal}{\kern0pt}\ {\isasymemptyset}\ {\isasymlongrightarrow}\ {\isacharparenleft}{\kern0pt}{\isasymforall}i{\isasymin}{\isacharbraceleft}{\kern0pt}{\isadigit{1}}{\isachardot}{\kern0pt}{\isachardot}{\kern0pt}m{\isacharbraceright}{\kern0pt}{\isachardot}{\kern0pt}\ S\ i\ {\isasyminter}\ R\ {\isacharequal}{\kern0pt}\ {\isasymemptyset}{\isacharparenright}{\kern0pt}}
\item \isa{k\ {\isasymin}\ {\isacharbraceleft}{\kern0pt}{\isadigit{1}}{\isachardot}{\kern0pt}{\isachardot}{\kern0pt}m{\isacharbraceright}{\kern0pt}\ {\isasymand}\ S\ k\ {\isasyminter}\ R\ {\isasymnoteq}\ {\isasymemptyset}\ {\isasymlongrightarrow}\ cost\ R\ {\isacharparenleft}{\kern0pt}min{\isacharunderscore}{\kern0pt}arg\ R\ m{\isacharparenright}{\kern0pt}\ {\isasymle}\ cost\ R\ k}
\end{enumerate}
\end{lem}

Unlike the proof of the load balancing problem, the desired approximation factor does not arise from general bounds. Instead we employ a function \isa{c} that returns the cost an individual element contributes to the set cover. Namely, at every iteration of the algorithm let \isa{c\ s\ {\isacharequal}{\kern0pt}\ cost\ R\ i} for all \isa{s\ {\isasymin}\ S\ i\ {\isasyminter}\ R} where \isa{i} is the subset picked by \isa{min{\isacharunderscore}{\kern0pt}arg}. As this function is only relevant to the proof of the algorithm, we can use existential quantification to lift it into the invariant. This function satisfies two important properties upon termination of the algorithm:
\begin{itemize}
\item \isa{W\ C\ {\isacharequal}{\kern0pt}\ {\isacharparenleft}{\kern0pt}{\isasymSum}s\ {\isasymin}\ U{\isachardot}{\kern0pt}\ c\ s{\isacharparenright}{\kern0pt}}
\item \isa{{\isasymforall}k\ {\isasymin}\ {\isacharbraceleft}{\kern0pt}{\isadigit{1}}{\isachardot}{\kern0pt}{\isachardot}{\kern0pt}m{\isacharbraceright}{\kern0pt}{\isachardot}{\kern0pt}\ {\isacharparenleft}{\kern0pt}{\isasymSum}s\ {\isasymin}\ S\ k{\isachardot}{\kern0pt}\ c\ s{\isacharparenright}{\kern0pt}\ {\isasymle}\ H\ {\isacharbar}{\kern0pt}S\ k{\isacharbar}{\kern0pt}\ {\isacharasterisk}{\kern0pt}\ w\ k{\isachardoublequote}{\kern0pt}}
\end{itemize}
The former may be rather intuitive, but reasoning about the latter is more difficult. Kleinberg and Tardos provide an informal proof that argues about the sum of all element costs in the subset at an arbitrary step of the algorithm where an element $j$ of the subset is being covered. Doing so requires indexing and reordering the elements from $1$ to $d$ where $d$ is the cardinality of the subset. We can then show the bound by splitting the subset in such a way that elements already covered come before the element(s) that will be covered in this step, and elements that have not yet been covered come after. Assuming \isa{k\ {\isasymin}\ {\isacharbraceleft}{\kern0pt}{\isadigit{1}}{\isachardot}{\kern0pt}{\isachardot}{\kern0pt}m{\isacharbraceright}{\kern0pt}}, we can see:
\begin{enumerate}
\item At least the elements from index $j$ to $d$ (cardinality) are not covered yet\\
      (i.e., \ $|\isa{S\ k\ {\isasyminter}\ R}| \geq d - j + 1$).
\item Therefore $\isa{cost\ R\ k} = \frac{\isa{w\ k}}{\isa{{\isacharbar}{\kern0pt}S\ k\ {\isasyminter}\ R{\isacharbar}{\kern0pt}}} \leq \frac{\isa{w\ k}}{d - j + 1}$
\item Covered elements up to index $j$ were covered by subsets picked by \isa{min{\isacharunderscore}{\kern0pt}arg}
\item Therefore \isa{cost\ R\ i\ {\isasymle}\ cost\ R\ k} because of Lemma~\ref{lem:min_arg}.3
\item $\sum_{s \in \isa{S\ k}} \isa{c\ s} = \sum_{j = 1}^d c~s_{k_j} \leq \sum_{j = 1}^d \frac{\isa{w\ k}}{d - j + 1}
       = \frac{\isa{w\ k}}d + \frac{\isa{w\ k}}{d - 1} + \dots + \frac{\isa{w\ k}}1 = \isa{H\ d\ {\isacharasterisk}{\kern0pt}\ w\ k}$
\end{enumerate}
This proof is concise and the final statement readily implies the desired approximation factor.
This is a nice textbook proof, but a direct formalization is very unpleasant:
it requires indexing elements in the order in which they were covered by the algorithm.
We give an invariant-based proof that builds on the same proof idea but does not require any indexing.
This is the invariant:
\begin{quoteni}
\isa{inv\ {\isacharcolon}{\kern0pt}{\isacharcolon}{\kern0pt}\ nat\ set\ {\isasymRightarrow}\ {\isacharprime}{\kern0pt}a\ set\ {\isasymRightarrow}\ bool}\\
\isa{inv\ C\ R\ {\isacharequal}{\kern0pt}\ {\isacharparenleft}{\kern0pt}sc\ C\ {\isacharparenleft}{\kern0pt}U\ {\isacharminus}{\kern0pt}\ R{\isacharparenright}{\kern0pt}\ {\isasymand}\ R\ {\isasymsubseteq}\ U\ {\isasymand}\ c{\isacharunderscore}{\kern0pt}exists\ C\ R{\isacharparenright}{\kern0pt}}
\\
\isa{c{\isacharunderscore}{\kern0pt}exists\ {\isacharcolon}{\kern0pt}{\isacharcolon}{\kern0pt}\ nat\ set\ {\isasymRightarrow}\ {\isacharprime}{\kern0pt}a\ set\ {\isasymRightarrow}\ bool}\\
\isa{c{\isacharunderscore}{\kern0pt}exists\ C\ R\ {\isacharequal}{\kern0pt}\isanewline
{\isacharparenleft}{\kern0pt}{\isasymexists}c{\isachardot}{\kern0pt}\ W\ C\ {\isacharequal}{\kern0pt}\ {\isacharparenleft}{\kern0pt}{\isasymSum}s{\isasymin}U\ {\isacharminus}{\kern0pt}\ R{\isachardot}{\kern0pt}\ c\ s{\isacharparenright}{\kern0pt}\ {\isasymand}\isanewline
\isaindent{{\isacharparenleft}{\kern0pt}{\isasymexists}c{\isachardot}{\kern0pt}\ }{\isacharparenleft}{\kern0pt}{\isasymforall}i{\isachardot}{\kern0pt}\ {\isadigit{0}}\ {\isasymle}\ c\ i{\isacharparenright}{\kern0pt}\ {\isasymand}\isanewline
\isaindent{{\isacharparenleft}{\kern0pt}{\isasymexists}c{\isachardot}{\kern0pt}\ }{\isacharparenleft}{\kern0pt}{\isasymforall}k{\isasymin}{\isacharbraceleft}{\kern0pt}{\isadigit{1}}{\isachardot}{\kern0pt}{\isachardot}{\kern0pt}m{\isacharbraceright}{\kern0pt}{\isachardot}{\kern0pt}\isanewline
\isaindent{{\isacharparenleft}{\kern0pt}{\isasymexists}c{\isachardot}{\kern0pt}\ {\isacharparenleft}{\kern0pt}\ \ \ }{\isacharparenleft}{\kern0pt}{\isasymSum}s{\isasymin}S\ k\ {\isasyminter}\ {\isacharparenleft}{\kern0pt}U\ {\isacharminus}{\kern0pt}\ R{\isacharparenright}{\kern0pt}{\isachardot}{\kern0pt}\ c\ s{\isacharparenright}{\kern0pt}\ {\isasymle}\ {\isacharparenleft}{\kern0pt}$\sum_{j\ {\isacharequal}{\kern0pt}\ {\isacharbar}{\kern0pt}S\ k\ {\isasyminter}\ R{\isacharbar}{\kern0pt}\ {\isacharplus}{\kern0pt}\ {\isadigit{1}}}^{{\isacharbar}{\kern0pt}S\ k{\isacharbar}{\kern0pt}}$\ {\isadigit{1}}\ {\isacharslash}{\kern0pt}\ j{\isacharparenright}{\kern0pt}\ {\isacharasterisk}{\kern0pt}\ w\ k{\isacharparenright}{\kern0pt}{\isacharparenright}{\kern0pt}}
\end{quoteni}
Using $\isa{U\ {\isacharminus}{\kern0pt}\ R}$ we can argue about the elements that have already been covered by the algorithm. The relation between the proof above and the upper bound in the last conjunct of \isa{c{\isacharunderscore}{\kern0pt}exists} is less apparent, but we can perform an index shift that makes the relation more obvious. Let $\isa{d\ {\isacharequal}{\kern0pt}\ {\isacharbar}{\kern0pt}S\ k{\isacharbar}{\kern0pt}}$, then:
\begin{align*}
\sum_{j = \isa{{\isacharbar}{\kern0pt}S\ k\ {\isasyminter}\ R{\isacharbar}{\kern0pt}\ {\isacharplus}{\kern0pt}\ {\isadigit{1}}}}^{d}\frac{\isa{w\ k}}j &= \sum_{j=1}^{\isa{d\ {\isacharminus}{\kern0pt}\ {\isacharbar}{\kern0pt}S\ k\ {\isasyminter}\ R{\isacharbar}{\kern0pt}}}\frac{\isa{w\ k}}{\isa{d\ {\isacharminus}{\kern0pt}\ j\ {\isacharplus}{\kern0pt}\ {\isadigit{1}}}}
= \sum_{j=1}^{\isa{{\isacharbar}{\kern0pt}S\ k\ {\isasyminter}\ {\isacharparenleft}{\kern0pt}U\ {\isacharminus}{\kern0pt}\ R{\isacharparenright}{\kern0pt}{\isacharbar}{\kern0pt}}}\frac{\isa{w\ k}}{\isa{d\ {\isacharminus}{\kern0pt}\ j\ {\isacharplus}{\kern0pt}\ {\isadigit{1}}}}
\end{align*}
This is closer to the sum we see in the proof by Kleinberg and Tardos, but complicates the invariant proof as one now has to argue about the bounds of the sum as well as the content of it, while our definition in \isa{c{\isacharunderscore}{\kern0pt}exists} leaves the content of the sum as $\frac1j$.

We now come to the proof of the invariant. Initialization is trivial if we pick $\isa{c\ {\isacharequal}{\kern0pt}\ {\isacharparenleft}{\kern0pt}{\isasymlambda}{\isacharunderscore}{\kern0pt}{\isachardot}{\kern0pt}\ {\isadigit{0}}{\isacharparenright}{\kern0pt}}$. We may now assume the invariant and $\isa{R\ {\isasymnoteq}\ {\isasymemptyset}}$. Let $i$ be the subset picked by \isa{min{\isacharunderscore}{\kern0pt}arg}, moreover \isa{R\isactrlsub g\ {\isacharcolon}{\kern0pt}{\isacharequal}{\kern0pt}\ R\ {\isacharminus}{\kern0pt}\ S\ i} and \isa{C\isactrlsub g\ {\isacharcolon}{\kern0pt}{\isacharequal}{\kern0pt}\ C\ {\isasymunion}\ {\isacharbraceleft}{\kern0pt}i{\isacharbraceright}{\kern0pt}}. First note $\isa{{\isadigit{0}}\ {\isacharless}{\kern0pt}\ m}$, as $\isa{R\ {\isacharequal}{\kern0pt}\ {\isasymemptyset}}$ otherwise (\isa{R\ {\isasymsubseteq}\ U} and $\isa{U\ {\isacharequal}{\kern0pt}\ {\isasymUnion}\ {\isacharparenleft}{\kern0pt}S\ {\isacharbackquote}{\kern0pt}\ {\isacharbraceleft}{\kern0pt}{\isadigit{1}}{\isachardot}{\kern0pt}{\isachardot}{\kern0pt}m{\isacharbraceright}{\kern0pt}{\isacharparenright}{\kern0pt}}$), hence $\isa{i\ {\isasymin}\ {\isacharbraceleft}{\kern0pt}{\isadigit{1}}{\isachardot}{\kern0pt}{\isachardot}{\kern0pt}m{\isacharbraceright}{\kern0pt}}$ (Lemma~\ref{lem:min_arg}.1). Correctness is preserved as $\isa{S\ i}$ is a subset of $U$, thus $\isa{sc\ C\isactrlsub g\ {\isacharparenleft}{\kern0pt}U\ {\isacharminus}{\kern0pt}\ R\isactrlsub g{\isacharparenright}{\kern0pt}}$. Since we are only removing elements from \isa{R}, \isa{R\isactrlsub g} remains a subset of $U$ as well. Before we prove that \isa{c{\isacharunderscore}{\kern0pt}exists\ C\isactrlsub g\ R\isactrlsub g} holds, note that the set cover actually strictly grows in an iteration of the algorithm. We know that $\isa{{\isasymexists}k\ {\isasymin}\ {\isacharbraceleft}{\kern0pt}{\isadigit{1}}{\isachardot}{\kern0pt}{\isachardot}{\kern0pt}m{\isacharbraceright}{\kern0pt}{\isachardot}{\kern0pt}\ S\ k\ {\isasyminter}\ R\ {\isasymnoteq}\ {\isasymemptyset}}$ because $\isa{R\ {\isasymnoteq}\ {\isasymemptyset}}$ and $\isa{R\ {\isasymsubseteq}\ U}$. Combined with Lemma~\ref{lem:min_arg}.2 we know that $\isa{S\ i\ {\isasyminter}\ R\ {\isasymnoteq}\ {\isasymemptyset}}$. Therefore new elements are covered in this iteration and by extension a subset was picked that was not in the cover before ($\isa{i\ {\isasymnotin}\ C}$), thus \textit{weight-eq}: $\isa{W\ C\isactrlsub g\ {\isacharequal}{\kern0pt}\ W\ C\ {\isacharplus}{\kern0pt}\ w\ i}$. We define the cost function $\isa{c\isactrlsub g\ {\isacharequal}{\kern0pt}\ {\isacharparenleft}{\kern0pt}{\isasymlambda}x{\isachardot}{\kern0pt}\ \textsf{if}\ x\ {\isasymin}\ S\ i\ {\isasyminter}\ R\ \textsf{then}\ cost\ R\ i\ \textsf{else}\ c\ x{\isacharparenright}{\kern0pt}}$, where \isa{c} is the cost function obtained using \isa{c{\isacharunderscore}{\kern0pt}exists}, and see that
\begin{align*}
\sum_{s \in U - R_g}\isa{c\isactrlsub g\ s} &= \sum_{s \in U - R}\isa{c\isactrlsub g\ s} + \sum_{s \in \isa{S\ i\ {\isasyminter}\ R}}\isa{c\isactrlsub g\ s} \tag*{by \isa{R\ {\isasymsubseteq}\ U}}\\
&= \sum_{s \in U - R}\isa{c\ s} + \isa{{\isacharbar}{\kern0pt}S\ i\ {\isasyminter}\ R{\isacharbar}{\kern0pt}cost\ R\ i} \tag*{by \isa{c\isactrlsub g}-def}\\
&= \sum_{s \in U - R}\isa{c\ s} + \isa{{\isacharbar}{\kern0pt}S\ i\ {\isasyminter}\ R{\isacharbar}{\kern0pt}}\frac{\isa{w\ i}}{\isa{{\isacharbar}{\kern0pt}S\ i\ {\isasyminter}\ R{\isacharbar}{\kern0pt}}} \tag*{by \isa{cost}-def}\\
&= \sum_{s \in U - R}\isa{c\ s} + \isa{w\ i} \tag*{by \isa{S\ i\ {\isasyminter}\ R\ {\isasymnoteq}\ {\isasymemptyset}}}\\
&= \isa{W\ C} + \isa{w\ i} = \isa{W\ C\isactrlsub g} \tag*{by \isa{inv}-def and \textit{weight-eq}}
\end{align*}
As $\isa{cost\ R\ i}$ is always positive, we know that $\isa{{\isasymforall}i{\isachardot}{\kern0pt}\ {\isadigit{0}}\ {\isasymle}\ c\isactrlsub g\ i}$ is preserved. Finally we come to the last conjunct of \isa{c{\isacharunderscore}{\kern0pt}exists}.
Assume $\isa{k\ {\isasymin}\ {\isacharbraceleft}{\kern0pt}{\isadigit{1}}{\isachardot}{\kern0pt}{\isachardot}{\kern0pt}m{\isacharbraceright}{\kern0pt}}$ and let \isa{A\ {\isacharequal}{\kern0pt}\ S\ k\ {\isasyminter}\ S\ i\ {\isasyminter}\ R} be the elements of subset \isa{S\ k} that will be covered in this iteration. We can transform the lower bound of the right sum as follows \textit{lbr}: \isa{{\isacharbar}{\kern0pt}S\ k\ {\isasyminter}\ R\isactrlsub g{\isacharbar}{\kern0pt}\ {\isacharplus}{\kern0pt}\ {\isadigit{1}}\ {\isacharequal}{\kern0pt}\ {\isacharbar}{\kern0pt}S\ k\ {\isasyminter}\ R{\isacharbar}{\kern0pt}\ {\isacharminus}{\kern0pt}\ {\isacharbar}{\kern0pt}A{\isacharbar}{\kern0pt}\ {\isacharplus}{\kern0pt}\ {\isadigit{1}}}. This allows for the following transformation:
\begin{align*}
&\sum_{s \in \isa{S\ k\ {\isasyminter}\ {\isacharparenleft}{\kern0pt}U\ {\isacharminus}{\kern0pt}\ R\isactrlsub g{\isacharparenright}{\kern0pt}}}\isa{c\isactrlsub g\ s} = \sum_{s \in \isa{S\ k\ {\isasyminter}\ {\isacharparenleft}{\kern0pt}U\ {\isacharminus}{\kern0pt}\ R{\isacharparenright}{\kern0pt}}}\isa{c\isactrlsub g\ s} + \sum_{s \in \isa{A}}\isa{c\isactrlsub g\ s} \tag*{by \isa{R\ {\isasymsubseteq}\ U}}\\
&=    \sum_{s \in \isa{S\ k\ {\isasyminter}\ {\isacharparenleft}{\kern0pt}U\ {\isacharminus}{\kern0pt}\ R{\isacharparenright}{\kern0pt}}}\isa{c\ s} + \isa{{\isacharbar}{\kern0pt}A{\isacharbar}{\kern0pt}cost\ R\ i} \tag*{by \isa{c\isactrlsub g}-def}\\
&\leq \isa{w\ k} \sum_{j = \isa{{\isacharbar}{\kern0pt}S\ k\ {\isasyminter}\ R{\isacharbar}{\kern0pt}\ {\isacharplus}{\kern0pt}\ {\isadigit{1}}}}^{\isa{{\isacharbar}{\kern0pt}S\ k{\isacharbar}{\kern0pt}}}\frac1j + \isa{{\isacharbar}{\kern0pt}A{\isacharbar}{\kern0pt}cost\ R\ k} \tag*{by \isa{inv}-def and Lem.~\ref{lem:min_arg}.3}\\
&=    \isa{w\ k} \left(\sum_{j = \isa{{\isacharbar}{\kern0pt}S\ k\ {\isasyminter}\ R{\isacharbar}{\kern0pt}\ {\isacharplus}{\kern0pt}\ {\isadigit{1}}}}^{\isa{{\isacharbar}{\kern0pt}S\ k{\isacharbar}{\kern0pt}}}\frac1j + \isa{{\isacharbar}{\kern0pt}A{\isacharbar}{\kern0pt}}\frac1{\isa{{\isacharbar}{\kern0pt}S\ k\ {\isasyminter}\ R{\isacharbar}{\kern0pt}}}\right) \tag*{by \isa{cost}-def}\\
&=    \isa{w\ k} \left(\sum_{j = \isa{{\isacharbar}{\kern0pt}S\ k\ {\isasyminter}\ R{\isacharbar}{\kern0pt}\ {\isacharplus}{\kern0pt}\ {\isadigit{1}}}}^{\isa{{\isacharbar}{\kern0pt}S\ k{\isacharbar}{\kern0pt}}}\frac1j + \sum_{j=\isa{{\isacharbar}{\kern0pt}S\ k\ {\isasyminter}\ R{\isacharbar}{\kern0pt}\ {\isacharminus}{\kern0pt}\ {\isacharbar}{\kern0pt}A{\isacharbar}{\kern0pt}\ {\isacharplus}{\kern0pt}\ {\isadigit{1}}}}^{\isa{{\isacharbar}{\kern0pt}S\ k\ {\isasyminter}\ R{\isacharbar}{\kern0pt}}}\frac1{\isa{{\isacharbar}{\kern0pt}S\ k\ {\isasyminter}\ R{\isacharbar}{\kern0pt}}}\right)\\
&\leq \isa{w\ k} \left(\sum_{j = \isa{{\isacharbar}{\kern0pt}S\ k\ {\isasyminter}\ R{\isacharbar}{\kern0pt}\ {\isacharplus}{\kern0pt}\ {\isadigit{1}}}}^{\isa{{\isacharbar}{\kern0pt}S\ k{\isacharbar}{\kern0pt}}}\frac1j + \sum_{j=\isa{{\isacharbar}{\kern0pt}S\ k\ {\isasyminter}\ R{\isacharbar}{\kern0pt}\ {\isacharminus}{\kern0pt}\ {\isacharbar}{\kern0pt}A{\isacharbar}{\kern0pt}\ {\isacharplus}{\kern0pt}\ {\isadigit{1}}}}^{\isa{{\isacharbar}{\kern0pt}S\ k\ {\isasyminter}\ R{\isacharbar}{\kern0pt}}}\frac1j\right)\\
&=    \isa{w\ k} \sum_{j = \isa{{\isacharbar}{\kern0pt}S\ k\ {\isasyminter}\ R{\isacharbar}{\kern0pt}\ {\isacharminus}{\kern0pt}\ {\isacharbar}{\kern0pt}A{\isacharbar}{\kern0pt}\ {\isacharplus}{\kern0pt}\ {\isadigit{1}}}}^{\isa{{\isacharbar}{\kern0pt}S\ k{\isacharbar}{\kern0pt}}}\frac1j
 =    \isa{w\ k} \sum_{j = \isa{{\isacharbar}{\kern0pt}S\ k\ {\isasyminter}\ R\isactrlsub g{\isacharbar}{\kern0pt}\ {\isacharplus}{\kern0pt}\ {\isadigit{1}}}}^{\isa{{\isacharbar}{\kern0pt}S\ k{\isacharbar}{\kern0pt}}}\frac1j \tag*{by \textit{lbr}}
\end{align*}
Unlike the proof in \cite{KleinbergT06} we only had to argue about the cost of \isa{A} here, as the cost of previous elements is already proved by the assumed invariant.

Finally we show how the invariant implies the approximation factor upon termination of the algorithm:
\begin{lem}
\isa{inv\ C\ {\isasymemptyset}\ {\isasymand}\ sc\ C{\isacharprime}{\kern0pt}\ U\ {\isasymlongrightarrow}\ W\ C\ {\isasymle}\ H\ d\isactrlsup {\isacharasterisk}{\kern0pt}\ {\isacharasterisk}{\kern0pt}\ W\ C{\isacharprime}{\kern0pt}}
\end{lem}
\noindent
From \isa{inv\ C\ {\isasymemptyset}}, \isa{c{\isacharunderscore}{\kern0pt}exists} and $\isa{R\ {\isacharequal}{\kern0pt}\ {\isasymemptyset}}$ we can infer \textit{cost-eq}: $\isa{W\ C\ {\isacharequal}{\kern0pt}\ {\isacharparenleft}{\kern0pt}{\isasymSum}s{\isasymin}U{\isachardot}{\kern0pt}\ c\ s{\isacharparenright}{\kern0pt}}$ and \textit{h-bound}: $\isa{{\isasymforall}k\ {\isasymin}\ {\isacharbraceleft}{\kern0pt}{\isadigit{1}}{\isachardot}{\kern0pt}{\isachardot}{\kern0pt}m{\isacharbraceright}{\kern0pt}{\isachardot}{\kern0pt}\ {\isacharparenleft}{\kern0pt}{\isasymSum}s\ {\isasymin}\ S\ k{\isachardot}{\kern0pt}\ c\ s{\isacharparenright}{\kern0pt}\ {\isasymle}\ H\ {\isacharbar}{\kern0pt}S\ k{\isacharbar}{\kern0pt}\ {\isacharasterisk}{\kern0pt}\ w\ k}$
and can derive the claim:
\begin{align*}
\isa{W\ C} &= \sum_{s \in U}\isa{c\ s} \tag*{by \textit{cost-eq}}\\
      &\leq \sum_{k \in \isa{C{\isacharprime}{\kern0pt}}} \sum_{s \in \isa{S\ k}}\isa{c\ s} \tag*{by \isa{{\isasymforall}i{\isachardot}{\kern0pt}\ {\isadigit{0}}\ {\isasymle}\ c\ i}}\\
      &\leq \sum_{k \in \isa{C{\isacharprime}{\kern0pt}}} \isa{H\ {\isacharbar}{\kern0pt}S\ k{\isacharbar}{\kern0pt}\ {\isacharasterisk}{\kern0pt}\ w\ k} \tag*{by \textit{h-bound}}\\
      &\leq \sum_{k \in \isa{C{\isacharprime}{\kern0pt}}} \isa{H\ d\isactrlsup {\isacharasterisk}{\kern0pt}\ {\isacharasterisk}{\kern0pt}\ w\ k} \tag*{by \isa{d\isactrlsup {\isacharasterisk}{\kern0pt}}-def}\\
      &= \isa{H\ d\isactrlsup {\isacharasterisk}{\kern0pt}}\sum_{k \in \isa{C{\isacharprime}{\kern0pt}}} \isa{w\ k} = \isa{H\ d\isactrlsup {\isacharasterisk}{\kern0pt}\ {\isacharasterisk}{\kern0pt}\ W\ C{\isacharprime}{\kern0pt}}
\end{align*}%
\end{isamarkuptext}\isamarkuptrue%
\isadelimtheory
\endisadelimtheory
\isatagtheory
\endisatagtheory
{\isafoldtheory}%
\isadelimtheory
\endisadelimtheory
\end{isabellebody}%

%% file: Paper_BP.tex
\begin{isabellebody}%
\setisabellecontext{Paper{\isacharunderscore}{\kern0pt}BP}%
\isadelimtheory
\endisadelimtheory
\isatagtheory
\endisatagtheory
{\isafoldtheory}%
\isadelimtheory
\endisadelimtheory
\begin{isamarkuptext}%
\section{Bin Packing}
We consider the linear time $\frac32$-approximation algorithm for the bin packing problem proposed by Berghammer and Reuter \cite{BerghammerR03}.
%
We are given a finite, non-empty set of objects \isa{U\ {\isacharcolon}{\kern0pt}{\isacharcolon}{\kern0pt}\ {\isacharprime}{\kern0pt}a\ set}, whose \isa{weights} are given by a function $\isa{w\ {\isacharcolon}{\kern0pt}{\isacharcolon}{\kern0pt}\ {\isacharprime}{\kern0pt}a\ {\isasymRightarrow}\ real}$. Note that in this paper \isa{nat}s are
implicitly converted to \isa{real}s if needed.
The weight of an object in \isa{U} is strictly greater than zero, but bounded by a maximum capacity $\isa{c\ {\isacharcolon}{\kern0pt}{\isacharcolon}{\kern0pt}\ nat}$. The abbreviation $W(B) \equiv \sum_{u \in B} w(u)$ denotes the weight of a bin $\isa{B\ {\isasymsubseteq}\ U}$.
The set \isa{U} can be separated into \isa{small} and \isa{large} objects. An object \isa{u} is considered small if $w(u) \le \frac{c}{2}$, and large otherwise. We assume that all small objects in \isa{U} can be found in a set \isa{S}, and large objects in \isa{U} can be found in a set \isa{L}, such that $\isa{S\ {\isasymunion}\ L\ {\isacharequal}{\kern0pt}\ U}$ and $\isa{S\ {\isasyminter}\ L\ {\isacharequal}{\kern0pt}\ {\isasymemptyset}}$. Of course \isa{L} and \isa{S} could also be computed from \isa{U} in linear time. Variables \isa{U}, \isa{w}, \isa{c}, \isa{L}, and \isa{S} are fixed throughout this section.

A solution \isa{P} to the bin packing problem is defined as follows:
\begin{quoteni}
\isa{bp\ {\isacharcolon}{\kern0pt}{\isacharcolon}{\kern0pt}\ {\isacharprime}{\kern0pt}a\ set\ set\ {\isasymRightarrow}\ bool}\\
\isa{bp\ P\ {\isacharequal}{\kern0pt}\ {\isacharparenleft}{\kern0pt}partition{\isacharunderscore}{\kern0pt}on\ U\ P\ {\isasymand}\ {\isacharparenleft}{\kern0pt}{\isasymforall}B{\isasymin}P{\isachardot}{\kern0pt}\ W\ B\ {\isasymle}\ c{\isacharparenright}{\kern0pt}{\isacharparenright}{\kern0pt}}
\end{quoteni}
The predicate \isa{partition{\isacharunderscore}{\kern0pt}on\ {\isacharcolon}{\kern0pt}{\isacharcolon}{\kern0pt}\ {\isacharprime}{\kern0pt}a\ set\ {\isasymRightarrow}\ {\isacharprime}{\kern0pt}a\ set\ set\ {\isasymRightarrow}\ bool} is defined in the Isabelle/HOL library:
\isa{partition{\isacharunderscore}{\kern0pt}on\ U\ P} means that \isa{P} is a partition of the set \isa{U}.
Viewing every element of \isa{P} as a bin, \isa{bp\ P} expresses that
all objects are contained in exactly one bin and the weight of every bin is bounded by \isa{c}.

The idea behind the algorithm proposed by Berghammer and Reuter is to split the solution \isa{P} into two partial solutions \isa{P\isactrlsub {\isadigit{1}}} and \isa{P\isactrlsub {\isadigit{2}}}. At every step of the algorithm we consider two bins \isa{B\isactrlsub {\isadigit{1}}} and \isa{B\isactrlsub {\isadigit{2}}} which we try to fill with remaining objects from \isa{V\ {\isasymsubseteq}\ U} that have not been assigned yet. If adding the object to \isa{B\isactrlsub {\isadigit{1}}} or \isa{B\isactrlsub {\isadigit{2}}} would cause it to exceed its maximum capacity, the bin is moved into the partial solution \isa{P\isactrlsub {\isadigit{1}}} or \isa{P\isactrlsub {\isadigit{2}}} respectively and cleared. Once there are no small objects left, the solution is the union of the partial solutions \isa{P\isactrlsub {\isadigit{1}}} and \isa{P\isactrlsub {\isadigit{2}}}, the bins \isa{B\isactrlsub {\isadigit{1}}} and \isa{B\isactrlsub {\isadigit{2}}} (if they still contain objects), and the remaining large objects, which each receive their own bin, as no two large objects can fit into a single bin. To ensure that no empty bins are added to the solution, we define:
\begin{quoteni}
\isa{{\isasymlbrakk}{\isasymcdot}{\isasymrbrakk}\ {\isacharcolon}{\kern0pt}{\isacharcolon}{\kern0pt}\ {\isacharprime}{\kern0pt}a\ set\ {\isasymRightarrow}\ {\isacharprime}{\kern0pt}a\ set\ set}\\
\isa{{\isasymlbrakk}B{\isasymrbrakk}\ {\isacharequal}{\kern0pt}\ {\isacharparenleft}{\kern0pt}\textsf{if}\ B\ {\isacharequal}{\kern0pt}\ {\isasymemptyset}\ \textsf{then}\ {\isasymemptyset}\ \textsf{else}\ {\isacharbraceleft}{\kern0pt}B{\isacharbraceright}{\kern0pt}{\isacharparenright}{\kern0pt}}
\end{quoteni}
The final union can now be written as $\isa{P\isactrlsub {\isadigit{1}}\ {\isasymunion}\ {\isasymlbrakk}B\isactrlsub {\isadigit{1}}{\isasymrbrakk}\ {\isasymunion}\ P\isactrlsub {\isadigit{2}}\ {\isasymunion}\ {\isasymlbrakk}B\isactrlsub {\isadigit{2}}{\isasymrbrakk}\ {\isasymunion}\ {\isacharbraceleft}{\kern0pt}{\isacharbraceleft}{\kern0pt}v{\isacharbraceright}{\kern0pt}\ {\isacharbar}{\kern0pt}\ v\ {\isasymin}\ V{\isacharbraceright}{\kern0pt}}$ where \isa{V} contains the remaining large elements. The algorithm can be specified by the following Hoare triple:
\begin{framedni}
\small{\isa{{\isacharbraceleft}{\kern0pt}True{\isacharbraceright}{\kern0pt}}\\
\isa{P\isactrlsub {\isadigit{1}}\ {\isacharcolon}{\kern0pt}{\isacharequal}{\kern0pt}\ {\isasymemptyset}{\isacharsemicolon}{\kern0pt}\ P\isactrlsub {\isadigit{2}}\ {\isacharcolon}{\kern0pt}{\isacharequal}{\kern0pt}\ {\isasymemptyset}{\isacharsemicolon}{\kern0pt}\ B\isactrlsub {\isadigit{1}}\ {\isacharcolon}{\kern0pt}{\isacharequal}{\kern0pt}\ {\isasymemptyset}{\isacharsemicolon}{\kern0pt}\ B\isactrlsub {\isadigit{2}}\ {\isacharcolon}{\kern0pt}{\isacharequal}{\kern0pt}\ {\isasymemptyset}{\isacharsemicolon}{\kern0pt}\ V\ {\isacharcolon}{\kern0pt}{\isacharequal}{\kern0pt}\ U{\isacharsemicolon}{\kern0pt}}\\
\isa{WHILE\ V\ {\isasyminter}\ S\ {\isasymnoteq}\ {\isasymemptyset}\ INV\ {\isacharbraceleft}{\kern0pt}inv\isactrlsub {\isadigit{3}}\ P\isactrlsub {\isadigit{1}}\ P\isactrlsub {\isadigit{2}}\ B\isactrlsub {\isadigit{1}}\ B\isactrlsub {\isadigit{2}}\ V{\isacharbraceright}{\kern0pt}\ DO}\\
\isa{IF\ B\isactrlsub {\isadigit{1}}\ {\isasymnoteq}\ {\isasymemptyset}\ THEN\ u\ {\isacharcolon}{\kern0pt}{\isacharequal}{\kern0pt}\ some\ {\isacharparenleft}{\kern0pt}V\ {\isasyminter}\ S{\isacharparenright}{\kern0pt}}\\
\isa{ELSE\ IF\ V\ {\isasyminter}\ L\ {\isasymnoteq}\ {\isasymemptyset}\ THEN\ u\ {\isacharcolon}{\kern0pt}{\isacharequal}{\kern0pt}\ some\ {\isacharparenleft}{\kern0pt}V\ {\isasyminter}\ L{\isacharparenright}{\kern0pt}}\\
\hphantom{\small{ELSE }}\isa{ELSE\ u\ {\isacharcolon}{\kern0pt}{\isacharequal}{\kern0pt}\ some\ {\isacharparenleft}{\kern0pt}V\ {\isasyminter}\ S{\isacharparenright}{\kern0pt}\ FI\ FI{\isacharsemicolon}{\kern0pt}}\\
\isa{V\ {\isacharcolon}{\kern0pt}{\isacharequal}{\kern0pt}\ V\ {\isacharminus}{\kern0pt}\ {\isacharbraceleft}{\kern0pt}u{\isacharbraceright}{\kern0pt}{\isacharsemicolon}{\kern0pt}}\\
\isa{IF\ W{\isacharparenleft}{\kern0pt}B\isactrlsub {\isadigit{1}}{\isacharparenright}{\kern0pt}\ {\isacharplus}{\kern0pt}\ w{\isacharparenleft}{\kern0pt}u{\isacharparenright}{\kern0pt}\ {\isasymle}\ c\ THEN\ B\isactrlsub {\isadigit{1}}\ {\isacharcolon}{\kern0pt}{\isacharequal}{\kern0pt}\ B\isactrlsub {\isadigit{1}}\ {\isasymunion}\ {\isacharbraceleft}{\kern0pt}u{\isacharbraceright}{\kern0pt}}\\
\isa{ELSE\ IF\ W{\isacharparenleft}{\kern0pt}B\isactrlsub {\isadigit{2}}{\isacharparenright}{\kern0pt}\ {\isacharplus}{\kern0pt}\ w{\isacharparenleft}{\kern0pt}u{\isacharparenright}{\kern0pt}\ {\isasymle}\ c\ THEN\ B\isactrlsub {\isadigit{2}}\ {\isacharcolon}{\kern0pt}{\isacharequal}{\kern0pt}\ B\isactrlsub {\isadigit{2}}\ {\isasymunion}\ {\isacharbraceleft}{\kern0pt}u{\isacharbraceright}{\kern0pt}}\\
\hphantom{\small{ELSE }}\isa{ELSE\ P\isactrlsub {\isadigit{2}}\ {\isacharcolon}{\kern0pt}{\isacharequal}{\kern0pt}\ P\isactrlsub {\isadigit{2}}\ {\isasymunion}\ {\isasymlbrakk}B\isactrlsub {\isadigit{2}}{\isasymrbrakk}{\isacharsemicolon}{\kern0pt}\ B\isactrlsub {\isadigit{2}}\ {\isacharcolon}{\kern0pt}{\isacharequal}{\kern0pt}\ {\isacharbraceleft}{\kern0pt}u{\isacharbraceright}{\kern0pt}\ FI{\isacharsemicolon}{\kern0pt}}\\
\hphantom{\small{ELSE }}\isa{P\isactrlsub {\isadigit{1}}\ {\isacharcolon}{\kern0pt}{\isacharequal}{\kern0pt}\ P\isactrlsub {\isadigit{1}}\ {\isasymunion}\ {\isasymlbrakk}B\isactrlsub {\isadigit{1}}{\isasymrbrakk}{\isacharsemicolon}{\kern0pt}\ B\isactrlsub {\isadigit{1}}\ {\isacharcolon}{\kern0pt}{\isacharequal}{\kern0pt}\ {\isasymemptyset}\ FI}\\
\isa{OD{\isacharsemicolon}{\kern0pt}}\\
\isa{P\ {\isacharcolon}{\kern0pt}{\isacharequal}{\kern0pt}\ P\isactrlsub {\isadigit{1}}\ {\isasymunion}\ {\isasymlbrakk}B\isactrlsub {\isadigit{1}}{\isasymrbrakk}\ {\isasymunion}\ P\isactrlsub {\isadigit{2}}\ {\isasymunion}\ {\isasymlbrakk}B\isactrlsub {\isadigit{2}}{\isasymrbrakk}\ {\isasymunion}\ {\isacharbraceleft}{\kern0pt}{\isacharbraceleft}{\kern0pt}v{\isacharbraceright}{\kern0pt}\ {\isacharbar}{\kern0pt}\ v\ {\isasymin}\ V{\isacharbraceright}{\kern0pt}}\\
\isa{{\isacharbraceleft}{\kern0pt}bp\ P\ {\isasymand}\ {\isacharparenleft}{\kern0pt}{\isasymforall}Q{\isachardot}{\kern0pt}\ bp\ Q\ {\isasymlongrightarrow}\ {\isacharbar}{\kern0pt}P{\isacharbar}{\kern0pt}\ {\isasymle}\ {\isadigit{3}}\ {\isacharslash}{\kern0pt}\ {\isadigit{2}}\ {\isacharasterisk}{\kern0pt}\ {\isacharbar}{\kern0pt}Q{\isacharbar}{\kern0pt}{\isacharparenright}{\kern0pt}{\isacharbraceright}{\kern0pt}}}
\end{framedni}
Berghammer and Reuter prove functional correctness using a simplified version of this algorithm where an arbitrary element of \isa{V} is assigned to \isa{u}. This allows for fewer case distinctions, as the first \isa{IF{\isacharminus}{\kern0pt}THEN{\isacharminus}{\kern0pt}ELSE} block can be ignored. One needs to find a loop invariant that implies functional correctness and prove that it is preserved in the following cases:
\begin{quoteni}
\textbf{Case 1} The object fits into \isa{B\isactrlsub {\isadigit{1}}}:
\begin{isabelle}%
inv\isactrlsub {\isadigit{1}}\ P\isactrlsub {\isadigit{1}}\ P\isactrlsub {\isadigit{2}}\ B\isactrlsub {\isadigit{1}}\ B\isactrlsub {\isadigit{2}}\ V\ {\isasymand}\ u\ {\isasymin}\ V\ {\isasymand}\ W\ B\isactrlsub {\isadigit{1}}\ {\isacharplus}{\kern0pt}\ w\ u\ {\isasymle}\ c\ {\isasymlongrightarrow}\isanewline
inv\isactrlsub {\isadigit{1}}\ P\isactrlsub {\isadigit{1}}\ P\isactrlsub {\isadigit{2}}\ {\isacharparenleft}{\kern0pt}B\isactrlsub {\isadigit{1}}\ {\isasymunion}\ {\isacharbraceleft}{\kern0pt}u{\isacharbraceright}{\kern0pt}{\isacharparenright}{\kern0pt}\ B\isactrlsub {\isadigit{2}}\ {\isacharparenleft}{\kern0pt}V\ {\isacharminus}{\kern0pt}\ {\isacharbraceleft}{\kern0pt}u{\isacharbraceright}{\kern0pt}{\isacharparenright}{\kern0pt}%
\end{isabelle}
\textbf{Case 2} The object fits into \isa{B\isactrlsub {\isadigit{2}}}:
\begin{isabelle}%
inv\isactrlsub {\isadigit{1}}\ P\isactrlsub {\isadigit{1}}\ P\isactrlsub {\isadigit{2}}\ B\isactrlsub {\isadigit{1}}\ B\isactrlsub {\isadigit{2}}\ V\ {\isasymand}\ u\ {\isasymin}\ V\ {\isasymand}\ W\ B\isactrlsub {\isadigit{2}}\ {\isacharplus}{\kern0pt}\ w\ u\ {\isasymle}\ c\ {\isasymlongrightarrow}\isanewline
inv\isactrlsub {\isadigit{1}}\ {\isacharparenleft}{\kern0pt}P\isactrlsub {\isadigit{1}}\ {\isasymunion}\ {\isasymlbrakk}B\isactrlsub {\isadigit{1}}{\isasymrbrakk}{\isacharparenright}{\kern0pt}\ P\isactrlsub {\isadigit{2}}\ {\isasymemptyset}\ {\isacharparenleft}{\kern0pt}B\isactrlsub {\isadigit{2}}\ {\isasymunion}\ {\isacharbraceleft}{\kern0pt}u{\isacharbraceright}{\kern0pt}{\isacharparenright}{\kern0pt}\ {\isacharparenleft}{\kern0pt}V\ {\isacharminus}{\kern0pt}\ {\isacharbraceleft}{\kern0pt}u{\isacharbraceright}{\kern0pt}{\isacharparenright}{\kern0pt}%
\end{isabelle}
\textbf{Case 3} The object fits into neither bin:
\begin{isabelle}%
inv\isactrlsub {\isadigit{1}}\ P\isactrlsub {\isadigit{1}}\ P\isactrlsub {\isadigit{2}}\ B\isactrlsub {\isadigit{1}}\ B\isactrlsub {\isadigit{2}}\ V\ {\isasymand}\ u\ {\isasymin}\ V\ {\isasymlongrightarrow}\isanewline
inv\isactrlsub {\isadigit{1}}\ {\isacharparenleft}{\kern0pt}P\isactrlsub {\isadigit{1}}\ {\isasymunion}\ {\isasymlbrakk}B\isactrlsub {\isadigit{1}}{\isasymrbrakk}{\isacharparenright}{\kern0pt}\ {\isacharparenleft}{\kern0pt}P\isactrlsub {\isadigit{2}}\ {\isasymunion}\ {\isasymlbrakk}B\isactrlsub {\isadigit{2}}{\isasymrbrakk}{\isacharparenright}{\kern0pt}\ {\isasymemptyset}\ {\isacharbraceleft}{\kern0pt}u{\isacharbraceright}{\kern0pt}\ {\isacharparenleft}{\kern0pt}V\ {\isacharminus}{\kern0pt}\ {\isacharbraceleft}{\kern0pt}u{\isacharbraceright}{\kern0pt}{\isacharparenright}{\kern0pt}%
\end{isabelle}
\end{quoteni}
Berghammer and Reuter \cite{BerghammerR03} define the following predicate as their loop invariant:
\begin{quoteni}
\isa{inv\isactrlsub {\isadigit{1}}\ P\isactrlsub {\isadigit{1}}\ P\isactrlsub {\isadigit{2}}\ B\isactrlsub {\isadigit{1}}\ B\isactrlsub {\isadigit{2}}\ V\ {\isacharequal}{\kern0pt}\ bp\ {\isacharparenleft}{\kern0pt}P\isactrlsub {\isadigit{1}}\ {\isasymunion}\ {\isasymlbrakk}B\isactrlsub {\isadigit{1}}{\isasymrbrakk}\ {\isasymunion}\ P\isactrlsub {\isadigit{2}}\ {\isasymunion}\ {\isasymlbrakk}B\isactrlsub {\isadigit{2}}{\isasymrbrakk}\ {\isasymunion}\ {\isacharbraceleft}{\kern0pt}{\isacharbraceleft}{\kern0pt}v{\isacharbraceright}{\kern0pt}\ {\isacharbar}{\kern0pt}\ v\ {\isasymin}\ V{\isacharbraceright}{\kern0pt}{\isacharparenright}{\kern0pt}}
\end{quoteni}
As it turns out, this invariant is too weak. Assume \isa{inv\isactrlsub {\isadigit{1}}\ P\isactrlsub {\isadigit{1}}\ P\isactrlsub {\isadigit{2}}\ B\isactrlsub {\isadigit{1}}\ B\isactrlsub {\isadigit{2}}\ V}.
Suppose \isa{P\isactrlsub {\isadigit{1}}} (alternatively \isa{P\isactrlsub {\isadigit{2}}}) already contains the non-empty bin \isa{B\isactrlsub {\isadigit{1}}}. Note that this does not violate
the invariant because \isa{P\isactrlsub {\isadigit{1}}\ {\isasymunion}\ {\isasymlbrakk}B\isactrlsub {\isadigit{1}}{\isasymrbrakk}\ {\isacharequal}{\kern0pt}\ P\isactrlsub {\isadigit{1}}}. Now, if the algorithm modifies \isa{B\isactrlsub {\isadigit{1}}} by adding an element from \isa{V} such that \isa{B\isactrlsub {\isadigit{1}}} becomes some \isa{B\isactrlsub {\isadigit{1}}{\isacharprime}{\kern0pt}} then \isa{B\isactrlsub {\isadigit{1}}\ {\isasyminter}\ B\isactrlsub {\isadigit{1}}{\isacharprime}{\kern0pt}\ {\isasymnoteq}\ {\isasymemptyset}} and \isa{B\isactrlsub {\isadigit{1}}\ {\isasymin}\ P\isactrlsub {\isadigit{1}}},
i.e., \isa{B\isactrlsub {\isadigit{1}}{\isacharprime}{\kern0pt}} is no longer disjoint from the elements of \isa{P}. The same issue arises with the added object \isa{u\ {\isasymin}\ V}, if \isa{{\isacharbraceleft}{\kern0pt}u{\isacharbraceright}{\kern0pt}} is already in \isa{P\isactrlsub {\isadigit{1}}} or \isa{P\isactrlsub {\isadigit{2}}}.
To account for such cases, we will require additional conjuncts:
\begin{quoteni}
\isa{inv\isactrlsub {\isadigit{1}}\ {\isacharcolon}{\kern0pt}{\isacharcolon}{\kern0pt}\ {\isacharprime}{\kern0pt}a\ set\ set\ {\isasymRightarrow}\ {\isacharprime}{\kern0pt}a\ set\ set\ {\isasymRightarrow}\ {\isacharprime}{\kern0pt}a\ set\ {\isasymRightarrow}\ {\isacharprime}{\kern0pt}a\ set\ {\isasymRightarrow}\ {\isacharprime}{\kern0pt}a\ set\ {\isasymRightarrow}\ bool}\\
\isa{inv\isactrlsub {\isadigit{1}}\ P\isactrlsub {\isadigit{1}}\ P\isactrlsub {\isadigit{2}}\ B\isactrlsub {\isadigit{1}}\ B\isactrlsub {\isadigit{2}}\ V\ {\isacharequal}{\kern0pt}\isanewline
{\isacharparenleft}{\kern0pt}bp\ {\isacharparenleft}{\kern0pt}P\isactrlsub {\isadigit{1}}\ {\isasymunion}\ {\isasymlbrakk}B\isactrlsub {\isadigit{1}}{\isasymrbrakk}\ {\isasymunion}\ P\isactrlsub {\isadigit{2}}\ {\isasymunion}\ {\isasymlbrakk}B\isactrlsub {\isadigit{2}}{\isasymrbrakk}\ {\isasymunion}\ {\isacharbraceleft}{\kern0pt}{\isacharbraceleft}{\kern0pt}v{\isacharbraceright}{\kern0pt}\ {\isacharbar}{\kern0pt}\ v\ {\isasymin}\ V{\isacharbraceright}{\kern0pt}{\isacharparenright}{\kern0pt}\ {\isasymand}\isanewline
\isaindent{{\isacharparenleft}{\kern0pt}}{\isasymUnion}\ {\isacharparenleft}{\kern0pt}P\isactrlsub {\isadigit{1}}\ {\isasymunion}\ {\isasymlbrakk}B\isactrlsub {\isadigit{1}}{\isasymrbrakk}\ {\isasymunion}\ P\isactrlsub {\isadigit{2}}\ {\isasymunion}\ {\isasymlbrakk}B\isactrlsub {\isadigit{2}}{\isasymrbrakk}{\isacharparenright}{\kern0pt}\ {\isacharequal}{\kern0pt}\ U\ {\isacharminus}{\kern0pt}\ V\ {\isasymand}\isanewline
\isaindent{{\isacharparenleft}{\kern0pt}}B\isactrlsub {\isadigit{1}}\ {\isasymnotin}\ P\isactrlsub {\isadigit{1}}\ {\isasymunion}\ P\isactrlsub {\isadigit{2}}\ {\isasymunion}\ {\isasymlbrakk}B\isactrlsub {\isadigit{2}}{\isasymrbrakk}\ {\isasymand}\isanewline
\isaindent{{\isacharparenleft}{\kern0pt}}B\isactrlsub {\isadigit{2}}\ {\isasymnotin}\ P\isactrlsub {\isadigit{1}}\ {\isasymunion}\ {\isasymlbrakk}B\isactrlsub {\isadigit{1}}{\isasymrbrakk}\ {\isasymunion}\ P\isactrlsub {\isadigit{2}}\ {\isasymand}\isanewline
\isaindent{{\isacharparenleft}{\kern0pt}}{\isacharparenleft}{\kern0pt}P\isactrlsub {\isadigit{1}}\ {\isasymunion}\ {\isasymlbrakk}B\isactrlsub {\isadigit{1}}{\isasymrbrakk}{\isacharparenright}{\kern0pt}\ {\isasyminter}\ {\isacharparenleft}{\kern0pt}P\isactrlsub {\isadigit{2}}\ {\isasymunion}\ {\isasymlbrakk}B\isactrlsub {\isadigit{2}}{\isasymrbrakk}{\isacharparenright}{\kern0pt}\ {\isacharequal}{\kern0pt}\ {\isasymemptyset}{\isacharparenright}{\kern0pt}}
\end{quoteni}
There are different ways to strengthen the original \isa{inv\isactrlsub {\isadigit{1}}}. We use the above additional conjuncts as they can be inserted in existing proofs with little modification, and their preservation in the invariant can be proved quite trivially. The first additional conjunct ensures that no element still in \isa{V} is already in a bin or partial solution. The second and third additional conjuncts ensure distinctness of the bins \isa{B\isactrlsub {\isadigit{1}}} and \isa{B\isactrlsub {\isadigit{2}}} with the remaining solution. The final conjunct ensures that the partial solutions with their added bins are disjoint from each other. It should be noted that the last conjunct is \emph{not} necessary to prove functional correctness. It will, however, be needed in later proofs, and as its preservation in this invariant for the simplified algorithm can be used in the proof of the full algorithm, one can save redundant case distinctions by proving it now. Another advantage of proving it now is that later invariants can remain identical to the invariants proposed in the paper.

We now prove the preservation of \isa{inv\isactrlsub {\isadigit{1}}} in all three cases. As we assume the invariant to hold before the execution of the loop body, we can see from the first additional conjunct \isa{{\isasymUnion}\ {\isacharparenleft}{\kern0pt}P\isactrlsub {\isadigit{1}}\ {\isasymunion}\ {\isasymlbrakk}B\isactrlsub {\isadigit{1}}{\isasymrbrakk}\ {\isasymunion}\ P\isactrlsub {\isadigit{2}}\ {\isasymunion}\ {\isasymlbrakk}B\isactrlsub {\isadigit{2}}{\isasymrbrakk}{\isacharparenright}{\kern0pt}\ {\isacharequal}{\kern0pt}\ U\ {\isacharminus}{\kern0pt}\ V} and the assumption \isa{u\ {\isasymin}\ V} that \textit{not-in:} \isa{{\isasymforall}B\ {\isasymin}\ P\isactrlsub {\isadigit{1}}\ {\isasymunion}\ {\isasymlbrakk}B\isactrlsub {\isadigit{1}}{\isasymrbrakk}\ {\isasymunion}\ P\isactrlsub {\isadigit{2}}\ {\isasymunion}\ {\isasymlbrakk}B\isactrlsub {\isadigit{2}}{\isasymrbrakk}{\isachardot}{\kern0pt}\ u\ {\isasymnotin}\ B} holds. This will be needed for all three cases. Now, we can begin with Case~1. We first show
\begin{quote}
\isa{bp\ {\isacharparenleft}{\kern0pt}P\isactrlsub {\isadigit{1}}\ {\isasymunion}\ {\isasymlbrakk}B\isactrlsub {\isadigit{1}}\ {\isasymunion}\ {\isacharbraceleft}{\kern0pt}u{\isacharbraceright}{\kern0pt}{\isasymrbrakk}\ {\isasymunion}\ P\isactrlsub {\isadigit{2}}\ {\isasymunion}\ {\isasymlbrakk}B\isactrlsub {\isadigit{2}}{\isasymrbrakk}\ {\isasymunion}\ {\isacharbraceleft}{\kern0pt}{\isacharbraceleft}{\kern0pt}v{\isacharbraceright}{\kern0pt}\ {\isacharbar}{\kern0pt}\ v\ {\isasymin}\ V\ {\isacharminus}{\kern0pt}\ {\isacharbraceleft}{\kern0pt}u{\isacharbraceright}{\kern0pt}{\isacharbraceright}{\kern0pt}{\isacharparenright}{\kern0pt}}
\end{quote}
One can see that this union does not contain the empty set. The object \isa{u} is now moved from a singleton set into \isa{B\isactrlsub {\isadigit{1}}}. Therefore, the union of all bins will again return \isa{U}. To show that this union remains pairwise disjoint, we can use \textit{not-in} and the second additional conjunct of \isa{inv\isactrlsub {\isadigit{1}}} to show that \isa{u} is not yet contained in the partial solution and \isa{B\isactrlsub {\isadigit{1}}} is \isa{distinct} from any other bin. Therefore, combined with the assumption that the union was pairwise disjoint before the modification, the union remains pairwise disjoint. To prove the preservation of the second conjunct of \isa{bp}, we need to show that the bin weights do not exceed their maximum capacity \isa{c}. The only bin that was changed in this step is \isa{B\isactrlsub {\isadigit{1}}}, which has increased its weight by \isa{w{\isacharparenleft}{\kern0pt}u{\isacharparenright}{\kern0pt}}. As we are in Case~1, we can assume that \isa{u} fits into \isa{B\isactrlsub {\isadigit{1}}}, \isa{W{\isacharparenleft}{\kern0pt}B\isactrlsub {\isadigit{1}}{\isacharparenright}{\kern0pt}\ {\isacharplus}{\kern0pt}\ w{\isacharparenleft}{\kern0pt}u{\isacharparenright}{\kern0pt}\ {\isasymle}\ c}. Therefore, this conjunct holds as well. Now, one only needs to show that the additional conjuncts are preserved. For the first additional conjunct, we can again use \textit{not-in} to show:
\begin{align*}
&& \isa{{\isasymUnion}\ {\isacharparenleft}{\kern0pt}P\isactrlsub {\isadigit{1}}\ {\isasymunion}\ {\isasymlbrakk}B\isactrlsub {\isadigit{1}}\ {\isasymunion}\ {\isacharbraceleft}{\kern0pt}u{\isacharbraceright}{\kern0pt}{\isasymrbrakk}\ {\isasymunion}\ P\isactrlsub {\isadigit{2}}\ {\isasymunion}\ {\isasymlbrakk}B\isactrlsub {\isadigit{2}}{\isasymrbrakk}{\isacharparenright}{\kern0pt}} &= \isa{U\ {\isacharminus}{\kern0pt}\ {\isacharparenleft}{\kern0pt}V\ {\isacharminus}{\kern0pt}\ {\isacharbraceleft}{\kern0pt}u{\isacharbraceright}{\kern0pt}{\isacharparenright}{\kern0pt}}\\
&\iff & \isa{{\isasymUnion}\ {\isacharparenleft}{\kern0pt}P\isactrlsub {\isadigit{1}}\ {\isasymunion}\ {\isasymlbrakk}B\isactrlsub {\isadigit{1}}{\isasymrbrakk}\ {\isasymunion}\ P\isactrlsub {\isadigit{2}}\ {\isasymunion}\ {\isasymlbrakk}B\isactrlsub {\isadigit{2}}{\isasymrbrakk}{\isacharparenright}{\kern0pt}\ {\isasymunion}\ {\isacharbraceleft}{\kern0pt}u{\isacharbraceright}{\kern0pt}} &= \isa{U\ {\isacharminus}{\kern0pt}\ {\isacharparenleft}{\kern0pt}V\ {\isacharminus}{\kern0pt}\ {\isacharbraceleft}{\kern0pt}u{\isacharbraceright}{\kern0pt}{\isacharparenright}{\kern0pt}}\tag*{by \textit{not-in}}\\
&\iff & \isa{{\isasymUnion}\ {\isacharparenleft}{\kern0pt}P\isactrlsub {\isadigit{1}}\ {\isasymunion}\ {\isasymlbrakk}B\isactrlsub {\isadigit{1}}{\isasymrbrakk}\ {\isasymunion}\ P\isactrlsub {\isadigit{2}}\ {\isasymunion}\ {\isasymlbrakk}B\isactrlsub {\isadigit{2}}{\isasymrbrakk}{\isacharparenright}{\kern0pt}\ {\isasymunion}\ {\isacharbraceleft}{\kern0pt}u{\isacharbraceright}{\kern0pt}} &= \isa{U\ {\isacharminus}{\kern0pt}\ V\ {\isasymunion}\ {\isacharbraceleft}{\kern0pt}u{\isacharbraceright}{\kern0pt}}\tag*{by \isa{u\ {\isasymin}\ U}}
\end{align*}
Using the first additional conjunct of the assumed invariant, one can see that this must hold. The remaining conjuncts
\begin{quote}
\isa{B\isactrlsub {\isadigit{1}}\ {\isasymunion}\ {\isacharbraceleft}{\kern0pt}u{\isacharbraceright}{\kern0pt}\ {\isasymnotin}\ P\isactrlsub {\isadigit{1}}\ {\isasymunion}\ P\isactrlsub {\isadigit{2}}\ {\isasymunion}\ {\isasymlbrakk}B\isactrlsub {\isadigit{2}}{\isasymrbrakk}}\\
\isa{B\isactrlsub {\isadigit{2}}\ {\isasymnotin}\ P\isactrlsub {\isadigit{1}}\ {\isasymunion}\ {\isasymlbrakk}B\isactrlsub {\isadigit{1}}\ {\isasymunion}\ {\isacharbraceleft}{\kern0pt}u{\isacharbraceright}{\kern0pt}{\isasymrbrakk}\ {\isasymunion}\ P\isactrlsub {\isadigit{2}}}\\
\isa{{\isacharparenleft}{\kern0pt}P\isactrlsub {\isadigit{1}}\ {\isasymunion}\ {\isasymlbrakk}B\isactrlsub {\isadigit{1}}\ {\isasymunion}\ {\isacharbraceleft}{\kern0pt}u{\isacharbraceright}{\kern0pt}{\isasymrbrakk}{\isacharparenright}{\kern0pt}\ {\isasyminter}\ {\isacharparenleft}{\kern0pt}P\isactrlsub {\isadigit{2}}\ {\isasymunion}\ {\isasymlbrakk}B\isactrlsub {\isadigit{2}}{\isasymrbrakk}{\isacharparenright}{\kern0pt}\ {\isacharequal}{\kern0pt}\ {\isasymemptyset}}
\end{quote}
can be automatically proved in Isabelle using \textit{not-in} and the assumption that the conjuncts of \isa{inv\isactrlsub {\isadigit{1}}\ P\isactrlsub {\isadigit{1}}\ P\isactrlsub {\isadigit{2}}\ B\isactrlsub {\isadigit{1}}\ B\isactrlsub {\isadigit{2}}\ V} held before the modification. The proof for Case~2 is almost identical to that of Case~1. The main difference is that the focus now lies on \isa{B\isactrlsub {\isadigit{2}}} and the fact that \isa{B\isactrlsub {\isadigit{1}}} is now emptied and the previous contents added to the partial solution \isa{P\isactrlsub {\isadigit{1}}}. One therefore has to show that 
\begin{quote}
\isa{bp\ {\isacharparenleft}{\kern0pt}P\isactrlsub {\isadigit{1}}\ {\isasymunion}\ {\isasymlbrakk}B\isactrlsub {\isadigit{1}}{\isasymrbrakk}\ {\isasymunion}\ {\isasymlbrakk}{\isasymemptyset}{\isasymrbrakk}\ {\isasymunion}\ P\isactrlsub {\isadigit{2}}\ {\isasymunion}\ {\isasymlbrakk}B\isactrlsub {\isadigit{2}}\ {\isasymunion}\ {\isacharbraceleft}{\kern0pt}u{\isacharbraceright}{\kern0pt}{\isasymrbrakk}\ {\isasymunion}\ {\isacharbraceleft}{\kern0pt}{\isacharbraceleft}{\kern0pt}v{\isacharbraceright}{\kern0pt}\ {\isacharbar}{\kern0pt}\ v\ {\isasymin}\ V\ {\isacharminus}{\kern0pt}\ {\isacharbraceleft}{\kern0pt}u{\isacharbraceright}{\kern0pt}{\isacharbraceright}{\kern0pt}{\isacharparenright}{\kern0pt}}
\end{quote}
holds. As \isa{{\isasymlbrakk}{\isasymemptyset}{\isasymrbrakk}} can be ignored, one can see that the act of emptying \isa{B\isactrlsub {\isadigit{1}}} and adding it to the partial solution will not otherwise affect the proof. The proof of \isa{bp} in Case~3 is trivial, as the modifications made in this step can simply be undone by applying the following steps:
\begin{align*}
&  \isa{P\isactrlsub {\isadigit{1}}\ {\isasymunion}\ {\isasymlbrakk}B\isactrlsub {\isadigit{1}}{\isasymrbrakk}\ {\isasymunion}\ {\isasymlbrakk}{\isasymemptyset}{\isasymrbrakk}\ {\isasymunion}\ {\isacharparenleft}{\kern0pt}P\isactrlsub {\isadigit{2}}\ {\isasymunion}\ {\isasymlbrakk}B\isactrlsub {\isadigit{2}}{\isasymrbrakk}{\isacharparenright}{\kern0pt}\ {\isasymunion}\ {\isasymlbrakk}{\isacharbraceleft}{\kern0pt}u{\isacharbraceright}{\kern0pt}{\isasymrbrakk}\ {\isasymunion}\ {\isacharbraceleft}{\kern0pt}{\isacharbraceleft}{\kern0pt}v{\isacharbraceright}{\kern0pt}\ {\isacharbar}{\kern0pt}\ v\ {\isasymin}\ V\ {\isacharminus}{\kern0pt}\ {\isacharbraceleft}{\kern0pt}u{\isacharbraceright}{\kern0pt}{\isacharbraceright}{\kern0pt}}\\
&= \isa{P\isactrlsub {\isadigit{1}}\ {\isasymunion}\ {\isasymlbrakk}B\isactrlsub {\isadigit{1}}{\isasymrbrakk}\ {\isasymunion}\ P\isactrlsub {\isadigit{2}}\ {\isasymunion}\ {\isasymlbrakk}B\isactrlsub {\isadigit{2}}{\isasymrbrakk}\ {\isasymunion}\ {\isacharbraceleft}{\kern0pt}{\isacharbraceleft}{\kern0pt}u{\isacharbraceright}{\kern0pt}{\isacharbraceright}{\kern0pt}\ {\isasymunion}\ {\isacharbraceleft}{\kern0pt}{\isacharbraceleft}{\kern0pt}v{\isacharbraceright}{\kern0pt}\ {\isacharbar}{\kern0pt}\ v\ {\isasymin}\ V\ {\isacharminus}{\kern0pt}\ {\isacharbraceleft}{\kern0pt}u{\isacharbraceright}{\kern0pt}{\isacharbraceright}{\kern0pt}}\tag*{by \isa{{\isasymlbrakk}{\isasymcdot}{\isasymrbrakk}}-def}\\
&= \isa{P\isactrlsub {\isadigit{1}}\ {\isasymunion}\ {\isasymlbrakk}B\isactrlsub {\isadigit{1}}{\isasymrbrakk}\ {\isasymunion}\ P\isactrlsub {\isadigit{2}}\ {\isasymunion}\ {\isasymlbrakk}B\isactrlsub {\isadigit{2}}{\isasymrbrakk}\ {\isasymunion}\ {\isacharbraceleft}{\kern0pt}{\isacharbraceleft}{\kern0pt}v{\isacharbraceright}{\kern0pt}\ {\isacharbar}{\kern0pt}\ v\ {\isasymin}\ V{\isacharbraceright}{\kern0pt}}\tag*{by \isa{u\ {\isasymin}\ V}}
\end{align*}
Now, one only needs to show that the remaining additional conjuncts hold. This can again be shown automatically using \textit{not-in} and the fact that \isa{inv\isactrlsub {\isadigit{1}}\ P\isactrlsub {\isadigit{1}}\ P\isactrlsub {\isadigit{2}}\ B\isactrlsub {\isadigit{1}}\ B\isactrlsub {\isadigit{2}}\ V} held before the modifications. Therefore, \isa{inv\isactrlsub {\isadigit{1}}} is preserved in all three cases.

To prove the approximation factor, we proceed as in \cite{BerghammerR03} and establish suitable lower bounds. The first lower bound
\begin{lem}\label{lem:L_lower_bound_card}
\begin{isabelle}%
bp\ P\ {\isasymlongrightarrow}\ {\isacharbar}{\kern0pt}L{\isacharbar}{\kern0pt}\ {\isasymle}\ {\isacharbar}{\kern0pt}P{\isacharbar}{\kern0pt}%
\end{isabelle}
\end{lem}
\noindent
holds because a bin can only contain at most one large object, and every large object needs to be in the solution. To prove this in Isabelle, we first make the observation that for every large object there exists a bin in \isa{P} in which it is contained. Therefore, we may obtain a function \isa{f} that returns this bin for every \isa{u\ {\isasymin}\ L}. Using the fact that any bin can hold at most one large object, we can show that this function has to be injective, as every large object must map to a unique bin. Hence, the number of large objects is equal to the number of bins \isa{f} maps to. Moreover, the image of \isa{f} has to be a subset of \isa{P}. Thus, the number of large objects has to be a lower bound on the number of bins in \isa{P}. 

As it turns out, the algorithm will ensure that there is always at least one large object in a bin for the first partial solution as long as large objects are available. This means we can assume that
\isa{V\ {\isasyminter}\ L\ {\isasymnoteq}\ {\isasymemptyset}\ {\isasymlongrightarrow}\ {\isacharparenleft}{\kern0pt}{\isasymforall}B{\isasymin}P\isactrlsub {\isadigit{1}}\ {\isasymunion}\ {\isasymlbrakk}B\isactrlsub {\isadigit{1}}{\isasymrbrakk}{\isachardot}{\kern0pt}\ B\ {\isasyminter}\ L\ {\isasymnoteq}\ {\isasymemptyset}{\isacharparenright}{\kern0pt}}.
Therefore, we can use the previous lower bound to show the following:
\begin{lem}\label{lem:L_bins_lower_bound_card}
\begin{isabelle}%
bp\ P\ {\isasymand}\ inv\isactrlsub {\isadigit{1}}\ P\isactrlsub {\isadigit{1}}\ P\isactrlsub {\isadigit{2}}\ B\isactrlsub {\isadigit{1}}\ B\isactrlsub {\isadigit{2}}\ V\ {\isasymand}\ {\isacharparenleft}{\kern0pt}{\isasymforall}B{\isasymin}P\isactrlsub {\isadigit{1}}\ {\isasymunion}\ {\isasymlbrakk}B\isactrlsub {\isadigit{1}}{\isasymrbrakk}{\isachardot}{\kern0pt}\ B\ {\isasyminter}\ L\ {\isasymnoteq}\ {\isasymemptyset}{\isacharparenright}{\kern0pt}\ {\isasymlongrightarrow}\isanewline
{\isacharbar}{\kern0pt}P\isactrlsub {\isadigit{1}}\ {\isasymunion}\ {\isasymlbrakk}B\isactrlsub {\isadigit{1}}{\isasymrbrakk}\ {\isasymunion}\ {\isacharbraceleft}{\kern0pt}{\isacharbraceleft}{\kern0pt}v{\isacharbraceright}{\kern0pt}\ {\isacharbar}{\kern0pt}\ v\ {\isasymin}\ V\ {\isasyminter}\ L{\isacharbraceright}{\kern0pt}{\isacharbar}{\kern0pt}\ {\isasymle}\ {\isacharbar}{\kern0pt}P{\isacharbar}{\kern0pt}%
\end{isabelle}
\end{lem}

Another easy lower bound is this one:
\begin{lem}\label{lem:sum_lower_bound_card}
\begin{quoteni}
\isa{bp\ P\ {\isasymlongrightarrow}\ {\isacharparenleft}{\kern0pt}}$\sum_{u \in U} w\,u$\isa{{\isacharparenright}{\kern0pt}\ {\isasymle}\ c\ {\isacharasterisk}{\kern0pt}\ {\isacharbar}{\kern0pt}P{\isacharbar}{\kern0pt}}
\end{quoteni}
\end{lem}

The next lower bound arises from the fact that an object is only ever put into \isa{B\isactrlsub {\isadigit{2}}}, and therefore \isa{P\isactrlsub {\isadigit{2}}}, if it would have caused \isa{B\isactrlsub {\isadigit{1}}} to overflow. As a result of this, we can define a bijective function \isa{f} that maps every bin of \isa{P\isactrlsub {\isadigit{1}}} to the object in \isa{P\isactrlsub {\isadigit{2}}\ {\isasymunion}\ {\isasymlbrakk}B\isactrlsub {\isadigit{2}}{\isasymrbrakk}} that would have caused the bin to overflow. We define:
\begin{quoteni}
\isa{bij{\isacharunderscore}{\kern0pt}exists\ {\isacharcolon}{\kern0pt}{\isacharcolon}{\kern0pt}\ {\isacharprime}{\kern0pt}a\ set\ set\ {\isasymRightarrow}\ {\isacharprime}{\kern0pt}a\ set\ {\isasymRightarrow}\ bool}\\
\isa{bij{\isacharunderscore}{\kern0pt}exists\ P\ V\ {\isacharequal}{\kern0pt}\ {\isacharparenleft}{\kern0pt}{\isasymexists}f{\isachardot}{\kern0pt}\ bij{\isacharunderscore}{\kern0pt}betw\ f\ P\ V\ {\isasymand}\ {\isacharparenleft}{\kern0pt}{\isasymforall}B{\isasymin}P{\isachardot}{\kern0pt}\ c\ {\isacharless}{\kern0pt}\ W\ B\ {\isacharplus}{\kern0pt}\ w\ {\isacharparenleft}{\kern0pt}f\ B{\isacharparenright}{\kern0pt}{\isacharparenright}{\kern0pt}{\isacharparenright}{\kern0pt}}
\end{quoteni}
From this, we can make the observation that the number of bins in \isa{P\isactrlsub {\isadigit{1}}} is a \emph{strict} lower bound on the number of bins of any correct bin packing \isa{P}:
\begin{lem}\label{lem:P1_lower_bound_card}
\begin{isabelle}%
bp\ P\ {\isasymand}\ inv\isactrlsub {\isadigit{1}}\ P\isactrlsub {\isadigit{1}}\ P\isactrlsub {\isadigit{2}}\ B\isactrlsub {\isadigit{1}}\ B\isactrlsub {\isadigit{2}}\ V\ {\isasymand}\ bij{\isacharunderscore}{\kern0pt}exists\ P\isactrlsub {\isadigit{1}}\ {\isacharparenleft}{\kern0pt}{\isasymUnion}\ {\isacharparenleft}{\kern0pt}P\isactrlsub {\isadigit{2}}\ {\isasymunion}\ {\isasymlbrakk}B\isactrlsub {\isadigit{2}}{\isasymrbrakk}{\isacharparenright}{\kern0pt}{\isacharparenright}{\kern0pt}\ {\isasymlongrightarrow}\isanewline
{\isacharbar}{\kern0pt}P\isactrlsub {\isadigit{1}}{\isacharbar}{\kern0pt}\ {\isacharplus}{\kern0pt}\ {\isadigit{1}}\ {\isasymle}\ {\isacharbar}{\kern0pt}P{\isacharbar}{\kern0pt}%
\end{isabelle}
\end{lem}
Unlike the proof outlined in \cite{BerghammerR03}, we begin with a case distinction on \isa{P\isactrlsub {\isadigit{1}}}. The reasoning behind this is that if \isa{P\isactrlsub {\isadigit{1}}} is empty, the strict nature of the lower bound cannot be shown from the calculation that Berghammer and Reuter make. Therefore, we consider the case where \isa{P\isactrlsub {\isadigit{1}}} is empty separately. If \isa{P\isactrlsub {\isadigit{1}}} is empty, our goal is to prove that 1 is a lower bound on the number of bins in \isa{P}. This follows from the fact that \isa{U} is non-empty, and therefore any correct bin packing must contain at least one bin. For the other case, we may now assume that \isa{P\isactrlsub {\isadigit{1}}} is non-empty. In the following proof, we will need the final conjunct of \isa{inv\isactrlsub {\isadigit{1}}}, $\isa{{\isacharparenleft}{\kern0pt}P\isactrlsub {\isadigit{1}}\ {\isasymunion}\ {\isasymlbrakk}B\isactrlsub {\isadigit{1}}{\isasymrbrakk}{\isacharparenright}{\kern0pt}\ {\isasyminter}\ {\isacharparenleft}{\kern0pt}P\isactrlsub {\isadigit{2}}\ {\isasymunion}\ {\isasymlbrakk}B\isactrlsub {\isadigit{2}}{\isasymrbrakk}{\isacharparenright}{\kern0pt}\ {\isacharequal}{\kern0pt}\ {\isasymemptyset}}$, which we can transform into \textit{disjoint}: $\isa{P\isactrlsub {\isadigit{1}}\ {\isasyminter}\ {\isacharparenleft}{\kern0pt}P\isactrlsub {\isadigit{2}}\ {\isasymunion}\ {\isasymlbrakk}B\isactrlsub {\isadigit{2}}{\isasymrbrakk}{\isacharparenright}{\kern0pt}\ {\isacharequal}{\kern0pt}\ {\isasymemptyset}}$. We also obtain the bijective function \isa{f} and observe that, as the object obtained from \isa{f} for a bin \isa{B\ {\isasymin}\ P\isactrlsub {\isadigit{1}}} caused \isa{B} to exceed its capacity, \textit{exceed}: \isa{c\ {\isacharless}{\kern0pt}\ W{\isacharparenleft}{\kern0pt}B{\isacharparenright}{\kern0pt}\ {\isacharplus}{\kern0pt}\ w{\isacharparenleft}{\kern0pt}f{\isacharparenleft}{\kern0pt}B{\isacharparenright}{\kern0pt}{\isacharparenright}{\kern0pt}} must hold. We can now perform the following calculation:
\begin{align*}
c|P_1| &=   \sum_{B \in P_1} c\\
          &<   \sum_{B \in P_1}  W(B) + \sum_{B \in P_1} w(f(B))\tag*{by \isa{P\isactrlsub {\isadigit{1}}\ {\isasymnoteq}\ {\isasymemptyset}} and \textit{exceed}}\\
          &=   \sum_{B \in P_1}  W(B) + \sum_{B \in \isa{P\isactrlsub {\isadigit{2}}\ {\isasymunion}\ {\isasymlbrakk}B\isactrlsub {\isadigit{2}}{\isasymrbrakk}}} W(B)\tag*{by \isa{f} bijective}\\
          &=   \sum_{B \in \isa{P\isactrlsub {\isadigit{1}}\ {\isasymunion}\ P\isactrlsub {\isadigit{2}}\ {\isasymunion}\ {\isasymlbrakk}B\isactrlsub {\isadigit{2}}{\isasymrbrakk}}} W(B)\tag*{by \textit{disjoint}}\\
          &\le \sum_{u \in U} w(u) \le c|P|\tag*{by \isa{inv\isactrlsub {\isadigit{1}}} and Lemma~\ref{lem:sum_lower_bound_card}}
\end{align*}
Therefore \isa{{\isacharbar}{\kern0pt}P\isactrlsub {\isadigit{1}}{\isacharbar}{\kern0pt}\ {\isacharless}{\kern0pt}\ {\isacharbar}{\kern0pt}P{\isacharbar}{\kern0pt}} and, by extension, \isa{{\isacharbar}{\kern0pt}P\isactrlsub {\isadigit{1}}{\isacharbar}{\kern0pt}\ {\isacharplus}{\kern0pt}\ {\isadigit{1}}\ {\isasymle}\ {\isacharbar}{\kern0pt}P{\isacharbar}{\kern0pt}}.

We only sketch the rest of the proof because it is almost identical to that in \cite{BerghammerR03}. First we need two extensions of \isa{inv\isactrlsub {\isadigit{1}}} to show the approximation ratio:
\begin{quoteni}
\isa{inv\isactrlsub {\isadigit{2}}\ P\isactrlsub {\isadigit{1}}\ P\isactrlsub {\isadigit{2}}\ B\isactrlsub {\isadigit{1}}\ B\isactrlsub {\isadigit{2}}\ V\ {\isacharequal}{\kern0pt}\isanewline
{\isacharparenleft}{\kern0pt}inv\isactrlsub {\isadigit{1}}\ P\isactrlsub {\isadigit{1}}\ P\isactrlsub {\isadigit{2}}\ B\isactrlsub {\isadigit{1}}\ B\isactrlsub {\isadigit{2}}\ V\ {\isasymand}\isanewline
\isaindent{{\isacharparenleft}{\kern0pt}}{\isacharparenleft}{\kern0pt}V\ {\isasyminter}\ L\ {\isasymnoteq}\ {\isasymemptyset}\ {\isasymlongrightarrow}\ {\isacharparenleft}{\kern0pt}{\isasymforall}B{\isasymin}P\isactrlsub {\isadigit{1}}\ {\isasymunion}\ {\isasymlbrakk}B\isactrlsub {\isadigit{1}}{\isasymrbrakk}{\isachardot}{\kern0pt}\ B\ {\isasyminter}\ L\ {\isasymnoteq}\ {\isasymemptyset}{\isacharparenright}{\kern0pt}{\isacharparenright}{\kern0pt}\ {\isasymand}\isanewline
\isaindent{{\isacharparenleft}{\kern0pt}}bij{\isacharunderscore}{\kern0pt}exists\ P\isactrlsub {\isadigit{1}}\ {\isacharparenleft}{\kern0pt}{\isasymUnion}\ {\isacharparenleft}{\kern0pt}P\isactrlsub {\isadigit{2}}\ {\isasymunion}\ {\isasymlbrakk}B\isactrlsub {\isadigit{2}}{\isasymrbrakk}{\isacharparenright}{\kern0pt}{\isacharparenright}{\kern0pt}\ {\isasymand}\ {\isadigit{2}}\ {\isacharasterisk}{\kern0pt}\ {\isacharbar}{\kern0pt}P\isactrlsub {\isadigit{2}}{\isacharbar}{\kern0pt}\ {\isasymle}\ {\isacharbar}{\kern0pt}{\isasymUnion}\ P\isactrlsub {\isadigit{2}}{\isacharbar}{\kern0pt}{\isacharparenright}{\kern0pt}}\smallskip\\
\isa{inv\isactrlsub {\isadigit{3}}\ P\isactrlsub {\isadigit{1}}\ P\isactrlsub {\isadigit{2}}\ B\isactrlsub {\isadigit{1}}\ B\isactrlsub {\isadigit{2}}\ V\ {\isacharequal}{\kern0pt}\ {\isacharparenleft}{\kern0pt}inv\isactrlsub {\isadigit{2}}\ P\isactrlsub {\isadigit{1}}\ P\isactrlsub {\isadigit{2}}\ B\isactrlsub {\isadigit{1}}\ B\isactrlsub {\isadigit{2}}\ V\ {\isasymand}\ B\isactrlsub {\isadigit{2}}\ {\isasymsubseteq}\ S{\isacharparenright}{\kern0pt}}
\end{quoteni}
The motivation for the last conjunct in \isa{inv\isactrlsub {\isadigit{2}}} is the following lower bound:
\begin{quoteni}
\isa{inv\isactrlsub {\isadigit{1}}\ P\isactrlsub {\isadigit{1}}\ P\isactrlsub {\isadigit{2}}\ B\isactrlsub {\isadigit{1}}\ B\isactrlsub {\isadigit{2}}\ V\ {\isasymand}\ {\isadigit{2}}\ {\isacharasterisk}{\kern0pt}\ {\isacharbar}{\kern0pt}P\isactrlsub {\isadigit{2}}{\isacharbar}{\kern0pt}\ {\isasymle}\ {\isacharbar}{\kern0pt}{\isasymUnion}\ P\isactrlsub {\isadigit{2}}{\isacharbar}{\kern0pt}\ {\isasymand}\ bij{\isacharunderscore}{\kern0pt}exists\ P\isactrlsub {\isadigit{1}}\ {\isacharparenleft}{\kern0pt}{\isasymUnion}\ {\isacharparenleft}{\kern0pt}P\isactrlsub {\isadigit{2}}\ {\isasymunion}\ {\isasymlbrakk}B\isactrlsub {\isadigit{2}}{\isasymrbrakk}{\isacharparenright}{\kern0pt}{\isacharparenright}{\kern0pt}\ {\isasymlongrightarrow}\isanewline
{\isadigit{2}}\ {\isacharasterisk}{\kern0pt}\ {\isacharbar}{\kern0pt}P\isactrlsub {\isadigit{2}}\ {\isasymunion}\ {\isasymlbrakk}B\isactrlsub {\isadigit{2}}{\isasymrbrakk}{\isacharbar}{\kern0pt}\ {\isasymle}\ {\isacharbar}{\kern0pt}P\isactrlsub {\isadigit{1}}{\isacharbar}{\kern0pt}\ {\isacharplus}{\kern0pt}\ {\isadigit{1}}}
\end{quoteni}

The main lower bound lemma (Theorem 4.1 in \cite{BerghammerR03}) is the following:
\begin{lem}
\isa{V\ {\isasyminter}\ S\ {\isacharequal}{\kern0pt}\ {\isasymemptyset}\ {\isasymand}\ inv\isactrlsub {\isadigit{2}}\ P\isactrlsub {\isadigit{1}}\ P\isactrlsub {\isadigit{2}}\ B\isactrlsub {\isadigit{1}}\ B\isactrlsub {\isadigit{2}}\ V\ {\isasymand}\ bp\ P\ {\isasymlongrightarrow}\isanewline
{\isacharbar}{\kern0pt}P\isactrlsub {\isadigit{1}}\ {\isasymunion}\ {\isasymlbrakk}B\isactrlsub {\isadigit{1}}{\isasymrbrakk}\ {\isasymunion}\ P\isactrlsub {\isadigit{2}}\ {\isasymunion}\ {\isasymlbrakk}B\isactrlsub {\isadigit{2}}{\isasymrbrakk}\ {\isasymunion}\ {\isacharbraceleft}{\kern0pt}{\isacharbraceleft}{\kern0pt}v{\isacharbraceright}{\kern0pt}\ {\isacharbar}{\kern0pt}\ v\ {\isasymin}\ V{\isacharbraceright}{\kern0pt}{\isacharbar}{\kern0pt}\ {\isasymle}\ {\isadigit{3}}\ {\isacharslash}{\kern0pt}\ {\isadigit{2}}\ {\isacharasterisk}{\kern0pt}\ {\isacharbar}{\kern0pt}P{\isacharbar}{\kern0pt}}
\end{lem}
From this lower bound the postcondition of the algorithm follows easily under
the assumption that \isa{inv\isactrlsub {\isadigit{2}}} holds at the end of the loop. This in turn follows
because \isa{inv\isactrlsub {\isadigit{3}}} can be shown to be a loop invariant.%
\end{isamarkuptext}\isamarkuptrue%
\isadelimtheory
\endisadelimtheory
\isatagtheory
\endisatagtheory
{\isafoldtheory}%
\isadelimtheory
\endisadelimtheory
\end{isabellebody}%

%% file: root.bbl
\begin{thebibliography}{ENRS20}

\bibitem[BHS16]{berghammer2016relational}
Rudolf Berghammer, Peter Höfner, and Insa Stucke.
\newblock Cardinality of relations and relational approximation algorithms.
\newblock {\em Journal of Logical and Algebraic Methods in Programming},
  85(2):269--286, 2016.

\bibitem[BHT19]{DBLP:conf/frocos/BotteschHT19}
Ralph Bottesch, Max~W. Haslbeck, and Ren{\'{e}} Thiemann.
\newblock Verifying an incremental theory solver for linear arithmetic in
  {Isabelle/HOL}.
\newblock In Andreas Herzig and Andrei Popescu, editors, {\em Frontiers of
  Combining Systems, FroCoS 2019}, volume 11715 of {\em Lecture Notes in
  Computer Science}, pages 223--239. Springer, 2019.
\newblock \href {https://doi.org/10.1007/978-3-030-29007-8\_13}
  {\path{doi:10.1007/978-3-030-29007-8\_13}}.

\bibitem[BK10]{BansalK10}
Nikhil Bansal and Subhash Khot.
\newblock Inapproximability of hypergraph vertex cover and applications to
  scheduling problems.
\newblock In Samson Abramsky, Cyril Gavoille, Claude Kirchner, Friedhelm {Meyer
  auf der Heide}, and Paul~G. Spirakis, editors, {\em Automata, Languages and
  Programming, {ICALP} 2010, Part {I}}, volume 6198 of {\em LNCS}, pages
  250--261. Springer, 2010.
\newblock \href {https://doi.org/10.1007/978-3-642-14165-2\_22}
  {\path{doi:10.1007/978-3-642-14165-2\_22}}.

\bibitem[BM03]{BerghammerM03}
Rudolf Berghammer and Markus M{\"{u}}ller{-}Olm.
\newblock Formal development and verification of approximation algorithms using
  auxiliary variables.
\newblock In Maurice Bruynooghe, editor, {\em Logic Based Program Synthesis and
  Transformation, {LOPSTR} 2003}, volume 3018 of {\em LNCS}, pages 59--74.
  Springer, 2003.
\newblock \href {https://doi.org/10.1007/978-3-540-25938-1\_6}
  {\path{doi:10.1007/978-3-540-25938-1\_6}}.

\bibitem[BR03]{BerghammerR03}
Rudolf Berghammer and Florian Reuter.
\newblock A linear approximation algorithm for bin packing with absolute
  approximation factor 3/2.
\newblock {\em Sci. Comput. Program.}, 48(1):67--80, 2003.
\newblock \href {https://doi.org/10.1016/S0167-6423(03)00011-X}
  {\path{doi:10.1016/S0167-6423(03)00011-X}}.

\bibitem[ENR20]{ijcar/EssmannNR20}
Robin E{\ss}mann, Tobias Nipkow, and Simon Robillard.
\newblock Verified approximation algorithms.
\newblock In Nicolas Peltier and Viorica Sofronie{-}Stokkermans, editors, {\em
  Automated Reasoning, {IJCAR} 2020, Part {II}}, volume 12167 of {\em Lecture
  Notes in Computer Science}, pages 291--306. Springer, 2020.
\newblock \href {https://doi.org/10.1007/978-3-030-51054-1\_17}
  {\path{doi:10.1007/978-3-030-51054-1\_17}}.

\bibitem[ENRS20]{Approximation_Algorithms-AFP}
Robin Eßmann, Tobias Nipkow, Simon Robillard, and Ujkan Sulejmani.
\newblock Verified approximation algorithms.
\newblock {\em Archive of Formal Proofs}, January 2020.
\newblock \url{http://isa-afp.org/entries/Approximation_Algorithms.html},
  Formal proof development.

\bibitem[HR97]{halldorsson1997greed}
Magn{\'u}s~M Halld{\'o}rsson and Jaikumar Radhakrishnan.
\newblock Greed is good: Approximating independent sets in sparse and
  bounded-degree graphs.
\newblock {\em Algorithmica}, 18(1):145--163, 1997.

\bibitem[KT06]{KleinbergT06}
Jon~M. Kleinberg and {\'{E}}va Tardos.
\newblock {\em Algorithm Design}.
\newblock Addison-Wesley, 2006.

\bibitem[NK14]{Concrete}
Tobias Nipkow and Gerwin Klein.
\newblock {\em Concrete Semantics with Isabelle/HOL}.
\newblock Springer, 2014.
\newblock \url{http://concrete-semantics.org}.

\bibitem[NPW02]{LNCS2283}
Tobias Nipkow, Lawrence Paulson, and Markus Wenzel.
\newblock {\em Isabelle/HOL --- A Proof Assistant for Higher-Order Logic},
  volume 2283 of {\em LNCS}.
\newblock Springer, 2002.

\bibitem[PK19]{Linear_Programming-AFP}
Julian Parsert and Cezary Kaliszyk.
\newblock Linear programming.
\newblock {\em Archive of Formal Proofs}, August 2019.
\newblock \url{http://isa-afp.org/entries/Linear_Programming.html}, Formal
  proof development.

\bibitem[Vaz03]{Vazirani}
Vijai Vazirani.
\newblock {\em Approximation Algorithms}.
\newblock Springer, 2003.

\bibitem[Wei81]{Wei}
V.K. Wei.
\newblock A lower bound for the stability number of a simple graph.
\newblock Technical Memorandum 81-11217-9, Bell Laboratories, 1981.

\end{thebibliography}
